\documentclass[floats,floatfix,amssymb,prd,twocolumn,superscriptaddress,nofootinbib,longbibliography]{revtex4-1}

\AtBeginDocument{%
    \newwrite\bibnotes
    \def\bibnotesext{Notes.bib}
    \immediate\openout\bibnotes=\jobname\bibnotesext
    \immediate\write\bibnotes{@CONTROL{REVTEX41Control}}
    \immediate\write\bibnotes{@CONTROL{%
    apsrev41Control,author="08",editor="1",pages="1",title="0",year="1"}}
     \if@filesw
     \immediate\write\@auxout{\string\citation{apsrev41Control}}%
    \fi
}%

\usepackage{graphicx,amssymb,amsmath,amsthm,amsfonts,epsfig,epsf}
\usepackage{float} % Para a opção "H" nas imagens
\usepackage[colorlinks=true,citecolor=purple,linktocpage]{hyperref}
\usepackage[usenames]{xcolor}
\usepackage{epstopdf}
\usepackage{bm}
\usepackage{dcolumn}
\usepackage{rotating}
\usepackage{hyperref}
\usepackage{xcolor}
\usepackage{longtable}

\usepackage{enumerate}
\usepackage{tensor,multirow}
\usepackage{url}
\usepackage{physunits} % to write unit names
\usepackage{mathrsfs} % to use \mathscr
\usepackage[normalem]{ulem}
\usepackage[export]{adjustbox}

\allowdisplaybreaks % para equações que não cabem numa página

\newcommand{\Mbh}{M_{\text{\tiny BH}}}
\newcommand{\rbh}{r_{\text{\tiny BH}}}
\newcommand{\Omegaorb}{\Omega_{\text{orb}}}

\usepackage{pifont}% para os checks e cruzes
\newcommand{\cmark}{\ding{51}}%
\newcommand{\xmark}{\ding{55}}%

\usepackage{xparse}
\ExplSyntaxOn
\NewDocumentCommand{\MeijerG}{smmmm}
 {
  \IfBooleanTF{#1}
   {
    \vic_meijerg:nnnnnn { #2 } { #3 } { #4 } { #5 } { small } { }
   }
   {
    \vic_meijerg:nnnnnn { #2 } { #3 } { #4 } { #5 } { } { \; }
   }
 }
\seq_new:N \l__vic_meijerg_args_in_seq
\seq_new:N \l__vic_meijerg_args_out_seq
\cs_new_protected:Nn \vic_meijerg:nnnnnn
 {
  \seq_set_split:Nnn \l__vic_meijerg_args_in_seq { | } { #3 }
  \seq_clear:N \l__vic_meijerg_args_out_seq  
  \seq_map_inline:Nn \l__vic_meijerg_args_in_seq
   {
    \seq_put_right:Nn \l__vic_meijerg_args_out_seq
     {
      \begin{#5matrix} ##1 \end{#5matrix}
     }
   }
  G\sp{#1}\sb{#2}
  \left(
  \seq_use:Nn \l__vic_meijerg_args_out_seq { #6\middle|#6 }
  #6\middle|#6
  #4
  \right)
 }
\ExplSyntaxOff

\begin{document}

\title{Tidal Love numbers of gravitational atoms}

\author{Ricardo Arana}
\affiliation{CENTRA, Departamento de F\'{\i}sica, Instituto Superior T\'ecnico -- IST, Universidade de Lisboa -- UL, Avenida Rovisco Pais 1, 1049-001 Lisboa, Portugal}
\author{Richard Brito}
\affiliation{CENTRA, Departamento de F\'{\i}sica, Instituto Superior T\'ecnico -- IST, Universidade de Lisboa -- UL, Avenida Rovisco Pais 1, 1049-001 Lisboa, Portugal}
\author{Gonçalo Castro}
\affiliation{CENTRA, Departamento de F\'{\i}sica, Instituto Superior T\'ecnico -- IST, Universidade de Lisboa -- UL, Avenida Rovisco Pais 1, 1049-001 Lisboa, Portugal}

\begin{abstract}
Ultralight bosonic fields can form condensates, or clouds, around spinning black holes. When this system is under the influence of a secondary massive body, its tidal response can be quantified in the tidal Love numbers (TLNs). Although TLNs vanish for black holes in vacuum, it has been shown that the same is not true for black holes immersed in matter environments. In this work, we compute the gravitational TLNs of black holes surrounded by scalar clouds, in the Newtonian limit. We show that they are nonvanishing, have a strong power-law dependence on the boson's mass, and are proportional to the scalar cloud's total mass. In particular, we find that, independently of the cloud's configuration, the TLNs from axisymmetric tides scale as $\propto r_c^{2l+1}$, for $r_c$ the cloud's ``radius'' and $l$ the  multipole order of the external tidal field. This differs by a factor $r_c$ from previous estimates based on scalar and vector tidal perturbations but is in perfect agreement with the behavior of TLNs in other matter systems. Furthermore, we show that the adiabatic tides approximation we employ is, in general, not appropriate for nonaxisymmetric tidal interactions.
\end{abstract}

\date{\today}

\maketitle
%%%%%%&&&&&&&
\section{Introduction}
\label{section:Introduction}
%%%%%%&&&&&&&
The quantities known as tidal Love numbers (TLNs) provide information as to how a self-gravitating object is deformed under the gravitational influence of another~\cite{Love:1911,Poisson:2014}. Crucially, they depend on the internal composition of the object that is being deformed and introduce corrections in the gravitational waveform emitted by a coalescing binary, in the late inspiral phase~\cite{Hinderer:2007,Flanagan:2007}. Extracting the TLNs from gravitational-wave (GWs) observations therefore provides a way to probe the nature of the objects in compact binary systems, the most notable example being the possibility to probe the equation of state of neutron stars from such observations (see Ref.~\cite{Chatziioannou:2020pqz} for a recent review).

Within this context, a remarkable property of vacuum black holes (BHs) in general relativity is the fact that their TLNs are zero in an asymptotically flat spacetime~\cite{Fang:2005,Binnington:2009,Pani:2015hfa,Landry:2015zfa,Chia:2020,Charalambous:2021}.\footnote{Note that it was recently shown that BHs in an asymptotically de Sitter geometry have nonzero TLNs~\cite{Nair:2024}. However, the corrections to the TLNs due to the nonzero cosmological constant measured in our Universe are extremely small for astrophysical BHs and unlikely to be measurable~\cite{Nair:2024}.} The measurement of a nonzero TLN in a dark compact object above the neutron star mass range therefore necessarily implies one of three possibilities: (i) the object is not a BH but rather some exotic compact object~\cite{Mendes:2016vdr,Cardoso:2017,Maselli:2017cmm,Sennett:2017etc,Cardoso:2019rvt,Herdeiro:2020kba}; (ii) General Relativity is not the correct description of gravity in the strong-field regime~\cite{Cardoso:2017,Cardoso:2018ptl,DeLuca:2022tkm}; (iii) the assumption that the BH can be taken to be living in vacuum is not valid~\cite{Baumann:2018,Duque:2019,DeLuca:2021,DeLuca:2022xlz,Katagiri:2023yzm,Cannizzaro:2024fpz}.  

In this work we focus on the third possibility, by considering the specific case of BHs surrounded by ultralight scalar fields, which may condense either through accretion~\cite{Barranco:2012qs,Cardoso:2022vpj,Cardoso:2022nzc} or through superradiant instabilities~\cite{Arvanitaki:2009fg,Brito:2014wla,East:2017,East:2018glu} (for an extended review see Ref.~\cite{Brito:2015oca}), and consequently form bosonic clouds. Since in the nonrelativistic limit these systems can be described by the Schr\"odinger equation, they have also been named as ``gravitational atoms''~\cite{Arvanitaki:2010sy,Baumann:2019eav}. Previous works on this subject suggested that the TLNs of these systems can be sufficiently large
to leave an observable signature in GW signals emitted by coalescing BH binaries~\cite{DeLuca:2021,DeLuca:2022xlz}. However, these works only considered scalar and vector tidal perturbations as a proxy for gravitational tidal perturbations. The only results available for the gravitational TLNs of boson clouds are based on dimensional analysis arguments~\cite{Baumann:2018,Baumann:2019ztm}. The main goal of this work is therefore to extend these results by computing for the first time, in a rigorous manner, the gravitational TLNs of boson clouds.  We will however restrict ourselves to the framework of Newtonian gravity in order to pave the way toward a fully relativistic calculation.

Besides TLNs, other signatures due to the deformation of the cloud in a binary system have also been studied in the literature, the most important ones being orbital resonances, dynamical friction or even tidal disruption of the cloud~\cite{Baumann:2018,Baumann:2019ztm,Cardoso:2020hca,Baumann:2021fkf,Tomaselli:2023ysb,Brito:2023pyl,Duque:2023seg,Tomaselli:2024bdd,Boskovic:2024fga}. A full understanding of the detectability of boson clouds in binary systems would need to take all these effects into account, including the impact of nonzero TLNs~\cite{Baumann:2018,Baumann:2019ztm,DeLuca:2021,DeLuca:2022xlz}. The TLNs we compute here are mostly relevant in the regime where the companion object can be considered to be ``outside the cloud''~\cite{Baumann:2018,DeLuca:2021,DeLuca:2022xlz}. Depending on the exact parameters of the binary, as the binary separation decreases the cloud can either be disrupted due to tidal interactions or the companion object will enter inside the cloud, at which point finite-size effects start being suppressed and dynamical friction becomes the leading signature of the cloud's presence~\cite{Baumann:2018,Baumann:2019ztm,DeLuca:2022xlz,Tomaselli:2024bdd}. In either case, such effects can be modeled by considering time-dependent TLNs that smoothly go to zero at high frequencies as was done in Ref.~\cite{DeLuca:2022tkm}. Therefore the computation of the static TLNs that we here consider constitutes just one of the necessary ingredients needed in order to build accurate gravitational waveforms for binary systems in which one or both of the components is endowed with a boson cloud.

%%%%%%&&&&&&&
\subsection{Outline of the paper}
%%%%%%&&&&&&&
The main body of this paper is divided as follows. In Sec.~\ref{section:TLN_GR} we review the theory of tidally deformed objects in general relativity, whereas Sec.~\ref{section:TLN_newton} discusses the relation between the relativistic TLNs and their Newtonian counterpart. In Sec.~\ref{section:formation} we then introduce the notion of gravitational atoms and explain the general formalism used in describing these systems.
We introduce the field equations for our problem at hand in Sec.~\ref{section:perturbations}, where we also describe our perturbative scheme and compute the background unperturbed solutions describing a gravitational atom.
In Sec.~\ref{section:solutions} we solve the perturbed field equations, whereas in Sec.~\ref{section:LoveNumbers} we discuss the main results obtained in this work, namely we obtain the Newtonian TLNs of this system. For the reader wishing to directly jump to those results, we refer to Eqs.~\eqref{specific1}--\eqref{specific3} where the TLNs of spherically symmetric and dipolar boson clouds are provided. Finally, in Sec.~\ref{section:conclusions} we discuss those results in view of previous works and present possible future directions.\par
In order to help the reader reproduce our calculations, more details are presented in the Appendices. Appendix~\ref{section:AppendixA} presents our conventions in the formalism of symmetric trace-free (STF) tensors and provides the necessary definitions in that context. Appendix~\ref{section:AppendixC} provides a comprehensive list of useful special functions and mathematical identities that we used in our calculations. Appendix~\ref{app:sep_variables} gives details on the approach we took to reduce the perturbed field equations to a system of ordinary differential equations using separation of variables. Appendix~\ref{app:BCs_gravpot} provides a derivation of the tidal potential produced by a secondary body moving in circular orbits, in the frequency-domain. 
Appendix~\ref{ssec:TLN_00_11} shows an explicit derivation of the tidal Love numbers for two specific cloud configurations of interest, namely a spherically symmetric and a dipolar cloud. Finally, Appendix~\ref{app:aux_func} provides the explicit form of some auxiliary functions that we use throughout the text and Appendices. 
For the reader's convenience, in~\cite{github} we also provide a publicly available \textit{Mathematica} package that can be used to compute the TLNs for any given choice of parameters and configurations.\par
Throughout this work we use geometrized units $G=c=1$.

%%%%%%&&&&&&&
\section{Framework}
%%%%%%&&&&&&&
\label{section:setup}
%%%%%%&&&&&&&
%%%%%%&&&&&&&
\subsection{Relativistic gravitational tidal Love numbers}\label{section:TLN_GR}
%%%%%%&&&&&&&
In order to introduce our theoretical setup, let us start by briefly reviewing the relativistic methods to compute tidal deformations of self-gravitating objects as described, for example, in Refs.~\cite{Hinderer:2007,Binnington:2009,Damour:2009vw}. Here we choose to use the same notation as Ref.~\cite{Binnington:2009}, but will work with the formalism developed by Thorne~\cite{Thorne:1980ru} (see also Appendix~\ref{section:AppendixA} for more details on the notation we employ).\par
Consider for simplicity a body of mass $M_b$ which, in the absence of any perturbations, is spherically symmetric such that its metric $g^{(b)}_{\mu\nu}$ in the vacuum region external to the body is described by the Schwarzschild metric.
Gravitational perturbations to this body can be split into even and odd parity sectors. Taking the perturbations to be induced by an external tidal field, one can define~\cite{Zhang:1986,Binnington:2009} electric-type tidal moments associated to the even sector $\mathcal{E}_{a_1\cdot\cdot\cdot a_l}\equiv [(l-2)!]^{-1}\langle C_{0a_10a_2;a_3\cdot\cdot\cdot a_l} \rangle$ and magnetic-type tidal moments associated to the odd sector $\mathcal{B}_{a_1\cdot\cdot\cdot a_l}\equiv [2(l+1)(l-2)!/3]^{-1}\langle \epsilon_{a_1 bc}C^{bc}_{a_20;a_3\cdot\cdot\cdot a_l} \rangle$, where $C_{a_1a_2a_3a_4}$ is the Weyl tensor, a semicolon denotes a covariant derivative, $\epsilon_{a_1 bc}$ is the Levi-Civita symbol and angular brackets denote the operation of taking the symmetric and trace-free part, meaning that the resulting tensors are symmetric in all indices and have vanishing trace for all possible contractions.\par
To linear order in perturbation theory, the tidal field will induce a proportional response in the mass and current multipole moments of the body.
For a spherically symmetric configuration, there are no couplings between parities, meaning that mass (current) multipole moments will have even (odd) parity and therefore only be proportional to electric-type (magnetic-type) tidal moments. One may then define, separately, electric-type TLNs $k^E$ and magnetic-type TLNs $k^B$.\par
As shown in Ref.~\cite{Binnington:2009}, in the asymptotic limit $r\gg M_b$, a static tidal perturbation induces perturbations to the 00-component of the body's metric, which we name $h_{00}$, that can be written as
\begin{equation}
    h_{00}=\sum_{l=2}^{\infty}\left[-\frac{2}{l(l-1)}r^l\sum_{m=-l}^l e_0(r)\mathcal{E}_{lm}Y_{lm}(\theta,\varphi)\right]\,,
\label{externalmetric}
\end{equation}
where $\mathcal{E}_{lm}$ are the components of the electric-type tidal moments $\mathcal{E}_{a_1\cdot\cdot\cdot a_l}$ in a scalar spherical harmonic basis $Y_{lm}(\theta,\varphi)$ and $e_0(r)=1+2k^E_{lm} (M_b/r)^{2l+1}$, with $k^E_{lm}$ the electric-type TLNs. As we discuss below, to leading-order in a weak field expansion, the body's gravitational potential is fully encoded in the total metric 00-component, $g_{00}=g^{(b)}_{00}+h_{00}$, which is only affected by even perturbations, as can be seen from Eq.~\eqref{externalmetric}. Therefore, in the Newtonian limit, $k^E_{lm}$ reduce to the Newtonian TLNs whereas no analogue to magnetic-type TLNs exists in Newtonian gravity. From the metric perturbation in Eq.~\eqref{externalmetric} one can identify the applied tidal field as the terms proportional to $r^l$, while the terms proportional to $r^{-l-1}$ can be associated with the body's response.\par

More generically, the body's response can be written in terms of induced mass multipole moments $M_{lm}$ (see Appendix~\ref{section:AppendixA} for details), such that the total asymptotic metric of the system in asymptotically Cartesian and mass centered (ACMC) coordinates is given by
\begin{widetext}
\begin{align}
\begin{split}
    g_{00}&=g^{(b)}_{00}+h_{00}\\
    &=-1+\frac{2M_b}{r}+\sum_{l=2}^{\infty}\sum_{m=-l}^l\left[-\frac{2}{l(l-1)}\mathcal{E}_{lm}r^l+\frac{2}{r^{l+1}}\sqrt{\frac{4\pi}{2l+1}}M_{lm}\right]Y_{lm}(\theta,\varphi)+\sum_{l=2}^{\infty}\left[\frac{2}{r^{l+1}}\mathcal{S}_{l-1}-\frac{2}{l(l-1)}r^l\, \mathcal{P}_{l-1}\right]\,,
\end{split}
\label{g00}
\end{align}
\end{widetext}
where $\mathcal{P}_{l}$ and $S_l$ are placeholder symbols which denote possible terms with an arbitrary dependence on the spherical harmonics with multipoles $0\leq l'\leq l$ and no radial dependence~\cite{Mayerson:2022}. Comparison with the expression above for $h_{00}$ allows us to define the relativistic gravitational electric-type TLNs as
\begin{equation}
    k^E_{lm}\equiv -\frac{l(l-1)}{2M_b^{2l+1}}\sqrt{\frac{4\pi}{2l+1}}\frac{M_{lm}}{\mathcal{E}_{lm}}\,.
\label{kE}
\end{equation}
It is relevant to note that here we follow the convention of Ref.~\cite{Cardoso:2017} which differs from the analogous TLNs defined in Refs.~\cite{Hinderer:2007,Binnington:2009} by a factor of $(M_b/R)^{2l+1}$ where $R$ is the radius of the body undergoing tidal influence. This convention was chosen because the radius of boson clouds is not a well-defined quantity, as we will discuss below. We also notice that the extra terms involving spherical harmonics with multipoles lower than $l$ are irrelevant for the discussion since the TLNs are only defined in terms of the functions multiplying $Y_{lm}$.
%%%%%%&&&&&&&
\subsection{Tidal Love numbers in the Newtonian limit}\label{section:TLN_newton}
%%%%%%&&&&&&&
In order to reduce Eq.~\eqref{g00} to the case of Newtonian gravity, one uses the weak-field approximation $g_{00}\simeq -1-2U_\text{\tiny N}$, which gives the equivalent Newtonian potential
\begin{widetext}
\begin{align}
\begin{split}
    U_{\text{\tiny N}}&=-\frac{M_b}{r}-\sum_{l=2}^{\infty}\sum_{m=-l}^l\left[-\frac{\mathcal{E}_{lm}}{l(l-1)}r^l+\frac{M_{lm}}{r^{l+1}}\sqrt{\frac{4\pi}{2l+1}}\right]Y_{lm}(\theta,\varphi)-\sum_{l=2}^{\infty}\left[\frac{1}{r^{l+1}}\mathcal{S}_{l-1}-\frac{1}{l(l-1)}r^l\, \mathcal{P}_{l-1}\right]\\
    &=-\frac{M_b}{r}+\sum_{l=2}^{\infty}\sum_{m=-l}^l\frac{1}{l(l-1)}\left[1+2k^E_{lm}\left(\frac{M_b}{R}\right)^{2l+1}\left(\frac{R}{r}\right)^{2l+1}\right]\mathcal{E}_{lm}r^lY_{lm}(\theta,\varphi)-\sum_{l=2}^{\infty}\left[\frac{1}{r^{l+1}}\mathcal{S}_{l-1}-\frac{1}{l(l-1)}r^l\, \mathcal{P}_{l-1}\right].
\end{split}
\label{UNewton}
\end{align}
\end{widetext}
By comparing with the appropriate potential multipole expansion in Newtonian gravity (see, for example, Eq. (1.2) in\footnote{Note that the potentials in that paper have a conventional opposite sign in the Newtonian potential with respect to this work.} Ref.~\cite{Binnington:2009}), we conclude that the electric-type relativistic gravitational TLNs reduce to the Newtonian gravitational TLNs in the Newtonian limit. However, to complete the equivalence, we must note that our definition considers Newtonian TLNs depending on the azimuthal number $m$, which is usually not done in the literature, given that for spherically symmetric bodies the static TLNs do not depend on $m$. Here we keep the $m$-dependence explicit since later on we will be interested in computing the TLNs of nonspherically symmetric configurations. Notice that by taking $k^E_{lm}\rightarrow k^E_{l0}$ in Eq.~\eqref{UNewton} the sum in $m$ commutes with the square brackets and we recover the usual STF decomposition $r^l\sum_m \mathcal{E}_{lm}Y_{lm}=\mathcal{E}_L x^L$.\par
Having established the correspondence between the relativistic electric-type and Newtonian TLNs, henceforth we shall drop the $E$ label and write only $k_{lm}$ for all TLNs computed in this paper, given that we will focus only on Newtonian TLNs. 

\subsection{Gravitational atoms}
\label{section:formation}
%%%%%%&&&&&&&
We wish to use the formalism above in order to compute the TLNs of a system composed of a complex\footnote{Although we focus our discussion on complex fields, at the Newtonian level there are no noticeable differences between real and complex scalar fields. The main notable difference between real and complex scalar fields is that, at the relativistic level, the latter admit truly stationary BH solutions surrounded by a scalar cloud~\cite{Herdeiro:2014goa} whereas scalar clouds composed of a real field necessarily slowly dissipate through GW emission~\cite{Yoshino:2013,Arvanitaki:2014wva,Brito:2017zvb}.} scalar field $\Phi$ with fundamental mass $m_b=\mu\hbar$ propagating on a single isolated Kerr BH of mass $\Mbh$. This field satisfies the Klein-Gordon equation $\Box\Phi=\mu^2\Phi$ on this geometry. 

In the small-coupling limit $\alpha\equiv\mu \Mbh\ll 1$, the Klein-Gordon equation can be solved analytically~\cite{Detweiler:1980,Baumann:2019eav}.
In fact, it admits solutions, independently of the BH spin, in a region sufficiently far from the BH's event horizon\footnote{This ``far-region'' is usually defined  as the region where $r\gg \Mbh$ but such that $r$ is not necessarily large when compared to $1/\mu$, see, e.g., Fig. 2 in Ref.~\cite{Baumann:2019eav}.} which, to leading-order in a small-$\alpha$ expansion, take a hydrogenlike form~\cite{Detweiler:1980,Baumann:2019eav}
\begin{equation}
    \Phi(t,r,\theta,\varphi)\approx \sum_{n,\ell,m}e^{-i\omega_{n\ell m}t}R_{n\ell}(r)Y_{\ell m}(\theta,\varphi)\,,
\label{scalar}
\end{equation}
where we considered the usual Boyer-Lindquist coordinates, which far from the BH's horizon reduce to a spherical coordinate system. Here $n\geq 0$, $\ell\geq 0$ and $-\ell\leq m\leq \ell$ are integer numbers, the radial eigenfunctions are given by
\begin{equation}
    R_{n\ell}(r)=C_{n\ell}\, r^{\ell}e^{-\frac{\Mbh \mu^2}{n+\ell+1}r} U\hspace{-0.1cm}\left(-n,2\ell+2,\frac{2\Mbh \mu^2}{n+\ell+1}r\right)\,,
\label{R}
\end{equation}
with $C_{n\ell}$ normalization constants and $U$ is the (Tricomi) confluent hypergeometric function\footnote{The radial eigenfunctions are usually written in terms of the generalized Laguerre polynomials $L^{(2\ell+1)}_n$, but we chose to use the relation $L^{(\beta)}_p(x)=(-1)^p U(-p,\beta+1,x)/p!$ to simplify future calculations.}. For concreteness here we work with normalization constants given by
\begin{equation}
    C_{n\ell}\equiv \frac{(-1)^n}{\sqrt{2n!(n+\ell+1)(n+2\ell+1)!}}\left(\frac{2\Mbh\mu^2}{n+\ell+1}\right)^{\ell+3/2}\,,
\end{equation}
such that $\int_0^{\infty} |R_{n\ell}(r)|^2 r^2 dr=1$.
In a Kerr BH spacetime the eigenfrequencies $\omega_{n\ell m}$ are generically complex, with real and imaginary parts which, to leading-order in the small-$\alpha$ limit, take the form~\cite{Detweiler:1980,Dolan:2007mj,Baumann:2019eav}
\begin{eqnarray}
   \text{Re}(\omega_{n\ell m}) &\approx& \mu-\frac{\mu}{2}\left(\frac{\Mbh\mu}{\ell+n+1}\right)^2\,, \label{omegaR}\\
    \text{Im}(\omega_{n\ell m}) &\propto& \left(\Mbh\mu\right)^{4\ell+5}\left(m\Omega_{\text{\tiny H}}-\text{Re}(\omega_{n\ell m})\right)\,,\label{omegaI}
\end{eqnarray}
where $\Omega_{\text{\tiny H}}\equiv a/(2\Mbh r_+)$ is the angular velocity of a Kerr BH with spin $a\Mbh$ at the event horizon $r_+\equiv \Mbh+\sqrt{\Mbh^2+a^2}$. We also notice that, at leading-order, $\text{Re}(\omega_{n\ell m})$ does not depend on $m$ and the BH spin. This dependence only appears at higher-order through a term proportional to $a m\mu \alpha^5$~\cite{Baumann:2018,Baumann:2019eav}. Finally, note that in Eq.~\eqref{scalar} the scalar field was expanded using scalar spherical harmonics, even though in a Kerr BH background the angular part should instead be decomposed using spin-0 spheroidal harmonics ${}_0S_{\ell m}(\theta,\varphi)$~\cite{Detweiler:1980,Dolan:2007mj}. However, in the small-$\alpha$ limit those can be expanded as ${}_0S_{\ell m}(\theta,\varphi)=Y_{\ell m}(\theta,\varphi)+\mathcal{O}\left(a^2(\omega^2_{n\ell m}-\mu^2)\right)=Y_{\ell m}(\theta,\varphi)+\mathcal{O}\left(a^2\alpha^4\right)$~\cite{Berti:2005gp,Dolan:2007mj} and therefore the angular dependence of the scalar field is very well-described by spherical harmonics even in a Kerr BH background.

For nonaxisymmetric modes with $m>0$, Eq.~\eqref{omegaI} tells us that when $\text{Re}(\omega_{n\ell m})<m\Omega_{\text{\tiny H}}$ the mode grows exponentially in time\footnote{The mode $(n,\ell,m)=(0,1,1)$ has special importance since it is the fastest growing mode~\cite{Detweiler:1980,Dolan:2007mj}.}, with an e-folding time $1/\text{Im}(\omega_{n\ell m})$. This instability can be linked to energy and angular momentum extraction from the spinning BH, due to a process known as BH superradiance. As the mode grows, the BH spins down such that the condition $\text{Re}(\omega_{n\ell m})\approx m\Omega_{\text{\tiny H}}$ will be asymptotically reached and the instability effectively stops, leaving behind a quasistationary state composed of a BH surrounded by a co-rotating scalar cloud~\cite{Brito:2014wla,East:2017,Herdeiro:2017phl}. In this process, up to $\sim 10\%$ of the BH's initial energy can be transferred to the scalar field~\cite{East:2018glu,Herdeiro:2021znw}. Although such states are not infinitely long-lived, given that one expects them to either decay through GW emission~\cite{Yoshino:2013,Arvanitaki:2014wva,Brito:2017zvb} (for real scalar fields) or to eventually be reabsorbed by the BH as modes with increasing azimuthal number $m$ keep extracting the BH's spin~\cite{Ficarra:2018rfu}, their lifetime can be extremely long for small enough $\Mbh\mu$ and therefore play an important role in astrophysics (see~\cite{Brito:2015oca} and references therein).

On the other hand, Eq.~\eqref{omegaI} also reveals that modes with $m\leq 0$ always decay exponentially in time independently of the BH spin. However, in the limit $\Mbh\mu \ll 1$, even those modes can be extremely long-lived given that $\text{Im}(\omega_{n\ell m})\Mbh\ll 1$. As such, these modes can also be relevant as transient states, as was seen for example in numerical relativity simulations of very different sets of problems~\cite{Barranco:2012qs,Barranco:2017aes,Cardoso:2022vpj,Cardoso:2022nzc}. Therefore in this work we will, for the most part, consider generic $(n,\ell,m)$ modes for the gravitational atom, keeping in mind that the modes one should consider in concrete applications will depend on how the cloud formed in the first place.

In either case, Eq.~\eqref{scalar} gives us the wave function of the scalar cloud from which one can introduce an estimate of its ``size''. Different estimates have been presented in the literature, in particular Ref.~\cite{Arvanitaki:2010sy} takes the particle approach in which the bosons associated to $\Phi$ are considered to be orbiting the BH in a (quasi)nonrelativistic Keplerian regime with orbital radius $r_c\sim (n+\ell+1)^2/(\Mbh\mu^2)$ while Ref.~\cite{Brito:2015oca} takes the quantum-mechanical view-point of calculating the expectation value of $r$ using Eq.~\eqref{R}, and obtains $\langle r\rangle=[3(n+\ell+1)^2-\ell(\ell+1)]/(2\Mbh\mu^2)$. Both estimates agree on the $\Mbh^{-1}\mu^{-2}$ dependence, which is simply the analogous of the Bohr radius for this system. In this work, we therefore assume the size of the cloud to be proportional to the Bohr radius $r_c=a_{n\ell}/(\Mbh\mu^2)$, where $a_{n\ell}$ is a constant to be chosen in each estimation.\par

\subsection{Tidally perturbing a gravitational atom}
%%%%%%&&&&&&&
\label{section:perturbations}
%%%%%%&&&&&&&
\subsubsection{Field equations in the Newtonian limit}
%%%%%%&&&&&&&
We now consider that a scalar cloud $\Phi$ of radius $r_c$ and total mass $M_c$, given by Eq.~\eqref{scalar}, has been formed around a BH and that the rotation of the system can be neglected with respect to the timescale of the tidal deformations (so as not to consider rotational effects on the tidal deformations). We take the Newtonian (or nonrelativistic) limit of the system such that the gravitational field is sourced by a Newtonian potential $U$ and, in Cartesian coordinates, the spacetime becomes
\begin{equation}
    ds^2= -(1+2U)dt^2+(1-2U)(dx^2+dy^2+dz^2)\,.
\end{equation}

The matter part of the system is described by the energy-momentum tensor of the scalar field 
\begin{equation}
    T^{\text{\tiny S}}_{\mu\nu}=\partial_{(\mu}\Phi^*\partial_{\nu)}\Phi-\frac{1}{2}g_{\mu\nu}(\partial_\alpha\Phi^*\partial^{\alpha}\Phi+\mu^2|\Phi|^2)\,.
\label{TS}
\end{equation}
In addition, we wish to model the presence of a BH, in such a way that in the absence of any external perturbations and in the Newtonian limit the scalar cloud can be described by Eq.~\eqref{scalar}. As we will show below, this can be done by modeling the BH as a point-particle located at the origin
\begin{equation}
    T^{\text{\tiny BH}}_{\mu\nu}=\Mbh\delta(r)\delta^0_{\mu}\delta^0_{\nu}.
\end{equation}

Let us now define an auxiliary field through $\Phi=e^{-i\mu t}\Psi/\sqrt{\mu}$ and perform the exact same calculations as Appendix A of Ref.~\cite{Annulli:2020} changing only $T^{\text{\tiny S}}_{\mu\nu}\rightarrow T^{\text{\tiny S}}_{\mu\nu}+T^{\text{\tiny BH}}_{\mu\nu}$. One can then easily see that $T^{\text{\tiny S}}_{tt}\simeq\mu|\Psi|^2$
and that the Einstein-Klein-Gordon system
\begin{align}
    \Box\Phi&=\mu^2\Phi,\\
    R_{tt}-\frac{1}{2}g_{tt}R&=8\pi(T^{\text{\tiny S}}_{tt}+T^{\text{\tiny BH}}_{tt}),
\end{align}
reduces itself to the Schr\"odinger-Poisson system
\begin{align}
    i\frac{\partial\Psi}{\partial t}&=-\frac{1}{2\mu}\nabla^2\Psi+\mu U\Psi, \label{Sch}\\
    \nabla^2 U&=4\pi\Mbh\delta(r)+4\pi\mu|\Psi|^2, \label{Poisson}
\end{align}
where we considered $|U|\ll 1$ and $|\partial_t\Psi|\ll \mu|\Psi|$. Equations~\eqref{Sch} and~\eqref{Poisson} are the Newtonian field equations of the system we will now perturb.
%%%%%%&&&&&&&
\subsubsection{Linear perturbation theory}
\label{section: LPT}
\begin{figure}[H]
  \centering
  \includegraphics[width=0.48\textwidth]{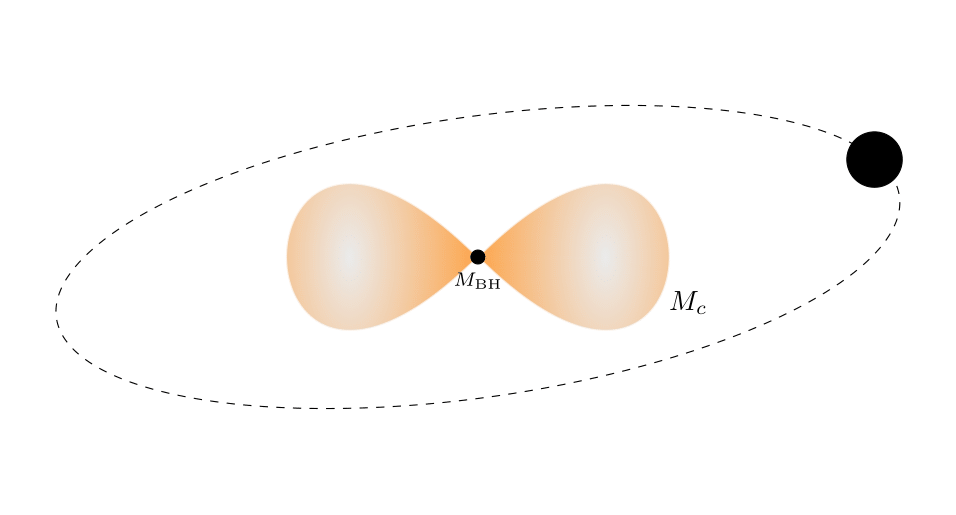}
  \caption{Gravitational atom with BH mass $M_{\text{BH}}$ and cloud mass $M_c$ in mode $\ell_i=m_i=1$ [see Eq.~\eqref{cloud}], suffering tidal perturbations. The secondary object undergoes circular orbits.}
  \label{fig:naxicloud}
\end{figure}

%%%%%%&&&&&&&
Let us introduce a secondary (or companion) body, creating a tidal field which induces a response in the gravitational atom. To visualize the problem at hand, Fig.~\ref{fig:naxicloud} shows a sketch of the total system for a specific nonaxisymmetric cloud configuration.

In order to obtain solutions of the Schr\"odinger-Poisson system, we turn to the use of linear perturbation theory, as is usually done in other contexts. 

The tidal field is assumed to be produced by an isolated body (i.e., with no accretion or release of matter) moving in a circular orbit\footnote{Considering circular orbits will help us make some symmetry considerations later on that simplify some of our calculations. However we do not expect the values of the static TLNs to depend on this assumption.} of frequency $\Omegaorb$. It may be any sort of object such as a point-particle, a star, a BH, etc., which is at a large enough distance from the gravitational atom such that the orbital timescale $\tau_{\text{orb}}$ is much larger than the timescale of any processes $\tau_{\text{int}}$ taking place inside each of the two bodies, hence $\tau_{\text{orb}}\gg\tau_{\text{int}}$ and one can consider only the exterior dynamics between them~\cite{Poisson:2014}.\par 
By working in the center-of-mass frame of the gravitational atom, such that the $z$ axis coincides with the orbital angular momentum vector, one can apply the one-body formalism and make use of Kepler's law $\Omegaorb^2=\Mbh/r^3_{\text{orb}}$ with $r_{\text{orb}}$ the orbital separation. In order to treat the gravitational influence of the companion in terms of a multipole expansion we consider that the orbital separation is larger than the size of the cloud $r_{\text{orb}}\gg r_c$. Using Kepler's law, this can also be written as $\Omegaorb\ll \mu\alpha^2$.
In addition, we restrict our calculations to radii $r_c\ll r\ll r_{\text{orb}}$ where the tidal moments can be clearly defined.
Finally, the small-coupling limit $\alpha\ll 1$ discussed in Sec.~\ref{section:formation} combined with the linear dependence of $r_c$ on $\Mbh/\alpha^2$ gives $r_c\gg\Mbh$.
In summary, the working assumptions of this model, which allow us to use linear perturbation theory, are
\begin{subequations}
\renewcommand{\theequation}{\theparentequation.\arabic{equation}} 
\begin{align}
%\centering
    &\Mbh\ll r_c\ll r\ll r_{\text{orb}}\,,&&\label{region}\\
    &\Omegaorb\ll\Mbh^2\mu^3\,,&&\label{Taylor}\\
    &\Omegaorb^2=\Mbh/r^3_{\text{orb}}\,.&&
\end{align}
\end{subequations}
Having these in mind, we introduce two bookkeeping parameters, $\epsilon$ and $\epsilon_p$, 
which will allow us to separate the different orders at which each process takes place in the system. We then consider an \textit{Ansatz} to solve the system of Eqs.~\eqref{Sch} and~\eqref{Poisson} given by (see Ref.~\cite{Brito:2023pyl} where a similar \textit{Ansatz} was considered in another context)
\begin{align}
    \Psi&=\epsilon\psi+\epsilon_p\delta\Psi,\label{scalar_expansion}\\
    U&=U_{\text{\tiny BH}}+\epsilon^2\delta U+\epsilon_p\delta U_{\text{T}}, \label{Uordereffects}
\end{align}
where we assume $|\epsilon_p\delta\Psi|\ll |\epsilon\psi|$, $|\epsilon^2\delta U|\ll |U_{\text{\tiny BH}}|$ and $|\epsilon_p\delta U_{\text{T}}|\ll |U_{\text{\tiny BH}}|$.
We consider that, in the absence of any external perturbations, the scalar field has an amplitude of order\footnote{For notational simplicity we use the bookkeeping parameters to indicate the order of the expansion, however one should bear in mind that the notation $\mathcal{O}(\epsilon)$ and $\mathcal{O}(\epsilon_p)$ indicates quantities of the order of the amplitude of $\psi$ and of the tidal field, respectively.} $\mathcal{O}(\epsilon)$ such that to leading-order the gravitational potential is entirely dictated by the BH's potential, $U_{\text{\tiny BH}}$. The term $\epsilon^2\delta U$ encodes the response of the potential due to the presence of the scalar field, which enters only at $\mathcal{O}(\epsilon^2)$ because the scalar field only enters at quadratic order in Poisson's equation~\eqref{Poisson}. On the other hand $\epsilon_p\delta U_{\text{T}}$ encodes both the tidal field and the response of the system to it. Finally $\epsilon_p\delta\Psi$ encodes information on how the scalar field configuration responds to the tidal field.\par
Substituting in Eqs.~\eqref{Sch} and \eqref{Poisson} one finds, up to linear order in $\epsilon$ and $\epsilon_p$:
\begin{align}
    \nabla^2 U_{\text{\tiny BH}}&=4\pi\Mbh\delta(r),\label{UBHequation}\\
    i\frac{\partial\psi}{\partial t}&=-\frac{1}{2\mu}\nabla^2\psi+\mu U_{\text{\tiny BH}}\psi,\label{psizero}\\
    \nabla^2\delta U&=4\pi\mu|\psi|^2,\label{br}\\
    \nabla^2\delta U_{\text T}&=4\pi\mu\epsilon(\psi^*\delta\Psi+\psi\delta\Psi^*),\label{Utidal}\\
    i\frac{\partial\delta\Psi}{\partial t}&=-\frac{1}{2\mu}\nabla^2\delta\Psi+\mu U_{\text{\tiny BH}}\delta\Psi+\epsilon\mu\psi\delta U_{\text{T}}.\label{deltaPsi}
\end{align}
Eq.~\eqref{UBHequation} is a spherically symmetric Poisson equation describing the potential of a point-particle with mass $\Mbh$, hence its solution is simply
\begin{equation}
    U_{\text{\tiny BH}}(r)=-\frac{\Mbh}{r}\,.
\label{UBH}
\end{equation}
On the other hand, Eq.~\eqref{psizero} is Schr\"odinger's equation with a Coulomb potential, which has exactly the same form as Schr\"odinger's equation for the hydrogen atom. Imposing regularity at $r=0$ and at $r\to \infty$, its eigenfunctions are precisely given by Eq.~\eqref{scalar}, which we introduced as describing scalar clouds around BHs in the small-$\alpha$ limit. This justifies our approximation of modeling the BH as a point-particle.
For simplicity, we will consider that the scalar cloud consists of just one mode $(n,\ell_i,m_i)$ instead of a linear combination of them. Thus
\begin{equation}
    \psi(t,r,\theta,\varphi)=e^{-iE_{n\ell_i}t}R_{n\ell_i}(r)Y_{\ell_i m_i}(\theta,\varphi),
\label{cloud}
\end{equation}
where
\begin{equation}
    E_{n\ell_i}= -\frac{\mu^3\Mbh^2}{2(n+\ell_i+1)^2}
\label{Eli}
\end{equation}
are the energy levels of each state which are obtained, as usual, by separating Eq.~\eqref{psizero} and computing the eigenvalues of the resulting radial equation under regular boundary conditions. Comparing with Eq.~\eqref{omegaR} one can see that $E_{n\ell_i}\approx\text{Re}(\omega_{n\ell m})-\mu$ to leading-order in a small-$\alpha$ approximation, consistent\footnote{Notice that we took out the factor $e^{i\mu t}$ coming from the definition of $\Psi$, when we factored out the high-frequency oscillations of $\Phi$.} with the fact that our Newtonian field equations provide a very good approximation to the Klein-Gordon equation in a BH background in this limit\footnote{The Newtonian field equations cannot capture the imaginary part of $\omega_{n\ell m}$, since this is related to the presence of an event horizon. However since we are assuming $\text{Im}(\omega_{n\ell m})\Mbh\ll 1$ or even $\text{Im}(\omega_{n\ell_i m_i})=0$, as is the case at the end of the superradiant instability phase, we can ignore the imaginary part.}.
The subscripts $i$ act merely as a label here, which will distinguish these functions from the perturbed ones.

From Eq.~\eqref{Eli} one can see that the condition~\eqref{Taylor} can also be interpreted as the requirement that $\tau_{\text{orb}}\gg\tau_{\text{int}}$. This can be seen from the fact that, for our system, $\tau_{\text{int}}$ can be taken to be the typical oscillation period of the configuration~\eqref{cloud} which is set by $\tau_{\text{int}}\propto 1/|E_{n\ell_i}|$. Since $\tau_{\text{orb}}\propto 1/\Omegaorb$ one sees that $\tau_{\text{orb}}\gg\tau_{\text{int}}$ is equivalent to~\eqref{Taylor}. 

One may also use Eq.~\eqref{cloud} to estimate the mass of the cloud, noting that $\mu|\Psi|^2$ is the energy density of the auxiliary scalar field in the Newtonian limit [see Eq.~\eqref{Poisson}]:
\begin{equation}
    M_c=\int \mu|\Psi|^2d^3x=\mu\epsilon^2+\mathcal{O}(\epsilon,\epsilon_p)+\mathcal{O}(\epsilon_p^2),
\label{cloudmass}
\end{equation}
where we normalized $\psi$ according to\footnote{In a relativistic framework, one would integrate starting at $r=\rbh$. However, here we may integrate from the origin since the BH has no dimensions in this point-particle approximation.} $\int|\psi|^2d^3x=1$. 
In this work, we will always use $M_c\simeq\mu\epsilon^2$.\par
The equation appearing at order $\mathcal{O}(\epsilon^2)$, Eq.~\eqref{br}, encodes information on the effect of the scalar cloud on the gravitational potential in the absence of tidal perturbations. However since it does not influence any of the other equations in the system, we will not solve it. 
In particular this implies that, to leading-order in $\epsilon$, the background potential is given by $U_{\text{\tiny BH}}(r)$ which is spherically symmetric. This allows us to have a well-defined separation between the multipole moments induced by the tidal field and the intrinsic multipoles of the gravitational field in the absence of tidal perturbations, i.e., all multipoles starting at $l\geq 2$ that we compute below will belong to the tidally perturbed system\footnote{As usual, the dipole moment vanishes by fixing the center-of-mass frame.}. 
Now, only Eqs.~\eqref{Utidal} and~\eqref{deltaPsi} need to be solved which will be the main purpose of the next section.
%%%%%%&&&&&&&
\section{Results}
%%%%%%&&&&&&&
\label{section:results}
%%%%%%&&&&&&&
\subsection{Solutions for the perturbed scalar field and potential}
\label{section:solutions}
%%%%%%&&&&&&&
In this section, we present the methods by which we solved Eqs.~\eqref{Utidal} and~\eqref{deltaPsi} and the corresponding solutions. For the sake of readability, here we only discuss the main results. More details can be found in the Appendices.\par
Following previous work~\cite{Annulli:2020,Brito:2023pyl}, we consider the following \textit{Ans\"atze} 
\begin{widetext}
\begin{align}
    \delta\Psi(t,r,\theta,\varphi)&=\int\frac{d\omega}{2\pi}\frac{1}{r}\sum_{\ell_j=0}^{\infty}\sum_{m_j=-\ell_j}^{\ell_j}\hspace{-0.2cm}e^{-iE_{n\ell_i}t}[\hat{Z}^{\ell_j m_j}_1(\omega,r)Y_{\ell_j m_j}(\theta,\varphi)e^{-i\omega t}+(\hat{Z}^*_2)^{\ell_j m_j}(\omega,r)Y^*_{\ell_j m_j}(\theta,\varphi)e^{i\omega t}],\label{ansatzPsi}\\
    \delta U_T(t,r,\theta,\varphi)&=\int\frac{d\omega}{2\pi}\sum_{l=2}^{\infty}\sum_{m=-l}^{l}[\hat{u}^{lm}(\omega,r)Y_{lm}(\theta,\varphi)e^{-i\omega t}+(\hat{u}^*)^{lm}(\omega,r)Y^*_{lm}(\theta,\varphi)e^{i\omega t}],\label{ansatzUT}
\end{align}
\end{widetext}
where $\omega\in \mathbb{R}$ and, once again, the subscript $j$ serves the purpose of a label to distinguish from the background scalar field solution [see Eq.~\eqref{cloud}]. Here, we should notice that unlike what is typically done when computing static TLNs, we do not start with $\omega=0$ from the onset of the calculations. We will instead keep $\omega\neq 0$ in the calculations and work in a small-frequency approximation, only taking the static limit at the end. The reason why we do this will become clearer below.

Following the steps in Appendix~\ref{app:sep_variables}, it is evident from Eqs.~\eqref{uhatlm}--\eqref{Z2hat} that an additional order separation can be done
\begin{align}
    \hat{Z}^{\ell_j m_j}_1(\omega,r)&=\epsilon (\hat{Z}_1)^{\ell_j m_j}_{(1)}(\omega,r),\label{Z1pert}\\
    (\hat{Z}^*_2)^{\ell_j m_j}(\omega,r)&=\epsilon (\hat{Z}^*_2)^{\ell_j m_j}_{(1)}(\omega,r),\label{Z2pert}\\
    \hat{u}^{lm}(\omega,r)&=\hat{u}^{lm}_{(0)}(\omega,r)+\epsilon^2\hat{u}^{lm}_{(2)}(\omega,r),\label{upert}
\end{align}
which at order $\mathcal{O}(\epsilon^0)$ results in
\begin{align}
    \mathcal{D}\hat{u}^{lm}_{(0)}&=0,\label{u0}\\
    \mathcal{D}(\hat{u}^*)^{lm}_{(0)}&=0,\label{u0*}
\end{align}
whereas the linear and quadratic terms in $\epsilon$ give the following system of ordinary differential equations:
\begin{widetext}
\begin{align}
    \mathcal{D}\hat{u}^{lm}_{(2)}&=\frac{4\pi\mu}{r}R_{n\ell_i}\hspace{-0.2cm}\sum_{k=0}^{\min(l,\ell_i)}\hspace{-0.1cm}\left[(C_1)^{l\ell_i k}_{m m_i}(\hat{Z}_1)_{(1)}^{|l-\ell_i|+2k,m+m_i}+(C_2)^{l\ell_i k}_{m m_i}(\hat{Z}_2)_{(1)}^{|l-\ell_i|+2k,m-m_i}\right],\label{u2}\\
    \mathcal{D}(\hat{u}^*)^{lm}_{(2)}&=\frac{4\pi\mu}{r}R_{n\ell_i}\hspace{-0.2cm}\sum_{k=0}^{\min(l,\ell_i)}\hspace{-0.1cm}\left[(C_4)^{l\ell_i k}_{m m_i}(\hat{Z}^*_1)_{(1)}^{|l-\ell_i|+2k,m+m_i}+(C_3)^{l\ell_i k}_{m m_i}(\hat{Z}^*_2)_{(1)}^{|l-\ell_i|+2k,m-m_i}\right],\label{u2*}\\
    \mathcal{L}_{+}(\hat{Z}_1)^{\ell_j m_j}_{(1)}&=2\mu^2r R_{n\ell_i}\left(\,\sum_{l\leq\ell_i}\sum_{k=0}^{l}(C_4)^{l\ell_i k}_{m_j-m_i,m_i}\hat{u}_{(0)}^{l,m_j-m_i}\delta_{\ell_j,\ell_i-l+2k}+\sum_{l>\ell_i}\sum_{k=0}^{\ell_i}(C_4)^{l\ell_i k}_{m_j-m_i,m_i}\hat{u}_{(0)}^{l,m_j-m_i}\delta_{\ell_j,l-\ell_i+2k}\right),\label{Z1ode}\\
     \mathcal{L}_{-}(\hat{Z}^*_2)^{\ell_j m_j}_{(1)}&=2\mu^2r R_{n\ell_i}\left(\,\sum_{l\leq\ell_i}\sum_{k=0}^{l}(C_2)^{l\ell_i k}_{m_j+m_i,m_i}(\hat{u}^*)_{(0)}^{l,m_j+m_i}\delta_{\ell_j,\ell_i-l+2k}+\sum_{l>\ell_i}\sum_{k=0}^{\ell_i}(C_2)^{l\ell_i k}_{m_j+m_i,m_i}(\hat{u}^*)_{(0)}^{l,m_j+m_i}\delta_{\ell_j,l-\ell_i+2k}\right)\,.\label{Z2ode}
\end{align}
\end{widetext}
Here $\mathcal{D}$ and $\mathcal{L}_{\pm}$ are linear differential operators given by
\begin{align}
    \mathcal{D}&\equiv\frac{d^2}{dr^2}+\frac{2}{r}\frac{d}{dr}-\frac{l(l+1)}{r^2},\label{D}\\
    \mathcal{L}_{\pm}&\equiv\frac{d^2}{dr^2}+\frac{2\mu^2\Mbh}{r}-\frac{\ell_j(\ell_j+1)}{r^2}+2\mu(E_{n\ell_i}\pm\omega)\,,\label{L}
\end{align}
whereas $(C_{\beta})^{l\ell_i k}_{m m_i}$ with $\beta=1,2,3,4$ are constants defined in Appendix~\ref{app:sep_variables}, see Eqs.~\eqref{C1}--\eqref{C4}.

Before solving this set of equations it is worth noting that, assuming real-valued boundary conditions, the system of Eqs.~\eqref{u0} and \eqref{u0*} implies that the functions $\hat{u}^{lm}_{(0)}$ are real-valued and, consequently, so are all the other functions, since in this case the source terms and the differential operators in Eqs.~\eqref{u2}--\eqref{Z2ode} are real-valued. 
Equation~\eqref{u0} simply corresponds to the radial part of Laplace's equation, which has the well-known solution $\hat{u}_{(0)}^{lm}(\omega,r)=A_{lm}(\omega)r^l+B_{lm}(\omega)r^{-l-1}$, with $A_{lm}(\omega), B_{lm}(\omega)$ constants to be set by applying appropriate boundary conditions.
Given its order in the perturbative scheme, we can identify this term with the tidal-field term of the perturbation $\delta U_{\text{T}}$ [see Eq.~\eqref{Uordereffects}], whereas the response to the tidal field is fully contained in $\hat{u}_{(2)}^{lm}(\omega,r)$. Therefore we set $B_{lm}=0$ and
\begin{equation}
    \hat{u}_{(0)}^{lm}(\omega,r)=A_{lm}(\omega)r^l.
\label{u0sol}
\end{equation}

Before proceeding further, an important observation is needed here. If one were to consider the most general way in which tidal interactions may occur, one would have to write $\hat{u}_{(2)}^{lm}$ as $\sum_{l'm'}\hat{u}_{(2)}^{lm,l'm'}$, where $lm$ would correspond to the induced multipoles and the sum over $l'm'$ to the tidal multipoles that induced them. These would then be restricted through some selection rules dependent on the system under study. For nonspherically symmetric clouds one can deduce that this sum would include terms $\{l'm'\}\neq\{lm\}$. This is in fact similar to what occurs when considering the gravitational response of a slowly spinning body to an external tidal field~\cite{Pani:2015hfa,Pani:2015nua,Landry:2015zfa,Pnigouras:2022zpx}. However, for the sake of simplicity, we have kept ourselves to the case where $\{l'm'\}=\{lm\}$, leaving the generalization for future work.

The consideration of circular orbits fixes the following identities (see Appendix~\ref{app:BCs_gravpot} for details):
\begin{align}
    A_{lm}(\omega)&=c_{lm}\delta(\omega-m\Omegaorb),\label{Alm}\\
    A_{l,-m}(-\omega)&=(-1)^m A_{lm}(\omega),\label{symAlm}\\
    c_{lm}&=-2\pi M_{\text{sec}} r_{\text{orb}}^{-(l+1)} W_{lm}\\
    c_{l,-m}&=(-1)^m c_{lm}\label{cl-m}
\end{align}
with $W_{lm}$ defined in Appendix~\ref{app:BCs_gravpot}, see Eq.~\eqref{Wlm}, and $M_{\text{sec}}$ the mass of the secondary body. Note that, given our assumption of circular orbits, the constants $c_{lm}$ vanish for odd values of $l+m$, so there will be no TLNs defined in those cases. One then finds that the following symmetry is satisfied:
\begin{equation}
    (\hat{Z}_1)_{(1)}^{\ell_j m_j}(\omega,r)=(-1)^{m_j}(\hat{Z}^*_2)_{(1)}^{\ell_j,-m_j}(-\omega,r)\,.
\label{symZ}
\end{equation}
This symmetry can be deduced by comparing Eqs.~\eqref{Z1ode} and~\eqref{Z2ode} together with the identity~\eqref{symAlm} and the symmetries of the constants $(C_{\beta})^{l\ell_i k}_{m m_i}$ found in Appendix~\ref{app:sep_variables}, namely Eqs.~\eqref{symmetry3} and~\eqref{symmetry4}.\par
The fact that all the radial functions are real, as well as Eq.~\eqref{symZ}, simplifies immensely the system of equations. From the four unsolved Eqs.~\eqref{u2}--\eqref{Z2ode}, only two require solving, which we choose to be Eqs.~\eqref{u2} and~\eqref{Z1ode}.\par
Given the linearity of the problem, the delta function appearing in Eq.~\eqref{Alm} will be present in all the radial functions through their coupling to $\hat{u}_{(0)}^{lm}$. This means that the solutions only have support at frequencies $\omega=m\Omegaorb$, and it is the first indication that there will be differences between the cases $m=0$ and $m\neq 0$. We will come back to this hypothesis at the end of the calculations. Having stated this, we are justified in writing
\begin{align}
    (\hat{Z}_1)_{(1)}^{\ell_j m_j}(\omega,r)&=(\hat{Z}_1)_{(1),s}^{\ell_j m_j}(\omega,r)\delta(\omega-(m_j-m_i)\Omegaorb),\label{Z1p}\\
    \hat{u}_{(2)}^{lm}(\omega,r)&=\hat{u}_{(2),s}^{lm}(\omega,r)\delta(\omega-m\Omegaorb),
\end{align}
which will allow us to solve the field equations as series expansions in some adimensional quantity involving $\omega$, to be determined.\par
Using Eq.~\eqref{symZ}, Eq.~\eqref{u2} may be simplified to
\begin{widetext}
\begin{align}
    \mathcal{D}\hat{u}^{lm}_{(2)}(\omega,r)&=\frac{4\pi\mu}{r}R_{n\ell_i}\hspace{-0.2cm}\sum_{k=0}^{\min(l,\ell_i)}\hspace{-0.1cm}\left[(C_1)^{l\ell_i k}_{m m_i}(\hat{Z}_1)_{(1)}^{|l-\ell_i|+2k,m+m_i}(\omega,r)+(-1)^{m+m_i}(C_2)^{l\ell_i k}_{m m_i}(\hat{Z}_1)_{(1)}^{|l-\ell_i|+2k,-m+m_i}(-\omega,r)\right],
\label{upertsimp}
\end{align}
\end{widetext}
therefore Eq.~\eqref{Z1ode} only needs to be solved for $\ell_j=|l-\ell_i|+2k$ and $m_j=\pm m+m_i$ since we only need to compute $(\hat{Z}_1)_{(1)}^{|l-\ell_i|+2k,\pm m+m_i}$. It is also easy to see, due to the Kronecker delta terms, that it may be solved independently (i.e. without the sums) for each source term, and the solution for each inhomogeneous equation can then be substituted in the equation for $\hat{u}^{lm}_{(2)}$. 

By resorting to the Green's function method, the solutions to Eq.~\eqref{Z1ode} for these values of $\ell_j,m_j$ with regular boundary conditions, are given by [separating out the delta functions with Eqs.~\eqref{Alm} and~\eqref{Z1p}]
\begin{widetext}
\begin{align}
\begin{split}
    (\hat{Z}_1)_{(1),s}^{|l-\ell_i|+2k,\pm m+m_i}(\omega,r)&=\frac{(\hat{Z}_{1,+})^{|l-\ell_i|+2k,\pm m+m_i}_{(1),s}(\omega,r)}{\mathcal{W}(\omega)}\int_0^r (\hat{Z}_{1,-})^{|l-\ell_i|+2k,\pm m+m_i}_{(1),s}(\omega,r')(\mathcal{S}_Z)^{l\ell_i k}_{\pm m, m_i}(r')dr'\\
    &+\frac{(\hat{Z}_{1,-})^{|l-\ell_i|+2k,\pm m+m_i}_{(1),s}(\omega,r)}{\mathcal{W}(\omega)}\int_r^\infty (\hat{Z}_{1,+})^{|l-\ell_i|+2k,\pm m+m_i}_{(1),s}(\omega,r')(\mathcal{S}_Z)^{l\ell_i k}_{\pm m, m_i}(r')dr',
\end{split}
\label{Z11s}
\end{align}
\end{widetext}
where
\begin{equation}
    (\mathcal{S}_Z)^{l\ell_i k}_{\pm m, m_i}(r)\equiv 2\mu^2(C_4)^{l\ell_i k}_{\pm m,m_i}c_{l,\pm m}r^{l+1}R_{n\ell_i}(r),
\label{SZ}
\end{equation}
and $(\hat{Z}_{1,\pm})^{\ell_j m_j}_{(1),s}$ are two linearly independent solutions to the homogeneous equation, such that $(\hat{Z}_{1,+})^{\ell_j m_j}_{(1),s}$ is regular at infinity whereas $(\hat{Z}_{1,-})^{\ell_j m_j}_{(1),s}$ is regular at the origin. For arbitrary values of $\ell_j,m_j$ those are given by
\begin{align}
    (\hat{Z}_{1,-})^{\ell_j m_j}_{(1),s}&=h_1^{\ell_j m_j}(\omega) M_{\kappa,\ell_j+\frac{1}{2}}\left(\sqrt{-8\mu(E_{n\ell_i}+\omega)}r\right),\\
    (\hat{Z}_{1,+})^{\ell_j m_j}_{(1),s}&=h_2^{\ell_j m_j}(\omega) W_{\kappa,\ell_j+\frac{1}{2}}\left(\sqrt{-8\mu(E_{n\ell_i}+\omega)}r\right),
\end{align}
with $\kappa=2\mu^2\Mbh/\sqrt{-8\mu(E_{n\ell_i}+\omega)}$, $h_1^{\ell_j m_j},h_2^{\ell_j m_j}$ integration constants and $M,W$ Whittaker functions. Their Wronskian $\mathcal{W}(\omega)$ is given by (see Eq.~\eqref{WronskianMW} in Appendix~\ref{section:AppendixC})
\begin{align}
\begin{split}
    &\mathcal{W}\left[(\hat{Z}_{1,-})^{\ell_j m_j}_{(1)},(\hat{Z}_{1,+})^{\ell_j m_j}_{(1)}\right](\omega)\\&\equiv(\hat{Z}_{1,-})^{\ell_j m_j}_{(1)}\frac{d}{dr}(\hat{Z}_{1,+})^{\ell_j m_j}_{(1)}-(\hat{Z}_{1,+})^{\ell_j m_j}_{(1)}\frac{d}{dr}(\hat{Z}_{1,-})^{\ell_j m_j}_{(1)} \\
    &=-h_1^{\ell_j m_j}h_2^{\ell_j m_j}\frac{\Gamma(2+2\ell_j)\sqrt{8\mu |E_{n\ell_i}|}\sqrt{1+\omega/E_{n\ell_i}}}{\Gamma\left(\ell_j+1-(n+\ell_i+1)/\sqrt{1+\omega/E_{n\ell_i}}\right)}\,,\label{eq:WZ1Z2}
\end{split}
\end{align}
where we omitted the explicit dependence of $h_1^{\ell_j m_j}$ and $h_2^{\ell_j m_j}$ on $\omega$ for ease of notation.

Given assumption~\eqref{Taylor}, we can expand all solutions as a power series in $\omega/E_{n\ell_i}\ll 1$ since they only have support at frequencies $\omega\sim\Omegaorb$, as we discussed above. However, at this point one needs to be careful given that the value of the argument of the Gamma function appearing in the denominator of the Wronskian~\eqref{eq:WZ1Z2} is essential in choosing how to proceed in solving the problem. In particular, if $\ell_j> n+\ell_i$ one simply has
\begin{align}
&\Gamma\left(\ell_j+1-(n+\ell_i+1)/\sqrt{1+\tilde\omega}\right)\nonumber\\
&= \Gamma\left(\ell_j-n-\ell_i\right)+\mathcal{O}\left(\tilde\omega\right)\,,
\label{eq:Gammaljmaior}
\end{align}
where we defined $\tilde\omega=\omega/E_{n\ell_i}$. On the other hand, if $\ell_j\leq n+\ell_i$ one should instead expand the gamma function as a Laurent series\footnote{By defining $f(\tilde\omega)$ as the argument of the gamma function in Eq.~\eqref{eq:GammaLaurent}, the leading-order term in the Laurent series of $\Gamma(f(\tilde\omega))$ can be derived using the fact that the residue of $\Gamma(f(\tilde\omega))$ at $\tilde\omega=0$ is ${\rm Res}(\Gamma(f(\tilde\omega)),0)={\rm Res}(\Gamma(z),-\tilde n)/f'(0)$ where ${\rm Res}(\Gamma(z),-\tilde n)=(-1)^{\tilde n}/\tilde n!$ for $\tilde n\equiv \ell_i+n-\ell_j=0,1,2,\ldots$.}:  
\begin{align}
&\Gamma\left(\ell_j+1-(n+\ell_i+1)/\sqrt{1+\tilde\omega}\right)\nonumber\\
&= \frac{2(-1)^{\ell_i+\ell_j+n}}{(\ell_i+n+1)(\ell_i+n-\ell_j)! \tilde\omega}+\mathcal{O}\left(\tilde\omega^0\right)\,.
\label{eq:GammaLaurent}
\end{align}
Notice that, had we taken $\omega=0$ from the outset of the calculation, we would obtain that the Wronskian~\eqref{eq:WZ1Z2} would then be exactly zero when $\ell_j\leq n+\ell_i$, leading to an ill-defined solution for Eq.~\eqref{Z11s} in those cases. Considering nonzero frequencies (i.e., nonstatic tides) allowed us to proceed with the calculation at the expense of obtaining solutions for $\hat{Z}_1$ that can go as $\tilde\omega^{-1}$ to leading-order. We will discuss these terms further below, but for now let us proceed.\par

Since we need to compute the solutions for $\ell_j=|l-\ell_i|+2k$, the cases $2\leq l\leq\ell_i$ and $l>\ell_i$ need to be considered separately, remembering also that in the former case we only need to consider $0\leq k\leq l$ and in the latter $0\leq k\leq\ell_i$, due to the restriction $0\leq k\leq\min(l,\ell_i)$ in the source term of Eq.~\eqref{upertsimp}. 
Having these intervals in mind and the conditions leading to Eqs.~\eqref{eq:Gammaljmaior} and~\eqref{eq:GammaLaurent}, it is then clear that one needs to compare $\ell_j=|l-\ell_i|+2k$ with $n+\ell_i$, since they will lead to different calculations. It is therefore helpful to construct Tables~\ref{table:allowed1} and~\ref{table:allowed2}, which are the subdivisions of $k$ resulting from $\ell_j<n+\ell_i$, $\ell_j=n+\ell_i$, and $\ell_j>n+\ell_i$, while taking into account the relative values between $n,l$, and $\ell_i$.
\begin{table}[htb!]
\centering
\begin{tabular}{|c|c|c|c|}
\hline
$\boldsymbol{2\leq l\leq\ell_i}$     & $0\leq k<\frac{n+l}{2}$  & $k=\frac{n+l}{2}$ & $\frac{n+l}{2}<k\leq l$                                                                                                  \\ \hline
$0\leq n<l$ & \cmark & \cmark & \cmark \\ \hline
$n=l$       & \cmark & \cmark & \xmark                                                                                                        \\ \hline
$n>l$ & \cmark & \xmark & \xmark                                                                                                                \\ \hline
\end{tabular}
\caption{Allowed values for $k$  (corresponding to the entries marked with \cmark), depending on the relative values of $n,l$, when $2\leq l\leq\ell_i$. Note that the case $k=(n+l)/2$, $0\leq n<l$ is only possible when $n$ and $l$ have the same parity, given that $k$ is an integer number.}
\label{table:allowed1}
\end{table}

\begin{table}[htb!]
%\hspace{-1cm}
\centering
\begin{tabular}{|c|c|c|c|}
\hline
$\boldsymbol{l>\ell_i}$ & $0\leq k<L$ & $k=L$ & $L<k\leq\ell_i$ \\ \hline
$0\leq n<l$ and $\ell_i<\frac{l-n}{2}$ &     \xmark                           &            \xmark              &   \cmark                                 \\ \hline
           $0\leq n<l$ and $\ell_i=\frac{l-n}{2}$                      &           \xmark                     &    \cmark                      &       \cmark                             \\ \hline
              $0\leq n<l$ and $\ell_i>\frac{l-n}{2}$                   &              \cmark                  &    \cmark                      &            \cmark                        \\ \hline
                    $n=l$             &     \cmark                           &      \cmark                    &       \xmark                             \\ \hline
                       $n>l$          &     \cmark                           &  \xmark                        &      \xmark                              \\ \hline
\end{tabular}
\caption{Allowed values for $k$ (corresponding to the entries marked with \cmark), depending on the relative values of $n,l,\ell_i$, when $l>\ell_i$. In order to simplify the entries in the Table we defined $L=\ell_i+(n-l)/2$. Note that in this case $k=\ell_i+(n-l)/2$, $0\leq n<l$ and $\ell_i=(l-n)/2$ is only possible when $n$ and $l$ have the same parity.}
\label{table:allowed2}
\end{table}
Given all these considerations, in Tables~\ref{table:Z1s1} and~\ref{table:Z1s2} we show the schematic form that the solution~\eqref{Z11s} takes when expanding it in powers of $\tilde\omega$.
\begin{table}[htb!]
\centering
\begin{tabular}{|c|c|}
\hline
$\boldsymbol{2\leq l\leq\ell_i}$     & $(\hat{Z}_1)_{(1),s}^{\ell_i-l+2k,\pm m+m_i}\big/\left[(C_4)^{l\ell_i k}_{\pm m,m_i}c_{l,\pm m}\right]$                                                                                                    \\ \hline
$0\leq k<\frac{n+l}{2}$ & $F^{(-1)}_{<}(r)\left(\tilde\omega\right)^{-1}+F^{(0)}_{<}(r)+\mathcal{O}\left(\tilde\omega\right)$ \\ \hline
$k=\frac{n+l}{2}$       & $G^{(-1)}(r)\left(\tilde\omega\right)^{-1}+G^{(0)}(r)+\mathcal{O}\left(\tilde\omega\right)$                                                                                                         \\ \hline
$\frac{n+l}{2}<k\leq l$ & $H_{<}(r)+\mathcal{O}\left(\tilde\omega\right)$                                                                                                                \\ \hline
\end{tabular}
\caption{Schematic form of the solutions for $(\hat{Z}_1)_{(1),s}^{|l-\ell_i|+2k,\pm m+m_i}$ when $2\leq l\leq\ell_i$ for each value of $0\leq k\leq l$. We remind that $k=(n+l)/2$ is only possible when $n$ and $l$ have the same parity. The explicit expressions for the auxiliary functions $F^{(-1)}_{\lessgtr}(r)$, $F^{(0)}_{\lessgtr}(r)$, $G^{(-1)}(r)$, $G^{(0)}(r)$ and $H_{\lessgtr}(r)$, are given in Appendix~\ref{app:scalar_auxfunc}, see Eqs.~\eqref{F<}--\eqref{H>}.}
\label{table:Z1s1}
\end{table}
%\hspace{3mm}
\begin{table}[htb!]
\begin{tabular}{|c|c|}
\hline
$\boldsymbol{l>\ell_i}$     & $(\hat{Z}_1)_{(1),s}^{l-\ell_i+2k,\pm m+m_i}\big/\left[(C_4)^{l\ell_i k}_{\pm m,m_i}c_{l,\pm m}\right]$                                                                                                    \\ \hline
$0\leq k<\ell_i+\frac{n-l}{2}$ & $F^{(-1)}_{>}(r)\left(\tilde\omega\right)^{-1}+F^{(0)}_{>}(r)+\mathcal{O}\left(\tilde\omega\right)$ \\ \hline
$k=\ell_i+\frac{n-l}{2}$       & $G^{(-1)}(r)\left(\tilde\omega\right)^{-1}+G^{(0)}(r)+\mathcal{O}\left(\tilde\omega\right)$                                                                                                         \\ \hline
$\ell_i+\frac{n-l}{2}<k\leq \ell_i$ & $H_{>}(r)+\mathcal{O}\left(\tilde\omega\right)$                                                                                                                \\ \hline
\end{tabular}
\caption{Same as Table~\ref{table:Z1s1} but now for $l>\ell_i$ and for each value of $0\leq k\leq \ell_i$. In this case $k=\ell_i+(n-l)/2$ is only possible when $n$ and $l$ have the same parity.}
\label{table:Z1s2}
\end{table}

We now have all the information needed to solve Eq.~\eqref{upertsimp}, but before writing down the solutions an important observation is required. By looking at Tables~\ref{table:Z1s1} and~\ref{table:Z1s2}, and the source terms~\eqref{SZ}, one sees that
\begin{align}
    (\hat{Z}_1)_{(1),s}^{|l-\ell_i|+2k,\pm m+m_i}=\sum_{q=\eta_k}^{\infty}(C_4)^{l\ell_i k}_{\pm m,m_i}c_{l,\pm m}f_{(q)}(r)\frac{\omega^q}{E_{n\ell_i}^q},
\end{align}
where $\eta_k=-1,0$ depending on the value of $k$ and $f_{(q)}$ are radial functions resulting from performing the integrations in Eq.~\eqref{Z11s}. The right-hand side of Eq.~\eqref{upertsimp} may then be written as
\begin{widetext}
\begin{align}
\begin{split}
    &\frac{4\pi\mu}{r}R_{n\ell_i}\hspace{-0.2cm}\sum_{k=0}^{\min(l,\ell_i)}\hspace{-0.1cm}\left[(C_1)^{l\ell_i k}_{m m_i}(\hat{Z}_1)_{(1)}^{|l-\ell_i|+2k,m+m_i}(\omega,r)+(-1)^{m+m_i}(C_2)^{l\ell_i k}_{m m_i}(\hat{Z}_1)_{(1)}^{|l-\ell_i|+2k,-m+m_i}(-\omega,r)\right]\\
    &=\frac{4\pi\mu}{r}R_{n\ell_i}\hspace{-0.2cm}\sum_{k=0}^{\min(l,\ell_i)}\sum_{q=\eta_k}^{\infty}\left[(C_1)^{l\ell_i k}_{m m_i}(C_4)^{l\ell_i k}_{m,m_i}c_{lm}+(-1)^{m+m_i}(C_2)^{l\ell_i k}_{m m_i}(C_4)^{l\ell_i k}_{-m,m_i}c_{l,-m}(-1)^q\right]f_{(q)}(r)\frac{\omega^q}{E_{n\ell_i}^q}\delta(\omega-m\Omegaorb)\\
    &=\frac{4\pi\mu}{r}R_{n\ell_i}\hspace{-0.2cm}\sum_{k=0}^{\min(l,\ell_i)}\sum_{q=\eta_k}^{\infty}\left[(C_1)^{l\ell_i k}_{m m_i}(C_1)^{l\ell_i k}_{m,m_i}+(-1)^q(C_2)^{l\ell_i k}_{m m_i}(C_2)^{l\ell_i k}_{m m_i}\right]c_{lm}f_{(q)}(r)\frac{\omega^q}{E_{n\ell_i}^q}\delta(\omega-m\Omegaorb),
\end{split}
\label{cancelamento}
\end{align}
\end{widetext}
where we used Eq.~\eqref{cl-m}, jointly with the properties that can be found in Appendix~\ref{app:sep_variables}, namely Eqs.~\eqref{symmetry1} and~\eqref{symmetry4}.
Since $f_{(q)}$ has no dependence in $m$ or $m_i$, we may safely conclude that all the odd-powered terms in the frequency-expansion cancel out when $m=0$ or $m_i=0$ given that $\left[(C_1)^{l\ell_i k}_{m m_i}\right]^2=\left[(C_2)^{l\ell_i k}_{m m_i}\right]^2$ in those cases [cf. Eq.~\eqref{C1} and Eq.~\eqref{C2}]. In particular, any term $q=-1$ cancels out when $m=0$ or $m_i=0$. We shall later use this argument to write the TLNs exactly for $m=0$. It is also worth highlighting the particular cases $(n,\ell_i,m_i)=(0,0,0)$ and $(n,\ell_i,m_i)=(0,1,1)$, that we will use as particular examples later on. In the former case only the $k=0$ term contributes to~\eqref{cancelamento} and from Table~\ref{table:Z1s2} one can see that, to leading-order, $\hat{Z}_1$ goes as $\tilde\omega^0$. On the other hand, for $(n,\ell_i,m_i)=(0,1,1)$, one finds that both $k=0$ and $k=1$ contribute to~\eqref{cancelamento} and Table~\ref{table:Z1s2} tells us that for $l=2$ and $k=0$, $\hat{Z}_1$ goes as $\tilde\omega^{-1}$. This means that, in this case, the leading-order term in Eq.~\eqref{cancelamento} will go as $\tilde\omega^{-1}$ unless we only consider $m=0$.\par
In order to solve Eq.~\eqref{upertsimp}, we use the same method (with Green's functions) as before. The left-hand side is exactly the same as Eq.~\eqref{u0}, meaning the solutions of the homogeneous equation are also the same, which we name $(\hat{u}_{-})^{lm}_{(2)}(\omega,r)\equiv d_1^{lm}(\omega)r^l$ and $(\hat{u}_{+})^{lm}_{(2)}(\omega,r)\equiv d_2^{lm}(\omega)r^{-(l+1)}$ with arbitrary integration constants $d_1^{lm}$ and $d_2^{lm}$. The resulting Wronskian is
\begin{widetext}
\begin{align}
\begin{split}
    \mathcal{W}
    \left[(\hat{u}_{-})^{lm}_{(2)},(\hat{u}_{+})^{lm}_{(2)}\right](\omega,r)
    \equiv(\hat{u}_{-})^{lm}_{(2)}\frac{d}{dr}(\hat{u}_{+})^{lm}_{(2)}-(\hat{u}_{+})^{lm}_{(2)}\frac{d}{dr}(\hat{u}_{-})^{lm}_{(2)}=-d_1^{lm}(\omega)d_2^{lm}(\omega)\frac{2l+1}{r^2}\,.
\end{split}
\end{align}
Therefore
\begin{align}
%\begin{split}
    \hat{u}_{(2),s}^{lm}(\omega,r)=(\hat{u}_{+})^{lm}_{(2)}(\omega,r)\int_0^r\frac{(\hat{u}_{-})^{lm}_{(2)}(\omega,r')(\mathcal{S}_u)^{l\ell_i}_{m m_i}(\omega,r')}{\mathcal{W}(\omega,r')}dr'
    +(\hat{u}_{-})^{lm}_{(2)}(\omega,r)\int_r^{\infty}\frac{(\hat{u}_{+})^{lm}_{(2)}(\omega,r')(\mathcal{S}_u)^{l\ell_i}_{m m_i}(\omega,r')}{\mathcal{W}(\omega,r')}dr',
%\end{split}
\label{uhatsol}
\end{align}
with
%\begin{widetext}
\begin{align}
    (\mathcal{S}_u)^{l\ell_i}_{m m_i}(\omega,r)\equiv \frac{4\pi\mu}{r}R_{n\ell_i}\hspace{-0.2cm}\sum_{k=0}^{\min(l,\ell_i)}\hspace{-0.1cm}\left[(C_1)^{l\ell_i k}_{m m_i}(\hat{Z}_1)_{(1),s}^{|l-\ell_i|+2k,m+m_i}(\omega,r)+(-1)^{m+m_i}(C_2)^{l\ell_i k}_{m m_i}(\hat{Z}_1)_{(1),s}^{|l-\ell_i|+2k,-m+m_i}(-\omega,r)\right].\label{sourceu2}
\end{align}
\end{widetext}
These source terms differ depending on the relative values of $l$ and $\ell_i$, according to Tables~\ref{table:allowed1}--\ref{table:Z1s2}. Also note that, as usual in the computation of TLNs, we only need to obtain $\hat{u}_{(2)}^{lm}$ asymptotically as $r\to\infty$, and therefore we only need to explicitly compute the integral in the first term of Eq.~\eqref{uhatsol} taking the limit $r\rightarrow\infty$. The physical justification for this can be found in assumption~\eqref{region}. Putting all of this together, the asymptotic solution for $\hat{u}_{(2)}^{lm}(\omega,r)$ takes the generic form
\begin{equation}
     \hat{u}_{(2)}^{lm}(\omega,r)\sim r^{-(l+1)}\hat{u}_{(2),s X}^{lm}(\omega)\delta(\omega-m\Omegaorb)\,,\label{u2_schematic}
\end{equation}
where $\hat{u}_{(2),s X}^{lm}(\omega)$ are coefficients that do not depend on $r$. Their explicit expression for different values of $l,\ell_i$ and $n$ can be constructed using Eq.~\eqref{solucoesu2} in Appendix~\ref{app:gravpot_auxfunc}. 
%
%%%%%%&&&&&&&
\subsection{Newtonian tidal Love numbers}
\label{section:LoveNumbers}
%%%%%%&&&&&&&
\subsubsection{General results}
With the results obtained above, we can now extract the TLNs for this system. To do so, we just need to compare $g_{00}=-1-2U$ with Eq.~\eqref{g00}, where we recall that in our case $U$ is given by Eq.~\eqref{Uordereffects}. Since the tidal effects are encoded in the $\epsilon_p\delta U_{\text{T}}$ term, only this part of the potential contains the desired coefficients (and $\epsilon^2\delta U$, as we have mentioned, may remain undetermined since it does not affect the TLNs).\par
First, Eqs.~\eqref{Uordereffects}, \eqref{UBH}, \eqref{ansatzUT}, and~\eqref{upert} (along with $Y_{l,-m}=(-1)^m Y^*_{lm}$) give
%\begin{widetext}
\begin{align}
\begin{split}
    g_{00}=&
    -1+\frac{2\Mbh}{r}
    -2\epsilon^2\delta U-2\epsilon_p\sum_{l=2}^{\infty}\sum_{m=-l}^l Y_{lm}(\theta,\varphi)
    \\
    &\times \int_{-\infty}^\infty\frac{d\omega}{2\pi}\left[\left(\hat{u}^{lm}_{(0)}(\omega,r)+\epsilon^2\hat{u}^{lm}_{(2)}(\omega,r)\right)e^{-i\omega t}\right.\\
    &\left.+(-1)^m\left(\hat{u}^{l,-m}_{(0)}(\omega,r)+\epsilon^2\hat{u}^{l,-m}_{(2)}(\omega,r)\right)e^{i\omega t}\right]\,.
\end{split}
\label{g00lin}
\end{align}
%\end{widetext}
Then, by following Appendix~\ref{app:gravpot_auxfunc}, one can show that\footnote{In Appendix~\ref{app:gravpot_auxfunc} we show that this holds up to first order in the small-frequency expansion, however we expect this to hold at any order.}
\begin{align}\label{eq:u_sym}
\begin{split}
    &\int_{-\infty}^\infty\frac{d\omega}{2\pi}(-1)^m(\hat{u}^{l,-m}_{(0)}(\omega,r)+\epsilon^2\hat{u}^{l,-m}_{(2)}(\omega,r))e^{i\omega t}\\
    &=\int_{-\infty}^\infty\frac{d\omega}{2\pi}[(\hat{u}^{lm}_{(0)}(\omega,r)+\epsilon^2\hat{u}^{lm}_{(2)}(\omega,r))e^{-i\omega t}].
\end{split}
\end{align}
Therefore, using Eq.~\eqref{u2_schematic}, Eq.~\eqref{g00lin} may be simplified to
\begin{widetext}
\begin{align}
\begin{split}
    g_{00}&\sim -1+\frac{2\Mbh}{r}-2\epsilon^2\delta U-4\epsilon_p\sum_{l=2}^{\infty}\sum_{m=-l}^{l}\int_{-\infty}^\infty\frac{d\omega}{2\pi}[\hat{u}^{lm}_{(0)}(\omega,r)+\epsilon^2\hat{u}^{lm}_{(2)}(\omega,r)]e^{-i\omega t}Y_{lm}(\theta,\varphi)\\
    &=-1+\frac{2\Mbh}{r}-2\epsilon^2\delta U-4\epsilon_p\sum_{l=2}^{\infty}\sum_{m=-l}^{l}\int_{-\infty}^\infty\frac{d\omega}{2\pi}\left[c_{lm}r^l+\frac{\epsilon^2}{r^{l+1}}\hat{u}^{lm}_{(2),sX}(\omega)\right]e^{-i\omega t}\delta(\omega-m\Omegaorb)Y_{lm}(\theta,\varphi)\\
    &=-1+\frac{2\Mbh}{r}-2\epsilon^2\delta U-\frac{2\epsilon_p}{\pi}\sum_{l=2}^{\infty}\sum_{m=-l}^{l}\left[c_{lm}r^l+\frac{\epsilon^2}{r^{l+1}}\hat{u}^{lm}_{(2),sX}(m\Omegaorb)\right]e^{-im\Omegaorb t}Y_{lm}(\theta,\varphi)\,.\label{g00_final}
\end{split}
\end{align}
\end{widetext}
Comparing with Eqs.~\eqref{g00} and \eqref{kE}, we then conclude that the Newtonian static TLNs of the system may be determined from
\begin{align}
    k^{(n,\ell_i,m_i)}_{lm}&=\lim_{\Omegaorb\to 0}\frac{1}{2\Mbh^{2l+1}}\frac{\epsilon^2}{c_{lm}}\hat{u}^{lm}_{(2),sX}(m\Omegaorb),
\label{Kgeral}
\end{align}
so all that is left is to use the solutions in Eq.~\eqref{solucoesu2} for each cloud configuration.
When using Eq.~\eqref{solucoesu2}, note that the conditions $\ell_i \lessgtr (l-n)/2$ are equivalent to $l\gtrless n+2\ell_i$ and that the series starts at $l=2$, leading to many different specific cases. However, the main logic is to subdivide each of the three cases $n<\ell_i$, $n=\ell_i$, $n>\ell_i$ according to the possibilities of Eq.~\eqref{solucoesu2}. Given our assumption of circular orbits, one should also remember from Sec.~\ref{section:solutions} that the TLNs we compute here are only defined for even $l+m$, which only happens when $l$ and $m$ are both even or odd.

Moreover, as we already mentioned below Eq.~\eqref{cancelamento}, when $m=0$ or $m_i=0$ one has $\left[(C_1)^{l\ell_i k}_{m m_i}\right]^2=\left[(C_2)^{l\ell_i k}_{m m_i}\right]^2$ and the singular pieces of order $(\omega/E_{n\ell_i})^{-1}$ in the functions $\hat{u}^{lm}_{(2),s}$ cancel exactly. Additionally, when $m=0$ all the terms of order $\omega/E_{n\ell_i}$ or higher vanish (since for circular orbits they are being evaluated at $\omega=m\Omegaorb=0$) and $\hat{u}^{lm}_{(2),sX}$ becomes independent of $\Omegaorb$.

Given the very large number of possibilities for the parameters of the cloud and tidal perturbations, we do not write here the general results for the TLNs, but instead focus below on analytical results for two specific cloud configurations, namely $(n,\ell_i,m_i)=(0,0,0)$ and $(n,\ell_i,m_i)=(0,1,1)$, which we consider to be of physical interest. The TLNs for other cloud configurations can be computed using a publicly available \textit{Mathematica} package that can be found in Ref.~\cite{github}.\par

Let us however highlight two aspects that are generic for any choice of parameters: (i) the TLNs computed using Eq.~\eqref{Kgeral} are proportional to the scalar cloud's mass $M_c$, which follows directly from using $M_c\approx\mu\epsilon^2$ [see Eq.~\eqref{cloudmass}] to eliminate $\epsilon^2$ from the equations; (ii) the TLNs for axisymmetric tides have a $r_c^{2l+1}$ dependence on the cloud's radius, or equivalently, a $\alpha^{-4l-2}$ dependence on the coupling constant due to the relation $r_c\propto 1/(\Mbh\mu^2)$ discussed in Sec.~\ref{section:formation}. In fact, looking at the terms of order $\omega^0$ in Eqs.~\eqref{primeiraus}--\eqref{ultimaus} one sees that
\begin{align}
     \hat{u}^{l0}_{(2),sX}&\propto \mu^3\left(\sqrt{8\mu|E_{n\ell_i}|}\right)^{-2l-2}\propto\frac{1}{\mu^{4l+1}\Mbh^{2l+2}}\,,
\end{align}
where in the last step we used Eq.~\eqref{Eli}. Therefore using Eq.~\eqref{Kgeral} we get
\begin{align}
    k^{(n,\ell_i,m_i)}_{l0}&\propto \frac{1}{\Mbh^{2l+1}}\frac{M_c}{\mu}\frac{1}{\mu^{4l+1}\Mbh^{2l+2}}\nonumber\\
    &=\frac{M_c}{\Mbh}\frac{1}{\alpha^{4l+2}}\propto\frac{M_c r_c^{2l+1}}{\Mbh^{2l+2}}\,.
\label{Kscalling}
\end{align}
The proportionality factor in~\eqref{Kscalling} is highly dependent on the cloud's configuration and the tidal perturbation.
The strong dependence of the TLNs on the cloud's configuration can be seen in Fig.~\ref{fig:TLN} where we show the value\footnote{As we discuss in Appendix~\ref{app:gravpot_auxfunc}, we should note that we did not find analytical expressions for some integrals that appear in the computation of the TLNs. Therefore for the numbers shown in Fig.~\ref{fig:TLN} we used \textit{Mathematica}'s built-in function \texttt{NIntegrate} to compute the integrals numerically.} of $k^{(n,\ell_i,m_i=\ell_i)}_{20}$ multiplied by $\alpha^{10}\Mbh/M_c$ for $\ell_i=0,1,2,3$ and $n=0,1,2$. 
\begin{figure}[htb!]
\hspace{-10mm}
\includegraphics[width=0.47\textwidth,left]{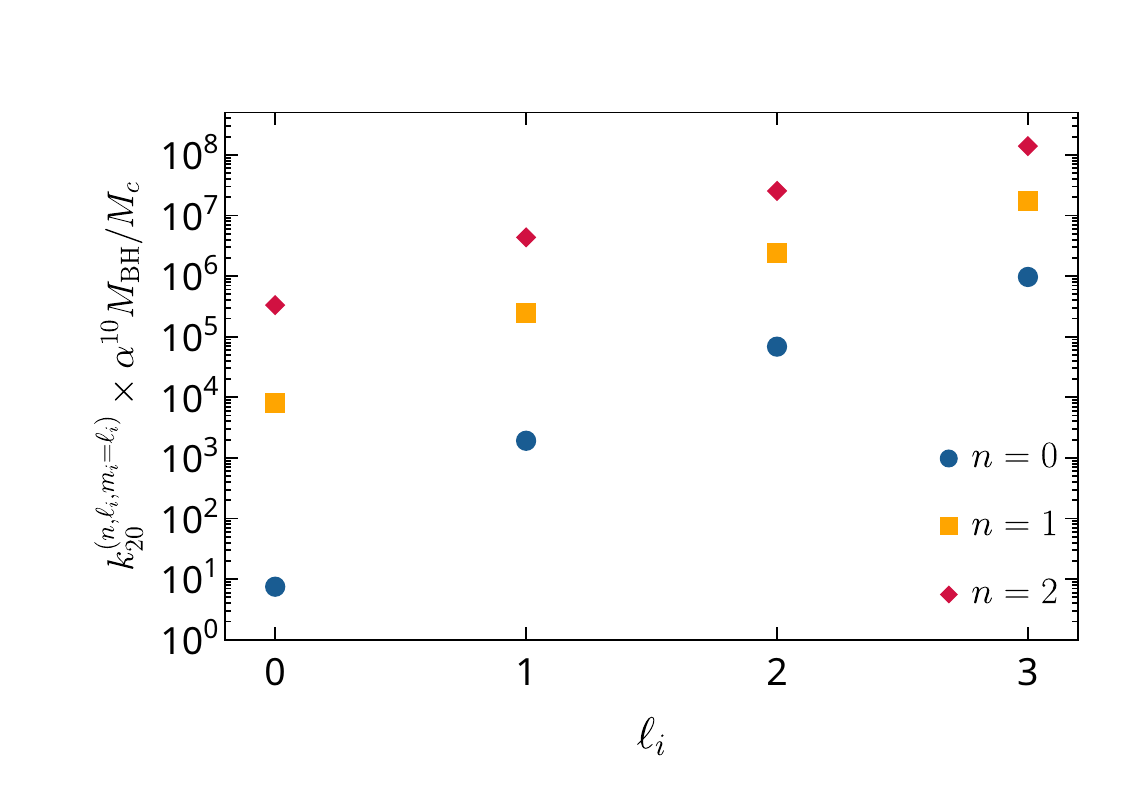}
  \caption{Value of the quadrupolar TLN $k^{(n,\ell_i,m_i=\ell_i)}_{20}$ (multiplied by $\alpha^{10}\Mbh/M_c$), for various choices of the cloud's angular number $\ell_i$ (x-axis) and overtone number $n$ (labeled points).}
  \label{fig:TLN}
\end{figure}
One can clearly see that the tidal deformability can increase quite drastically as one increases the cloud's angular number $\ell_i$ and overtone number $n$. We defer the discussion regarding a comparison of these results with previous works to the conclusions.

\subsubsection{Tidal Love numbers for $(n,\ell_i,m_i)=(0,0,0)$ and $(n,\ell_i,m_i)=(0,1,1)$}
Let us now focus on the two specific cloud configurations, namely $(n,\ell_i,m_i)=(0,0,0)$ and $(n,\ell_i,m_i)=(0,1,1)$. Let us call these the spherically symmetric and the dipolar configuration, respectively. As we mentioned in Sec.~\ref{section:formation}, the dipolar configuration can be formed around a spinning BH due to the superradiant instability and is therefore a case of particular interest. On the other hand, the spherically symmetric configuration cannot form through superradiance. It could however be potentially relevant as a transient state formed through accretion processes, see for example~\cite{Barranco:2012qs,Cardoso:2022vpj,Cardoso:2022nzc}. Moreover, it is the simplest cloud configuration and the calculations simplify considerably for this case. Therefore we find it useful to also mention it here.

To check the robustness of the general calculation described above, for these two configurations we also solved the field equations explicitly by substituting the values of $n,\ell_i,m_i$ at the beginning of the calculation, and obtained solutions which coincide with the expressions that can be obtained using the \textit{Mathematica} package in~\cite{github}. We checked this for any multipole $l$ in the case of the spherically symmetric configuration and for the quadrupole $l=2$ in the case of the dipolar configuration. The calculations for these specific cases can be found in Ref.~\cite{Arana_thesis}. We do not repeat them here given that they follow exactly the same procedure described for the general case above.

Since the procedure presented in this work allows us to compute the TLNs for any $l$, here we present the TLNs for all $l\geq 2$. We only show the final results and leave the details for Appendix~\ref{ssec:TLN_00_11}. By following Appendix~\ref{ssec:TLN_00_11}, the interested reader may learn how to apply the general results for any cloud configuration. 

For the spherically symmetric configuration we find
\begin{align}
\begin{split}
	&k^{(n=\ell_i=m_i=0)}_{lm}=\\
 &\frac{(l+2)\Gamma(4+4l)\Gamma(l)}{4^l\Gamma(3+3l)}\Big[{}_2 F_1(l,4+4l;3+3l;-1)\\
 &-2\, {}_2 F_1(l+1,4+4l;3+3l;-1)\Big]\frac{1}{\alpha^{4l+2}}\frac{M_c}{\Mbh}\,,
\end{split}
\label{specific1}
\end{align}
where ${}_2F_1$ is the hypergeometric function and we recall that, for circular orbits, the TLNs are only defined for even $l+m$, which only happens when $l$ and $m$ are both even or odd. Notice that, aside from this latter fact, the magnitude of the static TLNs does not depend on $m$ as expected for a spherically symmetric configuration. In particular, for a quadrupolar tide we find $k^{(n=\ell_i=m_i=0)}_{l=2,m}=15 M_c/(2\Mbh\alpha^{10})$.
On the other hand, for the dipolar configuration we need to distinguish between axisymmetric ($m=0$) and nonaxisymmetric ($m\neq 0$) tidal perturbations. Let us first consider the $m=0$ case for which we find 
\begin{align}
k_{l=2,m=0}^{(n=0,\ell_i=m_i=1)}=
\frac{1920}{\alpha^{10}}\frac{M_c}{\Mbh}\,,
\label{specific2}
\end{align}
and
\begin{align}
\begin{split}
	&k^{(n=0,\ell_i=m_i=1)}_{l>2,m=0}=\\
 &\frac{2(l+2)(-8-15l-l^2+4l^3+l^4)}{(l-2)(2l+1)}\frac{\Gamma(4+4l)\Gamma(l)}{\Gamma(3+3l)}\\
 &\times\Big[{}_2 F_1(l,4+4l;3+3l;-1)\\
 &-2\, {}_2 F_1(l+1,4+4l;3+3l;-1)\Big]\frac{1}{\alpha^{4l+2}}\frac{M_c}{\Mbh}\,,
\end{split}
\label{specific3}
\end{align} 
where we again recall that for circular orbits the TLN is only defined if $l$ is even since we are considering $m=0$. For the case $m\neq 0$ instead we find that the static TLN is not always well defined. For instance, in the case $l=2$ we have
\begin{align}
k_{l=2,m\neq 0}^{(n=0,\ell_i=m_i=1)}&=
\lim_{\Omegaorb\to 0}\left(\frac{54\mu^3\Mbh^2}{\Omegaorb}+1920+336m^2\right)\nonumber\\
&\times\frac{1}{\alpha^{10}}\frac{M_c}{\Mbh}
\label{specific2_mneq0}\,,
\end{align}
which clearly diverges due to the term $\mathcal{O}(\mu^3\Mbh^2/\Omegaorb)$. One can check that this behavior occurs quite generically (but not always) for $m\neq 0$, indicating that in those cases the TLNs do not have a well-defined static limit. We have not done a systematic study of all the cases in which this behavior occurs, however it is tempting to conjecture that this is related to the resonances first studied in Ref.~\cite{Baumann:2018}, but at this stage we have no way of providing a proof for this statement. Therefore, due to this problem, the \textit{Mathematica} package we provide in Ref.~\cite{github} can only be used to compute the TLNs for axisymmetric ($m=0$) tides, for which there is always a well-defined static limit.

\subsubsection{Validity of the perturbation scheme for axisymmetric tides}
To close this section, let us now give an estimate for when our perturbation scheme should be valid. Given the problems we mentioned above regarding nonaxisymmetric tides, we focus this discussion on axisymmetric tides.

Setting the bookkeeping parameters $\epsilon$ and $\epsilon_p$ to unity, our assumption that, to leading-order, the gravitational potential is
entirely dictated by the BH's potential translates to $|\delta U|\ll |U_{\text{\tiny BH}}|$ and $|\delta U_{\text{T}}|\ll |U_{\text{\tiny BH}}|$ [see Eq.~\eqref{Uordereffects}]. Here we recall that $\delta U$ encodes the response of the potential due to the presence of the scalar field whereas $\delta U_{\text{T}}$ encodes both the tidal field and the response of the system to it.

Although we did not compute $\delta U$ explicitly, we can do a very simple estimate by noticing that, at large distances, this term should go as $\delta U\sim -M_c/r$. Therefore the requirement $|\delta U|\ll |U_{\text{\tiny BH}}|$ simply translates to $M_c/\Mbh\ll 1$. On the other hand, we have seen that, at large enough distances, $\delta U_{\text{T}}$ can be schematically written as $\delta U_{\text{T}}\sim \delta U_{\text{tidal}}+\delta U_{\text{response}}$, where $\delta U_{\text{tidal}}$ is the tidal potential whereas $\delta U_{\text{response}}$ is the response of the system to this potential. Considering a tidal field with multipole $l$, the requirement $|\delta U_{\text{tidal}}|\ll |U_{\text{\tiny BH}}|$ together with the assumption that the tidal field is produced by a secondary object of mass $M_{\text{sec}}$ in a circular orbit gives the following condition:
\begin{equation}
\left(\frac{r}{r_{\text{orb}}}\right)^{l+1}\ll\frac{\Mbh}{M_{\text{sec}}}\,,\label{epsilonp}
\end{equation}
where we used the results of Appendix~\ref{app:BCs_gravpot}, namely Eq.~\eqref{Vext}. Since we work within the assumption that $r\ll r_{\text{orb}}$ [see condition~\eqref{region}] this condition is easily satisfied as long as the mass ratio $\Mbh/M_{\text{sec}}$ is not too small. Using Eq.~\eqref{g00_final} and~\eqref{Kgeral} we can also see that the requirement $|\delta U_{\text{response}}|\ll |U_{\text{\tiny BH}}|$ is satisfied if
\begin{align}
\begin{split}
&\frac{k_{l0} M_{\text{sec}}\Mbh^{2l+1}}{(r_{\text{orb}}r)^{l+1}}\ll\frac{\Mbh}{r}\\
\Leftrightarrow\,
&\frac{M_c}{\Mbh}\left(\frac{r_c}{r}\right)^{2l+1}\left(\frac{r}{r_{\text{orb}}}\right)^{l+1}\ll\frac{\Mbh}{M_{\text{sec}}}\,,
\end{split}
\end{align}
where in the last step we used Eq.~\eqref{Kscalling}. Since we work within the assumptions that $r_c\ll r \ll r_{\text{orb}}$ [see condition~\eqref{region}] and $M_c/\Mbh\ll 1$ (see discussion above), this condition is again easily satisfied as long as the mass ratio $\Mbh/M_{\text{sec}}$ is not too small.

%%%%%%%%%%%%%%%%%%%%%%%%%%%%%%%%%
\section{Discussion and Conclusions}
\label{section:conclusions}
%%%%%%%%%%%%%%%%%%%%%%%%%%%%%%%%%
The main goal of this work was to obtain estimates for the gravitational TLNs of a BH surrounded by a scalar cloud. To do so we resorted to a Newtonian approximation since in this framework we found it possible to determine the dependence of the TLNs on the coupling constant $\alpha\equiv\mu\Mbh$ exactly, as well as obtaining fully analytical results for some configurations. As we reviewed in Section~\ref{section:setup}, in the Newtonian limit the relativistic gravitational electric-type TLNs are equivalent to the Newtonian gravitational TLNs and so the results obtained here can be compared with future fully relativistic calculations when they become available.\par
Our results show that the TLNs for an axisymmetric tide with multipole $l$ have a power-law behavior on the coupling constant $\alpha$ as $\alpha^{-4l-2}$, independently of the cloud's configuration. This corresponds to a dependence on the scalar cloud's radius $r_c\propto \mu^{-2}\Mbh^{-1}$ as $r_c^{2l+1}$. Furthermore, the TLNs grow linearly with the cloud's total mass. The dependence on the coupling constant is in disagreement with the one found in Ref.~\cite{DeLuca:2021}, where TLNs of scalar clouds were also studied. There, it was found that the quadrupolar tides scale with $\alpha^{-8}$. However their framework differs from ours given that they only considered scalar and vector TLNs (i.e., tidal responses to scalar and vector field perturbations instead of gravitational perturbations). On the other hand, the scaling we find on the cloud's radius is in agreement with the prediction of Ref.~\cite{Baumann:2019ztm}, where the scaling of the TLNs with $r_c$ was estimated based on a dimensional analysis for quadrupolar gravitational tidal perturbations~\footnote{A previous estimate in Ref.~\cite{Baumann:2018} had predicted a scaling that was compatible with~\cite{DeLuca:2021}, however this estimate seems to have been corrected in Ref.~\cite{Baumann:2019ztm}, which agrees with our results.}. Our results are also compatible with the electric-type TLNs of other matter systems. For example, Ref.~\cite{Duque:2019} studied tidal gravitational perturbations of BHs surrounded by matter shells and found that the electric-type TLNs of this system scale with the shell's radius in the same fashion as the TLNs of scalar clouds. Similarly, Refs.~\cite{Mendes:2016vdr,Cardoso:2017,Sennett:2017etc} studied gravitational TLNs of boson stars and also found a scaling with the radius of these objects which agrees with the scaling we find. Since Refs.~\cite{Mendes:2016vdr,Cardoso:2017,Duque:2019,Baumann:2019ztm} all considered gravitational perturbations, the agreement in the cloud's radius scaling is, in our view, an indication that our results are robust. We do not have  at the moment a clear explanation as to why the scaling of the TLNs associated to electric-type gravitational tidal fields differ from the scaling found for scalar/vector TLNs found in Ref.~\cite{DeLuca:2021}. However, we notice that an important difference is the fact that, whereas the electric-type gravitational perturbations that we considered here couple to matter perturbations (in our case, the perturbations to the cloud configuration), the scalar and vector tidal perturbations considered in Ref.~\cite{DeLuca:2021} do not. Hence, the scalar/vector TLNs they find are purely due to perturbations in the spacetime curvature induced by the presence of the cloud (and therefore they do not have a Newtonian analog). This situation is in fact similar to magnetic-type gravitational perturbations (that we recall, do not have a Newtonian analog as well, see Sec.~\ref{section:TLN_GR}) of the matter systems considered in Refs.~\cite{Cardoso:2017,Duque:2019}, which do not couple to matter perturbations. Interestingly, the magnetic-type TLNs computed in Refs.~\cite{Cardoso:2017,Duque:2019} scale with the radius $r_0$ of the matter system as $r_0^{2l}$, which is in agreement with the scaling found in~\cite{DeLuca:2021} for scalar/vector TLNs. A deeper understanding of why the coupling of electric-type gravitational tidal perturbations to matter perturbations leads to a different scaling is an interesting problem for future work.
\par
We also considered nonaxisymmetric tidal deformations and found that, in this case, the TLNs can grow as $1/\Omegaorb$ in the adiabatic limit when the cloud is nonaxisymmetric, meaning that the static tides approximation breaks down for those cases. We conjecture that this could be related to the resonances first discussed in Ref.~\cite{Baumann:2018}, however a detailed understanding of this problem deserves further work.\par
This work can be extended in various ways. First of all, we did not study the detectability of the TLNs here obtained with GW detections. Such analysis was done in Refs.~\cite{DeLuca:2021,DeLuca:2022xlz}, suggesting that for $\alpha\sim \mathcal{O}(0.1)$ the quadrupolar TLNs of scalar clouds could in principle be measured through the observation of GW signals emitted by coalescing BH binaries with future GW detectors, such as LISA
and the Einstein Telescope. These results should however be revisited given that the scaling with $\alpha$ we found for the TLNs differs from Ref.~\cite{DeLuca:2021}. Given our results, we expect that the prospects for detection could be improved, since we found that the quadrupolar gravitational TLNs have enhancement by a factor $\alpha^{-2}$ with respect to the scalar and vector TLNs computed in~\cite{DeLuca:2021}. However we also caution that it is unclear at the moment if relativistic corrections to the TLNs are important enough to affect these results. This leads to a second extension of this work which is to perform a fully relativistic calculation of the gravitational TLNs (both electric-type and magnetic-type) of these systems. This could in principle be done using the formalism of Refs.~\cite{Brito:2023pyl,Katagiri:2024sep,Katagiri:2024oct}. Corrections due to the BH spin, that we neglected, could also lead to corrections to the TLNs~\cite{Pani:2015hfa,Pani:2015nua,Landry:2015zfa}. This could in principle also be computed using the formalism of Ref.~\cite{Brito:2023pyl} and resorting to a slowly rotating approximation as done in Refs.~\cite{Pani:2015hfa,Pani:2015nua,Landry:2015zfa} for the case of neutron stars. Finally, as we discussed in Sec.~\ref{section:solutions}, for nonspherically symmetric clouds, the cloud's response can also contain multipoles that differ from the multipole of the tidal field. Such terms, that we did not consider here, should be related to a different set of TLNs that are similar in nature to the spin-induced tidal Love numbers which exist for spinning bodies~\cite{Pani:2015hfa,Pani:2015nua,Landry:2015zfa,Landry:2017piv}. It would be interesting to compute these TLNs and to understand their impact for GW observations. 
%%%%%%%%%%%%%%%%%%%%%%%%%%%%%%%%%%
\begin{acknowledgments}
%%%%%%%%%%%%%%%%%%%%%%%%%%%%%%%%%%
R.B. acknowledges financial support provided by FCT---Fundação para a Ciência e a Tecnologia, I.P., under the Scientific Employment Stimulus---Individual Call---Grant No. \href{https://doi.org/10.54499/2020.00470.CEECIND/CP1587/CT0010}{2020.00470.CEECIND}. This work is supported by national funds through FCT, under the Project No.  \href{https://doi.org/10.54499/2022.01324.PTDC}{2022.01324.PTDC} and under FCT's ERC-Portugal program through the Project ``GravNewFields.''
%%%%%%%%%%%%%%%%%%%%%%%%%%%%%%%%%%
\end{acknowledgments}
%%%%%%%%%%%%%%%%%%%%%%%%%%%%%%%%%%

%%%%%%%%%%
\appendix
%%%%&&%

%%%%%
\section{Conventions regarding multipole moments and STF decomposition}
\label{section:AppendixA}
%%%%%
The previous works dealing with multipole expansions of the metric and tidal deformations have some differences in terms of conventions and notation, which we find useful to present here so that our formulation in Sec.~\ref{section:TLN_GR} may be better understood.\par
Following Ref.~\cite{Poisson:2014}, the decomposition of scalar spherical harmonics in the basis of STF tensors is written as
\begin{equation}
    Y_{lm}(\theta,\varphi)=\mathscr{Y}^{*\langle L\rangle}_{lm}n_{\langle L\rangle}
\label{dec}
\end{equation}
(as opposed to $Y_{lm}(\theta,\varphi)=\mathscr{Y}^{\langle L\rangle}_{lm}n_{\langle L\rangle}$ in Ref.~\cite{Thorne:1980ru}). Here $L$ denotes $l$ indices and angular brackets in expressions like $A^{\langle L\rangle}$ mean an STF tensor of rank $l$. Hence, in Eq.~\eqref{dec}, $\mathscr{Y}^{\langle L\rangle}_{lm}$ are constant STF tensors and $n_L\equiv n_{a_1}n_{a_2}\cdot\cdot\cdot n_{a_l}$ is a product of components of the unit radial vector $n_j=x_j/r$ with the Cartesian coordinates $(x^1,x^2,x^3)\equiv (x,y,z)$ and $r=\sqrt{x^2+y^2+z^2}$. The inverse relation to Eq.~\eqref{dec} is
\begin{equation}
    n^{\langle L\rangle}=\frac{4\pi l!}{(2l+1)!!}\sum_{m=-l}^l\mathscr{Y}^{\langle L\rangle}_{lm}Y_{lm}(\theta,\varphi)\,.
\end{equation}

Let $I_{\langle L\rangle}$ denote the STF equivalent of the mass multipole moments $I_{lm}$ for a given source of the gravitational field. These two quantities can be related by 
\begin{equation}
    I^{\langle L\rangle}=l!\sqrt{\frac{4\pi}{2l+1}}\sum_{m=-l}^l\mathscr{Y}^{*\langle L\rangle}_{lm}I_{lm}\,.
\label{ISTF}
\end{equation}
Here we notice that we fixed the overall $l$-dependent normalization factor such that we obtain the same expressions for $g_{00}$ as Refs.~\cite{Cardoso:2017,DeLuca:2021}, which differs from the normalization used in Eq. (4.6a) of Ref.~\cite{Thorne:1980ru}.
Using Eq. (2.26a) of Ref.~\cite{Thorne:1980ru}, one can invert Eq.~\eqref{ISTF} to write
\begin{equation}
    I_{lm}=\frac{4\pi}{(2l+1)!!}\sqrt{\frac{2l+1}{4\pi}}\mathscr{Y}^{\langle L\rangle}_{lm}I_{\langle L\rangle}.
\end{equation}

At this point, we can adopt the Geroch-Hansen normalization~\cite{Geroch:1970,Hansen:1974}, which means all expressions will be written in terms of the Geroch-Hansen multipoles\footnote{For a proof of this expression, see Ref.~\cite{Gursel:1983}.} $M^{\langle L\rangle}= (2l-1)!! I^{\langle L\rangle}$.\par
Consider now any stationary, asymptotically flat, vacuum spacetime. Then, in ACMC coordinates, according to our conventions, the gravitational field created by a body of mass $M_b$ results in~\cite{Thorne:1980ru,Gursel:1983}
\begin{equation}
    g_{00}=-1+\frac{2M_b}{r}+\sum_{l=2}^{\infty}\frac{1}{r^{l+1}}\left(\frac{2}{l!}M^{\langle L\rangle}n_{\langle L\rangle}+\mathcal{S}_{l-1}\right),
\end{equation}
in the asymptotic limit\footnote{The mass dipole terms vanish by virtue of choosing a reference frame whose spatial origin coincides with the center-of-mass of the body---the MC part of ACMC.}. Here $\mathcal{S}_l$ is a placeholder symbol which denotes an arbitrary dependence on the spherical harmonics with multipoles $0\leq l'\leq l$ but with no radial dependence~\cite{Mayerson:2022}. Inserting Eqs.~\eqref{ISTF} and~\eqref{dec} and absorbing a factor of $1/2$ in $\mathcal{S}_{l-1}$, we get
\begin{align}
\begin{split}
    g_{00}&=-1+\frac{2M_b}{r}\\
    &+\sum_{l=2}^{\infty}\frac{2}{r^{l+1}}\Bigg[\sqrt{\frac{4\pi}{2l+1}}\sum_{m=-l}^lM_{lm}Y_{lm}(\theta,\varphi)+\mathcal{S}_{l-1}\Bigg],
\end{split}
\label{AsMetric}
\end{align}
with $M_{lm}=(2l-1)!!I_{lm}$.
%%%%%
%%%%%
\section{Useful special functions and mathematical identities}
\label{section:AppendixC}
%%%%%
In this Appendix we provide a list of useful special functions and mathematical identities that we used in our calculations.

\subsection{Associated Legendre polynomials}
The associated Legendre polynomials
\begin{equation}
    P^m_l(x)=\frac{(-1)^m}{2^l l!}(1-x^2)^{m/2}\frac{d^{l+m}}{dx^{l+m}}(x^2-1)^l
\end{equation}
satisfy the identity
\begin{align}
    P^m_{l}(0)=\begin{cases}
			N_{lm}, & \text{if $l+m$ is even}\\
            0, & \text{if $l+m$ is odd}
		 \end{cases},
\label{Pl0}
\end{align}
with
\begin{align}
    N_{lm}\equiv (-1)^{\frac{l+m}{2}}(l+m)!\left[2^l\left(\frac{l+m}{2}\right)!\left(\frac{l-m}{2}\right)!\right]^{-1}.
\label{Nlm}
\end{align}
From the relation between the associated Legendre polynomials and the spherical harmonics we also have
\begin{equation}\label{Ylm*}
    Y^*_{lm}\left(\frac{\pi}{2},\varphi\right)=\sqrt{\frac{2l+1}{4\pi}\frac{(l-m)!}{(l+m)!}}P^m_l(0)e^{-im\varphi}.
\end{equation}
\subsection{Whittaker functions}
The Whittaker functions
\begin{align}
    M_{\kappa,\mu'}(z)&\equiv e^{-z/2}z^{\mu'+1/2}M\left(\mu'+\frac{1}{2}-\kappa,2\mu'+1,z\right),\\
    W_{\kappa,\mu'}(z)&\equiv e^{-z/2}z^{\mu'+1/2}U\left(\mu'+\frac{1}{2}-\kappa,2\mu'+1,z\right),
\end{align}
where $M(a,b,z),U(a,b,z)$ are (Kummer's and Tricomi's, respectively) confluent hypergeometric functions, have Wronskian~\cite{NIST:DLMF}
\begin{equation}
    \mathcal{W}(M_{\kappa,\mu'},W_{\kappa,\mu'})=-\frac{\Gamma(2\mu'+1)}{\Gamma\left(\mu'+\frac{1}{2}-\kappa\right)},
\label{WronskianMW}
\end{equation}
and can be inverted with~\cite{NIST:DLMF}
\begin{align}
    M(a,b,z)&=e^{z/2}z^{-b/2}M_{\frac{b}{2}-a,\frac{b}{2}-\frac{1}{2}}(z),\label{MtoM}\\
    U(a,b,z)&=e^{z/2}z^{-b/2}W_{\frac{b}{2}-a,\frac{b}{2}-\frac{1}{2}}(z).\label{UtoW}
\end{align}
When $a$ is a nonpositive integer, both $M$ and $U$ are polynomials in $z$~\cite{NIST:DLMF}
\begin{align}
    M(-m,b,z)&=\sum_{s=0}^m\binom{m}{s}\frac{(-z)^s}{(b)_s},
    \label{Mmenos}\\
    U(-m,b,z)&=(-1)^m\sum_{s=0}^m\binom{m}{s}(b+s)_{m-s}(-z)^s,
    \label{Umenos}
\end{align}
for $m=0,1,2,\ldots$. 
\subsection{Useful derivatives of Kummer's and Tricomi's confluent hypergeometric functions}
Using Eq.~(15) of Ref.~\cite{Ancarani:2008} one gets
\begin{equation}
    \frac{\partial}{\partial a}M(a,b,z)\Big|_{a=0}=\frac{z}{b}\,{}_2F_2(1,1;2,b+1;z)\,,
\label{derM}
\end{equation}
where ${}_2F_2$ is a generalized hypergeometric function. We now wish to prove the analogous result for Tricomi's function
\begin{equation}
    \frac{\partial}{\partial a}U(a,n,z)\Big|_{a=0}=\sum_{k=1}^{n-1}\binom{n-1}{-k+n-1}\Gamma(k)\, z^{-k}-\log(z),
\label{derU}
\end{equation}
with $n=1,2,\ldots$
We start by writing~\cite{weisstein3}
\begin{equation}
    U(a,a+n,z)=z^{-a}\sum_{k=0}^{n-1}\binom{n-1}{-k+n-1}(a)_k z^{-k},
\end{equation}
where again $n=1,2,\ldots$ Now, since
\begin{align}
    &\frac{\partial}{\partial a}U(a,a+n,z)\Big|_{a=0}\\
    &=\frac{\partial}{\partial a}U(a,n,z)\Big|_{a=0}+\frac{\partial}{\partial b}U(0,b,z)\Big|_{b=n},
\end{align}
and $U(0,b,z)=1$, then
\begin{align}
\begin{split}
    &\frac{\partial}{\partial a}U(a,n,z)\Big|_{a=0}=\frac{\partial}{\partial a}U(a,a+n,z)\Big|_{a=0}\\
    &=\Bigg[-z^{-a}\log(z)\sum_{k=0}^{n-1}\binom{n-1}{-k+n-1}(a)_k z^{-k}\\
    &+\sum_{k=0}^{n-1}\binom{n-1}{-k+n-1}z^{-a}(a)_k[\psi(a+k)-\psi(a)] z^{-k}\Bigg]\Bigg|_{a=0},
\end{split}
\end{align}
where $\psi$ is the digamma function (appearing from the derivative of the Pochhammer symbol). Using the identities
\begin{equation}
    (0)_k=\begin{cases}
        1, & k=0\\
        0, & k\neq 0
    \end{cases},
\end{equation}
and 
\begin{equation}
z^{-a}(a)_k[\psi(a+k)-\psi(a)]=\Gamma(k)+\mathcal{O}(a)\,,
\end{equation}
we find
%\begin{widetext}
\begin{align}
\begin{split}
    &\frac{\partial}{\partial a}U(a,n,z)\Big|_{a=0}
    =\Bigg[-z^{-a}\log(z)\\
    &+\sum_{k=1}^{n-1}\binom{n-1}{-k+n-1}[\Gamma(k)+\mathcal{O}(a)] z^{-k}\Bigg]\Bigg|_{a=0}\\
    &=-\log(z)+\sum_{k=1}^{n-1}\binom{n-1}{-k+n-1}\Gamma(k)\, z^{-k},
\end{split}
\end{align}
%\end{widetext}
which completes the proof.
\subsection{Useful integrals}
Using the lower and upper incomplete Gamma functions, one can show~\cite{weisstein1,weisstein2} that
\begin{align}
    \int_0^u x^{m}e^{-x}dx&=m!-\sum_{p=0}^{m}\frac{m!}{p!}u^pe^{-u},\quad m=0,1,2,...,\label{0u}\\
    \int_u^{\infty}x^m e^{-x}dx&=\sum_{p=0}^m\frac{m!}{p!}u^p e^{-u},\quad m=0,1,2,...\label{uinf}
\end{align}
and, consequently,
\begin{align}
    \int_0^{\infty}x^m e^{-x}dx&=m!,\quad m=0,1,2,...
\label{0inf}
\end{align}
\subsection{Wigner 3-j symbols: useful identities}
The symmetries of the Wigner 3-j symbols which were used in this work are (see for example page 1056 in Ref.~\cite{Messiah:1961})
\begin{align}
    \begin{pmatrix}
	\ell_1 & \ell_2 & \ell_3\\ 
	m_1 & m_2 & m_3
    \end{pmatrix}&=
    \begin{pmatrix}
	\ell_2 & \ell_3 & \ell_1\\ 
	m_2 & m_3 & m_1
    \end{pmatrix}\label{WignerColunas}\\
    \begin{pmatrix}
	\ell_1 & \ell_2 & \ell_3\\ 
	m_1 & m_2 & m_3
    \end{pmatrix}
    &=(-1)^{\ell_1+\ell_2+\ell_3}\begin{pmatrix}
	\ell_3 & \ell_2 & \ell_1\\ 
	m_3 & m_2 & m_1
    \end{pmatrix}\,.\label{WignerColunas2}\\
    \begin{pmatrix}
	\ell_1 & \ell_2 & \ell_3\\ 
	m_1 & m_2 & m_3
    \end{pmatrix}
    &=(-1)^{\ell_1+\ell_2+\ell_3}\begin{pmatrix}
	\ell_1 & \ell_2 & \ell_3\\ 
	-m_1 & -m_2 & -m_3
    \end{pmatrix}\,.\label{WignerColunas3}
\end{align}   
We also used the following explicit expressions for the Wigner 3-j symbols (see for example pages 1058-1059 in Ref.~\cite{Messiah:1961})
\begin{align}
&\begin{pmatrix}
	\ell & \ell & 0\\ 
	m & -m & 0
    \end{pmatrix}=\frac{(-1)^{\ell-m}}{\sqrt{2\ell+1}}\,,\label{WignerFrac}\\
%\end{align}
%\begin{align}
%\begin{split}
    &\begin{pmatrix}
	\ell_1 & \ell_2 & \ell_1+\ell_2\\ 
	m_1 & m_2 & -m_1-m_2
    \end{pmatrix} =(-1)^{\ell_1-\ell_2+m_1+m_2}\nonumber\\
    &\times\sqrt{\frac{(\ell_1+\ell_2+m_1+m_2)!(\ell_1+\ell_2-m_1-m_2)!}{(\ell_1+m_1)!(\ell_1-m_1)!(\ell_2+m_2)!(\ell_2-m_2)!}}\nonumber\\
    &\times\sqrt{\frac{(2\ell_1)!(2\ell_2)!}{(2\ell_1+2\ell_2+1)!}}\,,
%\end{split}
\label{WignerSoma}\\
%\end{align}
%\begin{align}
%\begin{split}
    &\begin{pmatrix}
	\ell_1 & \ell_2 & \ell_3\\ 
	0 & 0 & 0
    \end{pmatrix}=(-1)^p\sqrt{\frac{(2p-2\ell_1)!(2p-2\ell_2)!(2p-2\ell_3)!}{(2p+1)!}}\nonumber\\
    &\times\frac{p!}{(p-\ell_1)!(p-\ell_2)!(p-\ell_3)!}\delta_{\ell_1+\ell_2+\ell_3,2p}\,.
%\end{split}
\label{WignerRaiz}
\end{align}
%%%%%
%%%%%
\section{Separation of variables for the perturbed field equations}
\label{app:sep_variables}
%%%%%
In this Appendix we provide more details regarding the separation of variables method that we used to obtain the system of ordinary differential equations~\eqref{u0}--\eqref{Z2ode}.

The substitution of the \textit{Ansatz}~\eqref{ansatzPsi} and~\eqref{ansatzUT} in the field equations~\eqref{Utidal} and~\eqref{deltaPsi}, in addition to an inversion of the Fourier transforms and projection of the equations on the spherical harmonics give
\begin{align}
    \mathcal{D}\hat{u}^{lm}=\frac{4\pi\mu\epsilon}{r}R_{n\ell_i}&\sum_{\ell_j,m_j}[\hat{Z}^{\ell_j m_j}_1 I(i*,j,\cdot*)\nonumber\\
    &+\hat{Z}^{\ell_j m_j}_2 I(i,j,\cdot*)],\label{uhatlm}\\
%\end{align}
%\begin{align}
%\begin{split}
    \mathcal{D}(\hat{u}^*)^{lm}=\frac{4\pi\mu\epsilon}{r}R_{n\ell_i}&\sum_{\ell_j,m_j}[(\hat{Z}^*_1)^{\ell_j m_j}I(i,j*,\cdot)\nonumber\\
    &+(\hat{Z}^*_2)^{\ell_j m_j}I(i*,j*,\cdot)],\\
%\end{split}
%\end{align}
%\begin{align}
    \mathcal{L}_{+}\hat{Z}^{\ell_j m_j}_1= 2\epsilon\mu^2r R_{n\ell_i}&\sum_{l,m}\hat{u}^{lm}I(i,j*,\cdot),\label{Z1hat}\\
%\end{align}
%\begin{align}
    \mathcal{L}_{-}(\hat{Z}^*_2)^{\ell_j m_j}=  2\epsilon\mu^2r R_{n\ell_i}&\sum_{l,m}(\hat{u}^*)^{lm}I(i,j,\cdot*),\label{Z2hat}
\end{align}
where we used the orthonormality of the spherical-harmonic basis, and we recall that the linear differential operators are defined by
\begin{align}
    \mathcal{D}&\equiv\frac{d^2}{dr^2}+\frac{2}{r}\frac{d}{dr}-\frac{l(l+1)}{r^2},\\
    \mathcal{L}_{\pm}&\equiv\frac{d^2}{dr^2}+\frac{2\mu^2\Mbh}{r}-\frac{\ell_j(\ell_j+1)}{r^2}+2\mu(E_{n\ell_i}\pm\omega)\,.
\end{align}
The symbols such as $I(i*,j,\cdot)$ denote angular integrals of three spherical harmonics. The letters $i$ and $j$ label the spherical harmonics $Y_{\ell_i m_i}$ and $Y_{\ell_j m_j}$, respectively, whereas a dot labels $Y_{lm}$. The asterisk $*$ after a given label denotes a complex conjugation of the respective spherical harmonic. For example: $I(i*,j,\cdot)=\int d\Omega\, Y^*_{\ell_i m_i}Y_{\ell_j m_j}Y_{lm}$ whereas $I(i,j,\cdot*)=\int d\Omega\, Y_{\ell_i m_i}Y_{\ell_j m_j}Y^*_{lm}$. 

These angular integrals can all be written in terms of Wigner 3-j symbols using the relation (see, e.g., Ref.~\cite{Spiers:2023})
\begin{align}
\begin{split}
	I(1*,2,3)&=(-1)^{m_1}\sqrt{\frac{(2\ell_1+1)(2\ell_2+1)(2\ell_3+1)}{4\pi}}\\
	&\times
	\begin{pmatrix}
	\ell_1 & \ell_2 & \ell_3\\ 
	0 & 0 & 0
	\end{pmatrix}
	\begin{pmatrix}
	\ell_1 & \ell_2 & \ell_3\\ 
	-m_1 & m_2 & m_3
	\end{pmatrix}\,.
\end{split}
\label{wigner3j}
\end{align}
which satisfies the selection rules (meaning the integral above vanishes unless these rules are satisfied)
\begin{align}
	&-m_1+m_2+m_3=0,\label{rule1}\\
        &|\ell_1-\ell_2|\leq \ell_3\leq \ell_1+\ell_2,\label{rule2}\\
        &\ell_1+\ell_2+\ell_3=2p,\quad\text{for }p\in\mathbb{Z}.\label{rule3}
\end{align} 
By using the identity $Y_{\ell_1,-m_1}=(-1)^{m_1}Y^*_{\ell_1m_1}$, all angular integrals in Eqs.~\eqref{uhatlm}--\eqref{Z2hat} can be written in terms of Wigner 3-j symbols using Eq.~\eqref{wigner3j} and from which the selection rules directly follows.
Rule~\eqref{rule3} fixes the parity of $\ell_j$, and since $\ell_i,\ell_j,l$ are integers, rule~\eqref{rule2} implies that $\ell_j$ can take the values $\ell_j=\left\{|l-\ell_i|,|l-\ell_i|+2,...,l+\ell_i\right\}$ or, equivalently, $\ell_j=|l-\ell_i|+2k$ with $0\leq k\leq\min(l,\ell_i)$. The upper value of $k$ may be understood either from explicitly counting the number of possible values that $\ell_j$ can take depending on the parities of $l$ and $\ell_i$ ($\ell_j$ can take $\min(l,\ell_i)+1$ values) or by using $\min(l,\ell_i)=(l+\ell_i-|l-\ell_i|)/2$.
This deduction, along with the application of rule~\eqref{rule1} and Eq.~\eqref{wigner3j}, allows us to write
\begin{align}
    I(i*,j,\cdot*)&=\sum_{k=0}^{\min(l,\ell_i)}\hspace{-0.2cm}(C_1)^{l\ell_i k}_{m m_i}\delta_{\ell_j,|l-\ell_i|+2k}\delta_{m_j,m+m_i},\label{int1}\\
    I(i,j,\cdot*)&=\sum_{k=0}^{\min(l,\ell_i)}\hspace{-0.2cm}(C_2)^{l\ell_i k}_{m m_i}\delta_{\ell_j,|l-\ell_i|+2k}\delta_{m_j,m-m_i},\label{int2}\\
    I(i*,j*,\cdot)&=\sum_{k=0}^{\min(l,\ell_i)}\hspace{-0.2cm}(C_3)^{l\ell_i k}_{m m_i}\delta_{\ell_j,|l-\ell_i|+2k}\delta_{m_j,m-m_i},\label{int3}\\
    I(i,j*,\cdot)&=\sum_{k=0}^{\min(l,\ell_i)}\hspace{-0.2cm}(C_4)^{l\ell_i k}_{m m_i}\delta_{\ell_j,|l-\ell_i|+2k}\delta_{m_j,m+m_i},\label{int4}
\end{align}
with
\begin{align}
\begin{split}
    &(C_1)^{l\ell_i k}_{m m_i}\equiv (-1)^{m_i+m}\\
    &\times \sqrt{\frac{(2l+1)(2\ell_i+1)(2|l-\ell_i|+4k+1)}{4\pi}}\\
    &\times\begin{pmatrix}l & \ell_i & |l-\ell_i|+2k\\ 0 & 0 & 0\end{pmatrix}\begin{pmatrix}l & \ell_i & |l-\ell_i|+2k\\ -m & -m_i & m+m_i\end{pmatrix}, 
\end{split}
\label{C1}
\end{align}
\begin{align}
\begin{split}
    &(C_2)^{l\ell_i k}_{m m_i}\equiv (-1)^{m}\sqrt{\frac{(2l+1)(2\ell_i+1)(2|l-\ell_i|+4k+1)}{4\pi}}\\
    &\times\begin{pmatrix}l & \ell_i & |l-\ell_i|+2k\\ 0 & 0 & 0\end{pmatrix}\begin{pmatrix}l & \ell_i & |l-\ell_i|+2k\\ -m & m_i & m-m_i\end{pmatrix}, 
\end{split}
\label{C2}
\end{align}
\begin{align}
\begin{split}
    &(C_3)^{l\ell_i k}_{m m_i}\equiv (-1)^{m}\sqrt{\frac{(2l+1)(2\ell_i+1)(2|l-\ell_i|+4k+1)}{4\pi}}\\
    &\times\begin{pmatrix}l & \ell_i & |l-\ell_i|+2k\\ 0 & 0 & 0\end{pmatrix}\begin{pmatrix}l & \ell_i & |l-\ell_i|+2k\\ m & -m_i & -m+m_i\end{pmatrix}, 
\end{split}
\label{C3}
\end{align}
\begin{align}
\begin{split}
    &(C_4)^{l\ell_i k}_{m m_i}\equiv (-1)^{m_i+m}\\
    &\times\sqrt{\frac{(2l+1)(2\ell_i+1)(2|l-\ell_i|+4k+1)}{4\pi}}\\
    &\times\begin{pmatrix}l & \ell_i & |l-\ell_i|+2k\\ 0 & 0 & 0\end{pmatrix}\begin{pmatrix}l & \ell_i & |l-\ell_i|+2k\\ m & m_i & -m-m_i\end{pmatrix}. 
\end{split}
\label{C4}
\end{align}
The symmetries of the Wigner 3-j symbols [see Eq.~\eqref{WignerColunas3}] then give
\begin{align}
    (C_1)^{l\ell_i k}_{m m_i}&=(C_4)^{l\ell_i k}_{m m_i},\label{symmetry1}\\
    (C_2)^{l\ell_i k}_{m m_i}&=(C_3)^{l\ell_i k}_{m m_i},\label{symmetry2}\\
    (C_3)^{l\ell_i k}_{m, m_i}&=(-1)^{m_i}(C_1)^{l\ell_i k}_{-m, m_i},\label{symmetry3}\\
    (C_2)^{l\ell_i k}_{m, m_i}&=(-1)^{m_i}(C_4)^{l\ell_i k}_{-m, m_i}.\label{symmetry4}
\end{align}

Inserting the integrals~\eqref{int1}--\eqref{int4} in Eqs.~\eqref{uhatlm}--\eqref{Z2hat} and performing the expansions~\eqref{Z1pert}--\eqref{upert}, we then arrive at Eqs.~\eqref{u0}--\eqref{Z2ode} of the main text. 
%
%%%%%
\section{Tidal potential produced by a secondary body moving in circular orbits}\label{app:BCs_gravpot}
%%%%%
In this Appendix we discuss the conditions for determining the integration constant $A_{lm}(\omega)$ in Eq.~\eqref{u0sol} of the main text. To do so we will compute the tidal potential $V$ produced by a secondary body moving in circular orbits, in the frequency-domain.\par
In Newtonian gravity the gravitational potential satisfies Poisson's equation $\nabla^2 V=4\pi\rho$, where $\rho$ is the mass density sourcing $V$. The solution to Poisson's equation is given by
\begin{equation}
    V(t,\boldsymbol{x})=-\int\frac{\rho(t,\boldsymbol{x}')}{|\boldsymbol{x}-\boldsymbol{x}'|}d^3x',
\label{ExtPot}
\end{equation}
with the variable $\boldsymbol{x}'$ sweeping all source points.\par
Consider the identity (for a proof, see~\cite{Poisson:2014})
\begin{equation}
	\frac{1}{|\boldsymbol{x}-\boldsymbol{x}'|}=\sum_{l=0}^{\infty}\sum_{m=-l}^{l}\frac{4\pi}{2l+1}\frac{r^{l}_{<}}{r^{l+1}_{>}}Y^*_{lm}(\theta',\varphi')Y_{lm}(\theta,\varphi),
\label{Ident}
\end{equation}
with $(r,\theta,\varphi)$ and $(r',\theta',\varphi')$ the spherical polar coordinates corresponding to points $\boldsymbol{x}$ and $\boldsymbol{x}'$, respectively, and $r_{<}\equiv \min(r,r')$, $r_{>}\equiv \max(r,r')$. Remembering that our reference frame has the $z$ axis aligned with the orbital angular momentum of the system (see Sec.~\ref{section: LPT}) and that the orbit is circular, the coordinates of the secondary body are $(r',\theta',\varphi')=(r_{\text{orb}},\pi/2,\Omegaorb t)$. On the other hand, assumption~\eqref{region} means that the region in which the field equations are being solved satisfies $r<r_{\text{orb}}$, and therefore in this region $r_{<}=r$ and $r_{>}=r_{\text{orb}}$. Furthermore, since we are working in the center-of-mass frame of the body suffering the perturbation, which is a noninertial reference frame, the dipole term vanishes\footnote{There are different ways of showing this: Ref.~\cite{Poisson:2014} resorts to Euler's equation while Ref.~\cite{Baumann:2018} makes an expansion in mass ratios.}. On the other hand, the monopole term is constant, and irrelevant in this context. Hence, we may write the series starting at $l=2$. 

Therefore, in our region of interest, Eqs.~\eqref{ExtPot} and~\eqref{Ident} give
\begin{align}
    V(t,\boldsymbol{x})=-\sum_{l=2}^{\infty}&\sum_{m=-l}^l\frac{4\pi M_{\text{sec}}}{2l+1}\frac{r^l}{r_{\text{orb}}^{l+1}}\nonumber\\
    &\times Y^*_{lm}\left(\frac{\pi}{2},\Omegaorb t\right)Y_{lm}(\theta,\varphi),
\label{Vext}
\end{align}
with $M_{\text{sec}}=\int\rho(t,\boldsymbol{x}')d^3x'$ the mass of the secondary body. Finally, using Eq.~\eqref{Ylm*} we get
\begin{equation}
    V(t,\boldsymbol{x})=-M_{\text{sec}}\sum_{l=2}^{\infty}\sum_{m=-l}^l\frac{W_{lm}}{r_{\text{orb}}^{l+1}}r^lY_{lm}(\theta,\varphi)e^{-im\Omegaorb t},
\end{equation}
where
\begin{align}
\begin{split}
    W_{lm}&=(-1)^{\frac{l+m}{2}}\sqrt{\frac{4\pi}{2l+1}(l-m)!(l+m)!}\\
    &\times\left[2^l\left(\frac{l+m}{2}\right)!\left(\frac{l-m}{2}\right)!\right]^{-1},\ \text{if }l+m\text{ is even}
\end{split}
\label{Wlm}
\end{align}
whereas $W_{lm}=0$ if $l+m$ is odd. From this expression it follows that $W_{l,-m}=(-1)^m W_{lm}$. We can now change to the frequency-domain. The Fourier transform of this potential is given by
\begin{align}
\begin{split}
    &\hat{V}(\omega,\boldsymbol{x})=\int V(t,\boldsymbol{x})e^{i\omega t}dt\\
    &=-2\pi M_{\text{sec}}\sum_{l=2}^{\infty}\sum_{m=-l}^l\frac{W_{lm}}{r_{\text{orb}}^{l+1}}r^lY_{lm}(\theta,\varphi)\delta(\omega-m\Omegaorb).
\end{split}
\end{align}
Therefore, requiring that $\hat{u}_{(0)}^{lm}(\omega,r)=A_{lm}(\omega)r^l$ describes this tidal field, one finds
\begin{align}
    A_{lm}(\omega)&=c_{lm}\delta(\omega-m\Omegaorb),\\
    c_{lm}&=-2\pi M_{\text{sec}} r_{\text{orb}}^{-(l+1)} W_{lm}.
\end{align}
The constants $c_{lm}$ inherit the symmetries of $W_{lm}$, meaning they vanish for odd $l+m$ and $c_{l,-m}=(-1)^mc_{lm}$. Using the parity symmetry of the delta function then gives $A_{l,-m}(-\omega)=(-1)^m A_{lm}(\omega)$.
%
%%%%%
%%%%%
\section{Derivation of the tidal Love numbers for a spherically symmetric and a dipolar cloud}\label{ssec:TLN_00_11}
%%%%%
In this Appendix we use the general results of this work to obtain the TLNs for bosonic clouds with $(n,\ell_i,m_i)=(0,0,0)$ and $(n,\ell_i,m_i)=(0,1,1)$, which were given in Eqs.~\eqref{specific1}--\eqref{specific3}.
\vspace*{1pt}
\subsection{Spherically symmetric cloud}
Let us start with the case $n=\ell_i=m_i=0$. Applying Eq.~\eqref{Kgeral} and using the expressions for the coefficients $\hat{u}^{lm}_{(2),sX}$ found in Appendix~\ref{app:gravpot_auxfunc}, namely Eq.~\eqref{solucoesu2} jointly with~\eqref{usE}, we find
\begin{align}
\begin{split}
    k_{lm}^{(n=\ell_i=m_i=0)}&=\frac{1}{2^{2l+2}}\left[(C_1)^{l00}_{m0}\right]^2\mathcal{G}_0^{00l}\frac{1}{\alpha^{4l+2}}\frac{M_c}{\Mbh},
\end{split}
\label{TLNsph}
\end{align}
where we used $\left[(C_2)^{l00}_{m0}\right]^2=\left[(C_1)^{l00}_{m0}\right]^2$ jointly with Eq.~\eqref{Eli}, $M_c\simeq\mu\epsilon^2$ [see Eq.~\eqref{cloudmass}] and we note that, for circular orbits, the TLNs are only defined for even $l+m$. Now all that remains is to compute $(C_1)^{l00}_{m0}$ using Eq.~\eqref{C1} whereas the quantity $\mathcal{G}_0^{00l}$ is to be computed using Eq.~\eqref{Gcal}.
From Eqs.~\eqref{WignerColunas} and \eqref{WignerFrac} one simply has $(C_1)^{l00}_{m0}=(4\pi)^{-1/2}$. On the other hand, for $\mathcal{G}_0^{00l}$ we have
%\begin{widetext}
\begin{align}
\begin{split}
%\hspace{-1cm}
	\mathcal{G}^{00l}_0&=\frac{4\pi}{2l+1}\frac{\Gamma(l)}{\Gamma(2+2l)}\int_0^{\infty}e^{-y}y^{2l+2} \\
 &\times\Bigg[U(l,2l+2,y)\int_0^{y} e^{-x}x^{2l+2} M(l,2l+2,x)\, dx\\
 &+ M(l,2l+2,y)\int_{y}^{\infty}e^{-x} x^{2l+2} U(l,2l+2,x)\, dx\Bigg] dy.\\
\end{split}
\label{G000l}
\end{align}
%\end{widetext}
The indefinite integrals inside the square brackets can be computed explicitly\footnote{We used the \textit{Mathematica} software, version 14.0.}
\begin{widetext}
\begin{align}
%\begin{split}
	\int_0^{y} e^{-x}x^{2l+2} M(l,2l+2,x)\, dx&=4^{l}\frac{\Gamma(l+3/2)}{l+1}e^{-y/2}y^{l+\frac{1}{2}}\left[(4+8l+(2+4l)y+y^2)I_{l+\frac{1}{2}}\left(\frac{y}{2}\right)-(2y+y^2)I_{l-\frac{1}{2}}\left(\frac{y}{2}\right)\right]\,,\\
    \int_y^{\infty} e^{-x}x^{2l+2} U(l,2l+2,x)\, dx&=\frac{1}{2\sqrt{\pi}}e^{-y/2}y^{l+\frac{1}{2}}\left[(4+8l+(2+4l)y+y^2)K_{l+\frac{1}{2}}\left(\frac{y}{2}\right)+(2y+y^2)K_{l-\frac{1}{2}}\left(\frac{y}{2}\right)\right]\,,
%\end{split}
\end{align}
\end{widetext}
where $I_v(y)$ and $K_v(y)$ are modified Bessel functions of the first and second kind, respectively. To solve the definite integral in Eq.~\eqref{G000l} analytically with Mathematica we found it necessary to write the Kummer and Tricomi functions inside the integrand in terms of Whittaker functions using Eqs.~\eqref{MtoM} and~\eqref{UtoW}, and to use the following relations between the modified Bessel functions of the first and second kind and the Whittaker functions~\cite{weisstein4,weisstein5}
\begin{align}
	I_{l\pm\frac{1}{2}}\left(\frac{y}{2}\right)&=\frac{1}{2^{2l\pm 1}\Gamma\left(l+1\pm\frac{1}{2}\right)}y^{-\frac{1}{2}}M_{0,l\pm\frac{1}{2}}(y), \label{BesselI}\\
	K_{l\pm\frac{1}{2}}\left(\frac{y}{2}\right)&=\sqrt{\pi}y^{-\frac{1}{2}}W_{0,l\pm\frac{1}{2}}(y).\label{BesselK}
\end{align}
Using these relations we find the following analytical expression for $\mathcal{G}^{00l}_0$
\\
\\
\begin{align}
\begin{split}
	\mathcal{G}^{00l}_0&=16\pi(l+2)\frac{\Gamma(l)\Gamma(4+4l)}{\Gamma(3+3l)}\Big[{}_2 F_1(l,4+4l;3+3l;-1)\\
 &-2\, {}_2 F_1(l+1,4+4l;3+3l;-1)\Big].
\end{split}
\end{align}
Plugging this result, jointly with $(C_1)^{l00}_{m0}=(4\pi)^{-1/2}$, in Eq.~\eqref{TLNsph} we obtain Eq.~\eqref{specific1} of the main text.
\subsection{Dipolar cloud}
Consider now the state $(n,\ell_i,m_i)=(0,1,1)$. For this case we need to consider the cases $l=2$ and $l>2$ separately, as one can see by inspecting the different possible cases of Eq.~\eqref{solucoesu2}. Let us first look at $l=2$. Applying again~\eqref{Kgeral} but now using the expression in Eq.~\eqref{usF} for the coefficient $\hat{u}^{2m}_{(2),sX}$ we find
\begin{widetext}
\begin{align}\label{k2m_der}
\begin{split}
    k_{2m}^{(n=0,\ell_i=m_i=1)}
    &= \lim_{\Omegaorb\to 0}\Bigg\{-\frac{360\pi}{m}\left[(C_1)^{210}_{m1}(C_1)^{210}_{m1}-(C_2)^{210}_{m1}(C_2)^{210}_{m1}\right]\frac{\mu^3\Mbh^2}{\Omegaorb}\\
    &+\frac{1}{2}\left[(C_1)^{211}_{m1}(C_1)^{211}_{m1}+(C_2)^{211}_{m1}(C_2)^{211}_{m1}\right]\mathcal{G}_1^{012}\\
    &-\frac{1}{2}\left[(C_1)^{210}_{m1}(C_1)^{210}_{m1}+(C_2)^{210}_{m1}(C_2)^{210}_{m1}\right]\mathcal{F}^{012}\Bigg\}\frac{1}{\alpha^{10}}\frac{M_c}{\Mbh},\qquad\qquad\text{if $m\neq 0$}\\
    &=\Bigg((C_1)^{211}_{01}(C_1)^{211}_{01}\mathcal{G}_1^{012}-(C_1)^{210}_{01}(C_1)^{210}_{01}\mathcal{F}^{012}\Bigg)\frac{1}{\alpha^{10}}\frac{M_c}{\Mbh},\qquad\,\ \ \text{ if $m=0$},
\end{split}
\end{align}
\end{widetext}
where we remind the reader that for the case $m=0$ we used the fact that $\left[(C_2)^{21k}_{01}\right]^2=\left[(C_1)^{21k}_{01}\right]^2$. The coefficients $\mathcal{F}^{012}$ and $\mathcal{G}_1^{012}$ can be computed using  Eqs.~\eqref{Fcal} and~\eqref{Gcal}. Computing the integrals as was done in the previous section, we get 
\begin{align}
    \mathcal{F}^{012}&=1000560\pi+\frac{\pi^2}{90}
    \MeijerG
  {2,3}
  {4,4}
  {
   1,8,14,29/2\\
   14,14,7,29/2
  }
  {1},\\
  \mathcal{G}_1^{012}&=11648\pi,
\end{align}
where $
\MeijerG*
  {n,m}
  {q,p}
  {\cdot
  }
  {z}
$ is the Meijer G-function. Moreover, using Eqs.~\eqref{C1} and~\eqref{C2} jointly with the properties of the Wigner 3-j symbols, Eqs.~\eqref{WignerColunas2}--\eqref{WignerSoma}, we also find
\begin{align}
    (C_1)^{210}_{m1}&=-\sqrt{\frac{(m-1)(m-2)}{40\pi}},\\
    (C_2)^{210}_{m1}&=\sqrt{\frac{(m+1)(m+2)}{40\pi}},\\
    (C_1)^{211}_{m1}&=\frac{3}{2}\sqrt{\frac{(m+3)(m+4)}{210\pi}},\\
    (C_2)^{211}_{m1}&=-\frac{3}{2}\sqrt{\frac{(m-3)(m-4)}{210\pi}}.
\end{align}
Plugging everything back in Eq.~\eqref{k2m_der} results in Eqs.~\eqref{specific2} and~\eqref{specific2_mneq0} of the main text, for the $m=0$ and $m\neq 0$ cases, respectively.

On the other hand, an inspection of  Eq.~\eqref{solucoesu2} tells us that for the case $l>2$ we need to use Eq.~\eqref{usE} to compute the coefficients $\hat{u}^{lm}_{(2),sX}$. We notice that Eq.~\eqref{usE} does not contain any term proportional to $\omega^{-1}$, meaning that in this case we actually have well-defined static TLNs for any $m$. Nonetheless, for simplicity, let us focus on the case $m=0$. Again using Eq.~\eqref{Kgeral} we get
\begin{align}\label{KlDem}
\begin{split}
    k^{(n=0,\ell_i=m_i=1)}_{l>2,0}=&\left\{\left[(C_1)^{l10}_{01}\right]^2\mathcal{G}_0^{01l}+\left[(C_1)^{l11}_{01}\right]^2\mathcal{G}_1^{01l}\right\}\\
    &\times\frac{1}{\alpha^{4l+2}}\frac{M_c}{\Mbh}\,.
\end{split}
\end{align}
Notice that, given our choice of circular orbits, these TLNs are only defined for even $l$. The procedure is now completely analogous to the one used above in the spherically symmetric cloud case. Therefore, below we immediately present all necessary coefficients in their final form
%\begin{widetext}
\begin{align}
%\begin{split}
	\mathcal{G}^{01l}_0=&\frac{8\pi}{3}\frac{(l+2)(l+3)(2l-1)(-6-8l+3l^2+2l^3)}{2l+1}\nonumber\\
    &\times\frac{\Gamma(l-2)\Gamma(4l+4)}{\Gamma(3l+3)}\Big[{}_2 F_1(l,4+4l;3+3l;-1)\nonumber\\
    &-2\, {}_2 F_1(l+1,4+4l;3+3l;-1)\Big],\\
    \nonumber\\
    \nonumber\\
    \mathcal{G}^{01l}_1&=\frac{2\pi}{3}\frac{4+7l+2l^2}{(l+3)(2l+1)}\Gamma(l)\Gamma(4l+7)\nonumber\\
    &\times\Big[\frac{5}{\Gamma(3l+3)} {}_2 F_1(l,6+4l;3+3l;-1)\nonumber\\
    &-\frac{12(3l+4)}{\Gamma(3l+4)} {}_2 F_1(l+1,6+4l;3+3l;-1)\Big],\,
%\end{split}
\end{align}
%\end{widetext}
and
\begin{align}
    (C_1)^{l10}_{01}&=-\sqrt{\frac{3}{8\pi}}\sqrt{\frac{l(l-1)}{(2l-1)(2l+1)}}\,,\\
    (C_1)^{l11}_{01}&=\sqrt{\frac{3}{8\pi}}\sqrt{\frac{(l+1)(l+2)}{(2l+1)(2l+3)}}\,,
\end{align}
where we note that we used Eq.~\eqref{WignerRaiz} to compute the Wigner 3-j symbols. Finally, substituting back in Eq.~\eqref{KlDem} results in Eq.~\eqref{specific3} of the main text.
\\
%
%%%%%%%
\section{Explicit form of auxiliary functions}\label{app:aux_func}
%%%%%%%
In this Appendix we provide the explicit form for some auxiliary functions that we defined in order to write down the perturbed scalar field and gravitational potential in a simplified format. To simplify the expressions, in this Appendix we define the quantity $K_{n\ell_i}\equiv \sqrt{8\mu|E_{n\ell_i}|}$.

\subsection{Auxiliary functions for the perturbed scalar field}\label{app:scalar_auxfunc}
%%%%%
The auxiliary functions in Tables~\ref{table:Z1s1} and ~\ref{table:Z1s2} are
\begin{widetext}
{\small
\begin{align}
%\hspace{-1cm}
\begin{split}
    F^{(-1)}_{<}(r)&=\frac{4\mu^2(-1)^{l+1}\left(K_{n\ell_i}\right)^{\ell_i+2k-2l-1/2}}{\Gamma(2+2\ell_i-2l+4k)(n+\ell_i+1)(n+l-2k)!\sqrt{2n!(n+\ell_i+1)(n+2\ell_i+1)!}}e^{-K_{n\ell_i}r/2}r^{\ell_i-l+2k+1}\\
    &\times\Bigg\{U\left(-n-l+2k,2\ell_i-2l+4k+2,K_{n\ell_i}r\right)\sum_{s=0}^{2n+l-2k}\sum_{p=0}^s(-1)^{n+s}\binom{n+l-2k}{p}\binom{n}{s-p}\frac{(2\ell_i+2+s-p)_{n-s+p}}{(2\ell_i-2l+4k+2)_p}\\
    &\times\left[(2\ell_i+2k+s+2)!-\sum_{q=0}^{2\ell_i+2k+s+2}\frac{(2\ell_i+2k+s+2)!}{q!}\left(K_{n\ell_i}r\right)^q e^{-K_{n\ell_i}r}\right]\\
    &+M\left(-n-l+2k,2\ell_i-2l+4k+2,K_{n\ell_i}r\right)\sum_{s=0}^{2n+l-2k}\sum_{p=0}^s(-1)^{l+s}\binom{n+l-2k}{p}\binom{n}{s-p}\\
    &\times (2\ell_i-2l+4k+2+p)_{n+l-2k-p}(2\ell_i+2+s-p)_{n-s+p}\sum_{q=0}^{2\ell_i+2k+s+2}\frac{(2\ell_i+2k+s+2)!}{q!}\left(K_{n\ell_i}r\right)^q e^{-K_{n\ell_i}r}\Bigg\},
\end{split}
\label{F<}
\end{align}
\begin{align}
%\hspace{-1cm}
    F^{(0)}_{<}(r)&=\frac{2\mu^2(-1)^{l+1}\left(K_{n\ell_i}\right)^{\ell_i+2k-2l-1/2}}{\Gamma(2+2\ell_i-2l+4k)(n+\ell_i+1)(n+l-2k)!\sqrt{2n!(n+\ell_i+1)(n+2\ell_i+1)!}}e^{-K_{n\ell_i}r/2}r^{\ell_i-l+2k+1}\notag\\
    &\times\Bigg\{\frac{1}{n+\ell_i+1}U\left(-n-l+2k,2\ell_i-2l+4k+2,K_{n\ell_i}r\right)\left(2\ell_i-2l+4k+\frac{5}{2}-\frac{K_{n\ell_i}r}{2}+\psi(n+l-2k+1)\right)\notag\\
    &\times\sum_{s=0}^{2n+l-2k}\sum_{p=0}^s(-1)^{n+s}\binom{n+l-2k}{p}\binom{n}{s-p}\frac{(2\ell_i+2+s-p)_{n-s+p}}{(2\ell_i-2l+4k+2)_p}\Bigg[(2\ell_i+2k+s+2)!\notag\\
    &-\sum_{q=0}^{2\ell_i+2k+s+2}\frac{(2\ell_i+2k+s+2)!}{q!}\left(K_{n\ell_i}r\right)^q e^{-K_{n\ell_i}r}\Bigg]-\frac{1}{2(n+\ell_i+1)}U\left(-n-l+2k,2\ell_i-2l+4k+2,K_{n\ell_i}r\right)\notag\\
    &\times\sum_{s=0}^{2n+l-2k}\sum_{p=0}^s(-1)^{n+s}\binom{n+l-2k}{p}\binom{n}{s-p}\frac{(2\ell_i+2+s-p)_{n-s+p}}{(2\ell_i-2l+4k+2)_p}\Bigg[(2\ell_i+2k+s+3)!\notag\\
    &-\sum_{q=0}^{2\ell_i+2k+s+3}\frac{(2\ell_i+2k+s+3)!}{q!}\left(K_{n\ell_i}r\right)^q e^{-K_{n\ell_i}r}\Bigg]+\frac{-n-l+2k}{(2\ell_i-2l+4k+2)(n+\ell_i+1)}\notag\\
    &\times U\left(-n-l+2k,2\ell_i-2l+4k+2,K_{n\ell_i}r\right)\sum_{s=0}^{2n+l-2k-1}\sum_{p=0}^s(-1)^{n+s}\binom{n+l-2k-1}{p}\binom{n}{s-p}\notag\\
    &\times\frac{(2\ell_i+2+s-p)_{n-s+p}}{(2\ell_i-2l+4k+3)_p}\Bigg[(2\ell_i+2k+s+3)!-\sum_{q=0}^{2\ell_i+2k+s+3}\frac{(2\ell_i+2k+s+3)!}{q!}\left(K_{n\ell_i}r\right)^q e^{-K_{n\ell_i}r}\Bigg]\notag\\
    &+U\left(-n-l+2k,2\ell_i-2l+4k+2,K_{n\ell_i}r\right)I^{\ell_j=\ell_i-l+2k}_{dM}(r)+\Bigg[\frac{n+l-2k}{n+\ell_i+1}K_{n\ell_i}\, r\notag\\
    &\times U\left(-n-l+2k+1,2\ell_i-2l+4k+3,K_{n\ell_i}r\right)+\frac{\partial}{\partial a}U\left(a,2\ell_i-2l+4k+2,K_{n\ell_i}r\right)\Big|_{a=-n-l+2k}\Bigg]\notag\\
    &\times\sum_{s=0}^{2n+l-2k}\sum_{p=0}^s(-1)^{n+s}\binom{n+l-2k}{p}\binom{n}{s-p}\frac{(2\ell_i+2+s-p)_{n-s+p}}{(2\ell_i-2l+4k+2)_p}\Bigg[(2\ell_i+2k+s+2)!\notag\\
    &-\sum_{q=0}^{2\ell_i+2k+s+2}\frac{(2\ell_i+2k+s+2)!}{q!}\left(K_{n\ell_i}r\right)^q e^{-K_{n\ell_i}r}\Bigg]\notag\\
    &+\frac{1}{n+\ell_i+1}M\left(-n-l+2k,2\ell_i-2l+4k+2,K_{n\ell_i}r\right)\left(2\ell_i-2l+4k+\frac{5}{2}-\frac{K_{n\ell_i}r}{2}+\psi(n+l-2k+1)\right)\notag\\
    &\times\sum_{s=0}^{2n+l-2k}\sum_{p=0}^s(-1)^{l+s}\binom{n+l-2k}{p}\binom{n}{s-p}(2\ell_i-2l+4k+2+p)_{n+l-2k-p}(2\ell_i+2+s-p)_{n-s+p}\notag\\
    &\times\sum_{q=0}^{2\ell_i+2k+s+2}\frac{(2\ell_i+2k+s+2)!}{q!}\left(K_{n\ell_i}r\right)^q e^{-K_{n\ell_i}r}-\frac{1}{2(n+\ell_i+1)}M\left(-n-l+2k,2\ell_i-2l+4k+2,K_{n\ell_i}r\right)\notag\\
    &\times\sum_{s=0}^{2n+l-2k}\sum_{p=0}^s(-1)^{l+s}\binom{n+l-2k}{p}\binom{n}{s-p}(2\ell_i-2l+4k+2+p)_{n+l-2k-p}(2\ell_i+2+s-p)_{n-s+p}\notag\\
    &\times\sum_{q=0}^{2\ell_i+2k+s+3}\frac{(2\ell_i+2k+s+3)!}{q!}\left(K_{n\ell_i}r\right)^q e^{-K_{n\ell_i}r}+\Bigg[\frac{-n-l+2k}{n+\ell_i+1}K_{n\ell_i}\, r\notag\\
    &\times M\left(-n-l+2k+1,2\ell_i-2l+4k+3,K_{n\ell_i}r\right)+\frac{\partial}{\partial a}M\left(a,2\ell_i-2l+4k+2,K_{n\ell_i}r\right)\Big|_{a=-n-l+2k}\Bigg]\notag\\
    &\times\sum_{s=0}^{2n+l-2k}\sum_{p=0}^s(-1)^{l+s}\binom{n+l-2k}{p}\binom{n}{s-p}(2\ell_i-2l+4k+2+p)_{n+l-2k-p}(2\ell_i+2+s-p)_{n-s+p}\notag\\
    &\times \sum_{q=0}^{2\ell_i+2k+s+2}\frac{(2\ell_i+2k+s+2)!}{q!}\left(K_{n\ell_i}r\right)^q e^{-K_{n\ell_i}r}+\frac{n+l-2k}{n+\ell_i+1}M\left(-n-l+2k,2\ell_i-2l+4k+2,K_{n\ell_i}r\right)\notag\\
    &\times\sum_{s=0}^{2n+l-2k-1}\sum_{p=0}^s(-1)^{l+s+1}\binom{n+l-2k-1}{p}\binom{n}{s-p}(2\ell_i-2l+4k+3+p)_{n+l-2k-1-p}(2\ell_i+2+s-p)_{n-s+p}\notag\\
    &\times%\hspace{-0.5cm} 
    \sum_{q=0}^{2\ell_i+2k+s+3}\frac{(2\ell_i+2k+s+3)!}{q!}\left(K_{n\ell_i}r\right)^q e^{-K_{n\ell_i}r}+M\left(-n-l+2k,2\ell_i-2l+4k+2,K_{n\ell_i}r\right)I^{\ell_j=\ell_i-l+2k}_{dU}(r)\Bigg\},
\end{align}
\begin{align}
%\hspace{-0.5cm}
    G^{(-1)}(r)&=\frac{4\mu^2(-1)^{n+1}\left(K_{n\ell_i}\right)^{n+\ell_i-l-1/2}}{\Gamma(2+2n+2\ell_i)\sqrt{2n!(n+\ell_i+1)(n+2\ell_i+1)!}}\frac{\Gamma(n+2\ell_i+l+3)}{n+\ell_i+1}\frac{\Gamma(n+l+2)}{\Gamma(l+2)}e^{-K_{n\ell_i}r/2}r^{n+\ell_i+1},
\end{align}
\begin{align}
\begin{split}
%\hspace{-0.5cm}
    G^{(0)}(r)&=\frac{2\mu^2(-1)^n\left(K_{n\ell_i}\right)^{n+\ell_i-l-1/2}}{\Gamma(2+2n+2\ell_i)\sqrt{2n!(n+\ell_i+1)(n+2\ell_i+1)!}}e^{-K_{n\ell_i}r/2}r^{n+\ell_i+1}\\
    &\times\Bigg\{\left[-5-4n-4\ell_i+(2+2n+2\ell_i)\gamma+K_{n\ell_i}r\right]\frac{\Gamma(n+2\ell_i+l+3)\Gamma(n+l+2)}{(2+2n+2\ell_i)\Gamma(l+2)}+\frac{\Gamma(n+2\ell_i+l+4)\Gamma(n+l+3)}{(2+2n+2\ell_i)\Gamma(l+3)}\\
    &-\frac{I^{\ell_j=n+\ell_i}_{1}(r)}{2+2n+2\ell_i}-I^{\ell_j=n+\ell_i}_{2}(r)+(-1)^{n+1}\left[\sum_{s=1}^{2n+2\ell_i+1}\binom{2n+2\ell_i+1}{-s+2n+2\ell_i+1}\Gamma(s)\left(K_{n\ell_i}r\right)^{-s}-\log\left(K_{n\ell_i}r\right)\right]\\
    &\times\sum_{p=0}^n\binom{n}{p}(2\ell_i+2+p)_{n-p}(-1)^p\Bigg[(n+2\ell_i+l+p+2)!-\sum_{q=0}^{n+2\ell_i+l+p+2}\frac{(n+2\ell_i+l+p+2)!}{q!}\left(K_{n\ell_i}r\right)^q e^{-K_{n\ell_i}r}\Bigg]\\
    &+(-1)^{n+1}\frac{K_{n\ell_i}r}{2+2n+2\ell_i} {}_2F_2\left(1,1;2,3+2n+2\ell_i;K_{n\ell_i}r\right)\sum_{p=0}^n\binom{n}{p}(2\ell_i+2+p)_{n-p} (-1)^p\\
    &\times\sum_{q=0}^{n+2\ell_i+l+p+2}\frac{(n+2\ell_i+l+p+2)!}{q!}\left(K_{n\ell_i}r\right)^q e^{-K_{n\ell_i}r}\Bigg\},
\end{split}
\end{align}
\begin{align}
\begin{split}
    H_{<}(r)&=\frac{2\mu^2(-1)^{n+1}\left(K_{n\ell_i}\right)^{\ell_i-2l+2k-1/2}}{\Gamma(2+2\ell_i-2l+4k)\sqrt{2n!(n+\ell_i+1)(n+2\ell_i+1)!}}\frac{\Gamma(2k-n-l)}{\Gamma(2+2\ell_i-2l+4k)}e^{-K_{n\ell_i}r/2}r^{\ell_i-l+2k+1}\\
    &\times\Bigg[U\left(2k-n-l,2\ell_i-2l+4k+2,K_{n\ell_i}r\right) \int_0^{K_{n\ell_i}r} e^{-x}x^{2\ell_i+2k+2}\\
    &\times M(2k-n-l,2\ell_i-2l+4k+2,x)U(-n,2\ell_i+2,x)\,dx+ M\left(2k-n-l,2\ell_i-2l+4k+2,K_{n\ell_i}r\right)\\
    &\times\int_{K_{n\ell_i}r}^{\infty} e^{-x} x^{2\ell_i+2k+2}U(2k-n-l,2\ell_i-2l+4k+2,x) U(-n,2\ell_i+2,x)\,dx\Bigg],
\end{split}
\end{align}
\begin{align}
%\hspace{-1cm}
\begin{split}
    F^{(-1)}_{>}(r)&=\frac{4\mu^2(-1)^{l+1}\left(K_{n\ell_i}\right)^{2k-\ell_i-1/2}}{\Gamma(2+2l-2\ell_i+4k)(n+\ell_i+1)(n+2\ell_i-l-2k)!\sqrt{2n!(n+\ell_i+1)(n+2\ell_i+1)!}}e^{-K_{n\ell_i}r/2}r^{l-\ell_i+2k+1}\\
    &\times\Bigg\{U\left(-n-2\ell_i+l+2k,2l-2\ell_i+4k+2,K_{n\ell_i}r\right)\sum_{s=0}^{2n+2\ell_i-l-2k}\sum_{p=0}^s(-1)^{n+s}\binom{n+2\ell_i-l-2k}{p}\binom{n}{s-p}\\
    &\times\frac{(2\ell_i+2+s-p)_{n-s+p}}{(2l-2\ell_i+4k+2)_p}\left[(2l+2k+s+2)!-\sum_{q=0}^{2l+2k+s+2}\frac{(2l+2k+s+2)!}{q!}\left(K_{n\ell_i}r\right)^q e^{-K_{n\ell_i}r}\right]\\
    &+M\left(-n-2\ell_i+l+2k,2l-2\ell_i+4k+2,K_{n\ell_i}r\right)\sum_{s=0}^{2n+2\ell_i-l-2k}\sum_{p=0}^s(-1)^{l+s}\binom{n+2\ell_i-l-2k}{p}\binom{n}{s-p}\\
    &\times (2l-2\ell_i+4k+2+p)_{n+2\ell_i-2k-p} (2\ell_i+2+s-p)_{n-s+p}\sum_{q=0}^{2l+2k+s+2}\frac{(2l+2k+s+2)!}{q!}\left(K_{n\ell_i}r\right)^q e^{-K_{n\ell_i}r}\Bigg\},
\end{split}
\end{align}
\begin{align}
%\hspace{-1cm}
    F^{(0)}_{>}(r)&=\frac{2\mu^2(-1)^{l+1}\left(K_{n\ell_i}\right)^{2k-\ell_i-1/2}}{\Gamma(2+2l-2\ell_i+4k)(n+2\ell_i-l-2k)!\sqrt{2n!(n+\ell_i+1)(n+2\ell_i+1)!}}e^{-K_{n\ell_i}r/2}r^{l-\ell_i+2k+1}\notag\\
    &\times\Bigg\{\frac{1}{n+\ell_i+1}U\left(-n-2\ell_i+l+2k,2l-2\ell_i+4k+2,K_{n\ell_i}r\right)\Bigg(2l-2\ell_i+4k+\frac{5}{2}-\frac{K_{n\ell_i}r}{2}\notag\\
    &+\psi(n+2\ell_i-l-2k+1)\Bigg)\sum_{s=0}^{2n+2\ell_i-l-2k}\sum_{p=0}^s(-1)^{n+s}\binom{n+2\ell_i-l-2k}{p}\binom{n}{s-p}\frac{(2\ell_i+2+s-p)_{n-s+p}}{(2l-2\ell_i+4k+2)_p}\notag\\
    &\times\Bigg[(2l+2k+s+2)!-\sum_{q=0}^{2l+2k+s+2}\frac{(2l+2k+s+2)!}{q!}\left(K_{n\ell_i}r\right)^q e^{-K_{n\ell_i}r}\Bigg]\notag\\
    &-\frac{1}{2(n+\ell_i+1)}U\left(-n-2\ell_i+l+2k,2l-2\ell_i+4k+2,K_{n\ell_i}r\right)\sum_{s=0}^{2n+2\ell_i-l-2k}\sum_{p=0}^s(-1)^{n+s}\binom{n+2\ell_i-l-2k}{p}\notag\\
    &\times\binom{n}{s-p}\frac{(2\ell_i+2+s-p)_{n-s+p}}{(2l-2\ell_i+4k+2)_p}\Bigg[(2l+2k+s+3)!-\sum_{q=0}^{2l+2k+s+3}\frac{(2l+2k+s+3)!}{q!}\left(K_{n\ell_i}r\right)^q e^{-K_{n\ell_i}r}\Bigg]\notag\\
    &+\frac{-n-2\ell_i+l+2k}{(2l-2\ell_i+4k+2)(n+\ell_i+1)} U\left(-n-2\ell_i+l+2k,2l-2\ell_i+4k+2,K_{n\ell_i}r\right)\notag\\
    &\times\sum_{s=0}^{2n+2\ell_i-l-2k-1}\sum_{p=0}^s(-1)^{n+s}\binom{n+2\ell_i-l-2k-1}{p}\binom{n}{s-p}\frac{(2\ell_i+2+s-p)_{n-s+p}}{(2l-2\ell_i+4k+3)_p}\Bigg[(2l+2k+s+3)!\notag\\
    &-\sum_{q=0}^{2l+2k+s+3}\frac{(2l+2k+s+3)!}{q!} \left(K_{n\ell_i}r\right)^q e^{-K_{n\ell_i}r}\Bigg]+U\left(-n-2\ell_i+l+2k,2l-2\ell_i+4k+2,K_{n\ell_i}r\right)\notag\\
    &\times I^{\ell_j=l-\ell_i+2k}_{dM}(r)+\Bigg[\frac{n+2\ell_i-l-2k}{n+\ell_i+1}K_{n\ell_i}\, r\, U\left(-n-2\ell_i+l+2k+1,2l-2\ell_i+4k+3,K_{n\ell_i}r\right)\notag\\
    &+\frac{\partial}{\partial a}U\left(a,2l-2\ell_i+4k+2,K_{n\ell_i}r\right)\Big|_{a=-n-2\ell_i+l+2k}\Bigg]\sum_{s=0}^{2n+2\ell_i-l-2k}\sum_{p=0}^s(-1)^{n+s}\binom{n+2\ell_i-l-2k}{p}\binom{n}{s-p}\notag\\
    &\times\frac{(2\ell_i+2+s-p)_{n-s+p}}{(2l-2\ell_i+4k+2)_p}\Bigg[(2l+2k+s+2)!-\sum_{q=0}^{2l+2k+s+2}\frac{(2l+2k+s+2)!}{q!}\left(K_{n\ell_i}r\right)^q e^{-K_{n\ell_i}r}\Bigg]+\frac{1}{n+\ell_i+1}\notag\\
    &\times M\left(-n-2\ell_i+l+2k,2l-2\ell_i+4k+2,K_{n\ell_i}r\right)\left(2l-2\ell_i+4k+\frac{5}{2}-\frac{K_{n\ell_i}r}{2}+\psi(n+2\ell_i-l-2k+1)\right)\notag\\
    &\times\sum_{s=0}^{2n+2\ell_i-l-2k}\sum_{p=0}^s(-1)^{l+s}\binom{n+2\ell_i-l-2k}{p}\binom{n}{s-p}(2l-2\ell_i+4k+2+p)_{n+2\ell_i-l-2k-p}\notag\\
    &\times(2\ell_i+2+s-p)_{n-s+p}\sum_{q=0}^{2l+2k+s+2}\frac{(2l+2k+s+2)!}{q!}\left(K_{n\ell_i}r\right)^q e^{-K_{n\ell_i}r}-\frac{1}{2(n+\ell_i+1)}\notag\\
    &\times M\left(-n-2\ell_i+l+2k,2l-2\ell_i+4k+2,K_{n\ell_i}r\right)\sum_{s=0}^{2n+2\ell_i-l-2k}\sum_{p=0}^s(-1)^{l+s}\binom{n+2\ell_i-l-2k}{p}\binom{n}{s-p}\notag\\
    &\times (2l-2\ell_i+4k+2+p)_{n+2\ell_i-l-2k-p}(2\ell_i+2+s-p)_{n-s+p}\sum_{q=0}^{2l+2k+s+3}\frac{(2l+2k+s+3)!}{q!}\left(K_{n\ell_i}r\right)^q e^{-K_{n\ell_i}r}\notag\\
    &+\Bigg[\frac{-n-2\ell_i+l+2k}{n+\ell_i+1}K_{n\ell_i}\, r\, M\left(-n-2\ell_i+l+2k+1,2l-2\ell_i+4k+3,K_{n\ell_i}r\right)\notag\\
    &+\frac{\partial}{\partial a}M\left(a,2l-2\ell_i+4k+2,K_{n\ell_i}r\right)\Big|_{a=-n-2\ell_i+l+2k}\Bigg]\sum_{s=0}^{2n+2\ell_i-l-2k}\sum_{p=0}^s(-1)^{l+s}\binom{n+2\ell_i-l-2k}{p}\binom{n}{s-p}\notag\\
    &\times (2l-2\ell_i+4k+2+p)_{n+2\ell_i-l-2k-p}(2\ell_i+2+s-p)_{n-s+p}\sum_{q=0}^{2l+2k+s+2}\frac{(2l+2k+s+2)!}{q!}\left(K_{n\ell_i}r\right)^q e^{-K_{n\ell_i}r}\notag\\
    &+\frac{n+2\ell_i-l-2k}{n+\ell_i+1}M\left(-n-2\ell_i+l+2k,2l-2\ell_i+4k+2,K_{n\ell_i}r\right)\sum_{s=0}^{2n+2\ell_i-l-2k-1}\sum_{p=0}^s(-1)^{l+1+s}\binom{n+2\ell_i-l-2k-1}{p}\notag\\
    &\times\binom{n}{s-p}(2l-2\ell_i+4k+3+p)_{n+2\ell_i-l-2k-1-p}(2\ell_i+2+s-p)_{n-s+p}\hspace{-0.1cm}\sum_{q=0}^{2l+2k+s+3}\frac{(2l+2k+s+3)!}{q!}\left(K_{n\ell_i}r\right)^q e^{-K_{n\ell_i}r}\notag\\
    &+M\left(-n-2\ell_i+l+2k,2l-2\ell_i+4k+2,K_{n\ell_i}r\right)I^{\ell_j=l-\ell_i+2k}_{dU}(r)\Bigg\},
\end{align}
\begin{align}
\begin{split}
    H_{>}(r)&=\frac{2\mu^2(-1)^{n+1}\left(K_{n\ell_i}\right)^{2k-\ell_i-1/2}}{\sqrt{2n!(n+\ell_i+1)(n+2\ell_i+1)!}}\frac{\Gamma(l+2k-n-2\ell_i)}{\Gamma(2+2l-2\ell_i+4k)}e^{-K_{n\ell_i}r/2}r^{l-\ell_i+2k+1}\\
    &\times\Bigg[U\left(l+2k-n-2\ell_i,2l-2\ell_i+4k+2,K_{n\ell_i}r\right)\int_0^{K_{n\ell_i}r} e^{-x}x^{2l+2k+2} \\
    &\times M(l+2k-n-2\ell_i,2l-2\ell_i+4k+2,x) U(-n,2\ell_i+2,x)\,dx+ M\left(l+2k-n-2\ell_i,2l-2\ell_i+4k+2,K_{n\ell_i}r\right)\\
    &\times\int_{K_{n\ell_i}r}^{\infty} e^{-x} x^{2l+2k+2} U(l+2k-n-2\ell_i,2l-2\ell_i+4k+2,x)U(-n,2\ell_i+2,x)\,dx\Bigg],
\end{split}
\label{H>}
\end{align}
}
\end{widetext}
where $U,M$ are the confluent hypergeometric functions, curved brackets with integer subscripts denote Pochhammer symbols, $\gamma$ is the Euler-Mascheroni constant, $\psi$ represents the digamma function and ${}_2F_2$ is a generalized hypergeometric function. We were not able to compute the following integrals (which are present in the previous equations), generically
\begin{widetext}
\begin{align}
    I^{\ell_j<n+\ell_i}_{dM}(r)&\equiv\int_0^{K_{n\ell_i}r} e^{-x}x^{\ell_j+\ell_i+l+2}\frac{\partial}{\partial a}M(a,2\ell_j+2,x)\Big|_{a=\ell_j-n-\ell_i}U(-n,2\ell_i+2,x)dx,\\
    I^{\ell_j<n+\ell_i}_{dU}(r)&\equiv \int_{K_{n\ell_i}r}^{\infty} e^{-x}x^{\ell_j+\ell_i+l+2}\frac{\partial}{\partial a}U(a,2\ell_j+2,x)\Big|_{a=\ell_j-n-\ell_i}U(-n,2\ell_i+2,x)dx,\\
    I^{\ell_j=n+\ell_i}_1(r)&\equiv\int_0^{K_{n\ell_i}r} e^{-x}x^{\ell_j+\ell_i+l+3}{}_2F_2(1,1;2,2\ell_j+3;x)U(-n,2\ell_i+2,x)dx,\\
    I^{\ell_j=n+\ell_i}_2(r)&\equiv\int_{K_{n\ell_i}r}^{\infty}e^{-x}x^{\ell_j+\ell_i+l+2}\left[\sum_{s=1}^{2\ell_j+1}\binom{2\ell_j+1}{-s+2\ell_j+1}\Gamma(s)x^{-s}-\log(x)\right]U(-n,2\ell_i+2,x)dx.
\end{align}
\end{widetext}
As the notation indicates, $I_{dM},I_{dU}$ are only defined for $\ell_j<n+\ell_i$ and $I_1,I_2$ for $\ell_j=n+\ell_i$.\par
When calculating $F^{(-1)}_{<}$, the following integrals (which were determined with the help of Eqs.~\eqref{Mmenos}--\eqref{uinf}, as well as the Cauchy product) were used
\begin{widetext}
\begin{align}
    &\int_0^r e^{-K_{n\ell_i}r'/2}r'^{\ell_j+l+1}M\left(\ell_j-n-\ell_i,2\ell_j+2,K_{n\ell_i}r'\right)W_{n+\ell_i+1,\ell_i+\frac{1}{2}}\left(K_{n\ell_i}r'\right)dr'\notag\\
    &=\sum_{s=0}^{2n+\ell_i-\ell_j}\sum_{p=0}^s(-1)^{n+s}\binom{n+\ell_i-\ell_j}{p}\binom{n}{s-p}\frac{(2\ell_i+2+s-p)_{n-s+p}}{(2\ell_j+2)_p}\left(K_{n\ell_i}\right)^{-\ell_j-l-2}\Bigg[(\ell_j+\ell_i+l+s+2)!\notag\\
    &-\sum_{q=0}^{\ell_j+\ell_i+l+s+2}\frac{(\ell_j+\ell_i+l+s+2)!}{q!}\left(K_{n\ell_i}\right)^q r^q e^{-K_{n\ell_i}r}\Bigg],\qquad\ell_j<n+\ell_i;
\label{firstInt}
\end{align}
\begin{align}
\begin{split}
    &\int_r^{\infty} e^{-K_{n\ell_i}r'/2}r'^{\ell_j+l+1}U\left(\ell_j-n-\ell_i,2\ell_j+2,K_{n\ell_i}r'\right)W_{n+\ell_i+1,\ell_i+\frac{1}{2}}\left(K_{n\ell_i}r'\right)dr'\\
    &=\sum_{s=0}^{2n+\ell_i-\ell_j}\sum_{p=0}^s\sum_{q=0}^{\ell_j+\ell_i+l+s+2}\hspace{-0.2cm}(-1)^{\ell_i+\ell_j+s}\binom{n+\ell_i-\ell_j}{p}\binom{n}{s-p}(2\ell_j+2+p)_{n+\ell_i-\ell_j-p}(2\ell_i+2+s-p)_{n-s+p}\\
    &\times\left(K_{n\ell_i}\right)^{q-\ell_j-l-2}\frac{(\ell_j+\ell_i+l+s+2)!}{q!}r^q e^{-K_{n\ell_i}r},\qquad\ell_j<n+\ell_i.
\end{split}
\label{secondInt}
\end{align}
\end{widetext}
When calculating $F^{(0)}_{<}$, Eqs.~\eqref{firstInt} and \eqref{secondInt} were also required, in addition to the following integrals [again determined through Eqs.~\eqref{Mmenos}--\eqref{uinf}]
\begin{widetext}
\begin{align}
\begin{split}
    &\int_0^r e^{-K_{n\ell_i}r'/2}r'^{\ell_j+l+2}M\left(\ell_j-n-\ell_i,2\ell_j+2,K_{n\ell_i}r'\right)W_{n+\ell_i+1,\ell_i+\frac{1}{2}}\left(K_{n\ell_i}r'\right)dr'\\
    &=\sum_{s=0}^{2n+\ell_i-\ell_j}\sum_{p=0}^s(-1)^{n+s}\binom{n+\ell_i-\ell_j}{p}\binom{n}{s-p}\frac{(2\ell_i+2+s-p)_{n-s+p}}{(2\ell_j+2)_p}\left(K_{n\ell_i}\right)^{-\ell_j-l-3}\Bigg[(\ell_j+\ell_i+l+s+3)!\\
    &-\sum_{q=0}^{\ell_j+\ell_i+l+s+3}\frac{(\ell_j+\ell_i+l+s+3)!}{q!}\left(K_{n\ell_i}\right)^q r^q e^{-K_{n\ell_i}r}\Bigg],\qquad\ell_j<n+\ell_i;
\end{split}
\end{align}
\begin{align}
    &\int_0^r e^{-K_{n\ell_i}r'/2}r'^{\ell_j+l+2} M\left(\ell_j-n-\ell_i+1,2\ell_j+3,K_{n\ell_i}r'\right)W_{n+\ell_i+1,\ell_i+\frac{1}{2}}\left(K_{n\ell_i}r'\right)dr'\notag\\
    &=\sum_{s=0}^{2n+\ell_i-\ell_j-1}\sum_{p=0}^s(-1)^{n+s}\binom{n+\ell_i-\ell_j-1}{p}\binom{n}{s-p}\frac{(2\ell_i+2+s-p)_{n-s+p}}{(2\ell_j+3)_p}\left(K_{n\ell_i}\right)^{-\ell_j-l-3}\Bigg[(\ell_j+\ell_i+l+s+3)!\notag\\
    &-\sum_{q=0}^{\ell_j+\ell_i+l+s+3}\frac{(\ell_j+\ell_i+l+s+3)!}{q!}\left(K_{n\ell_i}\right)^q r^q e^{-K_{n\ell_i}r}\Bigg],\qquad\ell_j<n+\ell_i;
\end{align}
\begin{align}
\begin{split}
    &\int_r^{\infty}e^{-K_{n\ell_i}r'/2}r'^{\ell_j+l+2}U\left(\ell_j-n-\ell_i,2\ell_j+2,K_{n\ell_i}r'\right)W_{n+\ell_i+1,\ell_i+\frac{1}{2}}\left(K_{n\ell_i}r'\right)dr'\\
    &=\sum_{s=0}^{2n+\ell_i-\ell_j}\sum_{p=0}^s\sum_{q=0}^{\ell_j+\ell_i+l+s+3}\hspace{-0.2cm}(-1)^{\ell_i+\ell_j+s}\binom{n+\ell_i-\ell_j}{p}\binom{n}{s-p}(2\ell_j+2+p)_{n+\ell_i-\ell_j-p}(2\ell_i+2+s-p)_{n-s+p}\\
    &\times\left(K_{n\ell_i}\right)^{q-\ell_j-l-3}\frac{(\ell_j+\ell_i+l+s+3)!}{q!}r^q e^{-K_{n\ell_i}r},\qquad\ell_j<n+\ell_i;
\end{split}
\end{align}
\begin{align}
\begin{split}
    &\int_r^{\infty}e^{-K_{n\ell_i}r'/2}r'^{\ell_j+l+2}U\left(\ell_j-n-\ell_i+1,2\ell_j+3,K_{n\ell_i}r'\right)W_{n+\ell_i+1,\ell_i+\frac{1}{2}}\left(K_{n\ell_i}r'\right)dr'\\
    &=\sum_{s=0}^{2n+\ell_i-\ell_j-1}\sum_{p=0}^s\sum_{q=0}^{\ell_j+\ell_i+l+s+3}\hspace{-0.2cm}(-1)^{\ell_i+\ell_j+1+s}\binom{n+\ell_i-\ell_j-1}{p}\binom{n}{s-p}(2\ell_j+3+p)_{n+\ell_i-\ell_j-1-p}\\
    &\times(2\ell_i+2+s-p)_{n-s+p}\left(K_{n\ell_i}\right)^{q-\ell_j-l-3}\frac{(\ell_j+\ell_i+l+s+3)!}{q!}r^q e^{-K_{n\ell_i}r},\qquad\ell_j<n+\ell_i .
\end{split}
\end{align}
\end{widetext}
Finally, to derive $G^{(-1)}$ and $G^{(0)}$, the following integrals were used~\cite{Jeffrey:2007}
\begin{widetext}
\begin{align}
    \int_0^{\infty}e^{-K_{n\ell_i}r'/2}r'^{n+\ell_i+l+1}W_{n+\ell_i+1,\ell_i+\frac{1}{2}}\left(K_{n\ell_i}r'\right)dr'&=\left(K_{n\ell_i}\right)^{-n-\ell_i-l-2}\frac{\Gamma(n+2\ell_i+l+3)\Gamma(n+l+2)}{\Gamma(l+2)},\label{nexttolastInt}\\
    \int_0^{\infty}e^{-K_{n\ell_i}r'/2}r'^{n+\ell_i+l+2}W_{n+\ell_i+1,\ell_i+\frac{1}{2}}\left(K_{n\ell_i}r'\right)dr'&=\left(K_{n\ell_i}\right)^{-n-\ell_i-l-3}\frac{\Gamma(n+2\ell_i+l+4)\Gamma(n+l+3)}{\Gamma(l+3)},\label{lastInt}
\end{align}
\end{widetext}
along with Eqs.~\eqref{derM}, \eqref{derU}.
%%%%%
\subsection{Auxiliary functions for the perturbed gravitational potential}\label{app:gravpot_auxfunc}
%%%%%
%
The explicit form for the solution in Eq.~\eqref{u2_schematic} may be written as
\begin{widetext}
\begin{align}
    \hat{u}_{(2)}^{lm}(\omega,r)\sim\begin{cases}
        r^{-(l+1)}\hat{u}_{(2),sA}^{lm}(\omega)\delta(\omega-m\Omegaorb), & \text{if $2\leq l\leq \ell_i$, $0\leq n< l$ and $n$ and $l$ have different parity}\\
        r^{-(l+1)}\hat{u}_{(2),sB}^{lm}(\omega)\delta(\omega-m\Omegaorb), & \text{if $2\leq l\leq \ell_i$, $0\leq n< l$ and $n$ and $l$ have the same parity}\\
        r^{-(l+1)}\hat{u}_{(2),sC}^{lm}(\omega)\delta(\omega-m\Omegaorb), & \text{if $2\leq l\leq \ell_i$ and $n=l$}\\
        r^{-(l+1)}\hat{u}_{(2),sD}^{lm}(\omega)\delta(\omega-m\Omegaorb), & \text{if $2\leq l\leq \ell_i$ and $n>l$}\\
        r^{-(l+1)}\hat{u}_{(2),sE}^{lm}(\omega)\delta(\omega-m\Omegaorb), & \text{if $l>\ell_i$, $0\leq n<l$ and $\ell_i<\frac{l-n}{2}$}\\
        r^{-(l+1)}\hat{u}_{(2),sF}^{lm}(\omega)\delta(\omega-m\Omegaorb), & \text{if $l>\ell_i$, $0\leq n<l$ and $\ell_i=\frac{l-n}{2}$}\\
        r^{-(l+1)}\hat{u}_{(2),sG}^{lm}(\omega)\delta(\omega-m\Omegaorb), & \text{if $l>\ell_i$, $0\leq n<l$, $\ell_i>\frac{l-n}{2}$ and $n$ and $l$ have different parity}\\
        r^{-(l+1)}\hat{u}_{(2),sH}^{lm}(\omega)\delta(\omega-m\Omegaorb), & \text{if $l>\ell_i$, $0\leq n<l$, $\ell_i>\frac{l-n}{2}$ and $n$ and $l$ have the same parity}\\
        r^{-(l+1)}\hat{u}_{(2),sI}^{lm}(\omega)\delta(\omega-m\Omegaorb), & \text{if $l>\ell_i$ and $n=l$}\\
        r^{-(l+1)}\hat{u}_{(2),sJ}^{lm}(\omega)\delta(\omega-m\Omegaorb), & \text{if $l>\ell_i$ and $n>l$}
    \end{cases}\,,
\label{solucoesu2}
\end{align}
%\end{widetext}
%
where
%
%\begin{widetext}
\begin{align}
\begin{split}
    \hat{u}_{(2),sA}^{lm}(\omega)&=\sum_{k=0}^{(n+l-1)/2}\Bigg\{c_{lm}\left[(C_1)^{l\ell_i k}_{m m_i}(C_1)^{l\ell_i k}_{m m_i}-(C_2)^{l\ell_i k}_{m m_i}(C_2)^{l\ell_i k}_{m m_i}\right]\mu^3\left(K_{n\ell_i}\right)^{-2l-2}\mathcal{A}_k^{n\ell_i l}\left(\frac{\omega}{E_{n\ell_i}}\right)^{-1}\\
    &+c_{lm}\left[(C_1)^{l\ell_i k}_{m m_i}(C_1)^{l\ell_i k}_{m m_i}+(C_2)^{l\ell_i k}_{m m_i}(C_2)^{l\ell_i k}_{m m_i}\right]\mu^3\left(K_{n\ell_i}\right)^{-2l-2}\mathcal{B}_k^{n\ell_i l}\Bigg\}\\
    &+\sum_{k=(n+l+1)/2}^{l}\hspace{-0.5cm}c_{lm}\left[(C_1)^{l\ell_i k}_{m m_i}(C_1)^{l\ell_i k}_{m m_i}+(C_2)^{l\ell_i k}_{m m_i}(C_2)^{l\ell_i k}_{m m_i}\right]\mu^3\left(K_{n\ell_i}\right)^{-2l-2}\mathcal{C}_k^{n\ell_i l}+\mathcal{O}\left(\frac{\omega}{E_{n\ell_i}}\right),
\end{split}
\label{primeiraus}
\end{align}

\begin{align}
\begin{split}
    \hat{u}_{(2),sB}^{lm}(\omega)&=\sum_{k=0}^{(n+l)/2-1}\Bigg\{c_{lm}\left[(C_1)^{l\ell_i k}_{m m_i}(C_1)^{l\ell_i k}_{m m_i}-(C_2)^{l\ell_i k}_{m m_i}(C_2)^{l\ell_i k}_{m m_i}\right]\mu^3\left(K_{n\ell_i}\right)^{-2l-2}\mathcal{A}_k^{n\ell_i l}\left(\frac{\omega}{E_{n\ell_i}}\right)^{-1}\\
    &+c_{lm}\left[(C_1)^{l\ell_i k}_{m m_i}(C_1)^{l\ell_i k}_{m m_i}+(C_2)^{l\ell_i k}_{m m_i}(C_2)^{l\ell_i k}_{m m_i}\right]\mu^3\left(K_{n\ell_i}\right)^{-2l-2}\mathcal{B}_k^{n\ell_i l}\Bigg\}\\
    &+c_{lm}\left[(C_1)^{l\ell_i (n+l)/2}_{m m_i}(C_1)^{l\ell_i (n+l)/2}_{m m_i}-(C_2)^{l\ell_i (n+l)/2}_{m m_i}(C_2)^{l\ell_i (n+l)/2}_{m m_i}\right]\frac{8\pi\mu^3}{2l+1}\frac{\left(K_{n\ell_i}\right)^{-2l-2}}{n!(n+\ell_i+1)(n+2\ell_i+1)!}\\
    &\times\frac{1}{(n+\ell_i+1)\Gamma(2+2n+2\ell_i)}\left[\frac{\Gamma(n+2\ell_i+l+3)\Gamma(n+l+2)}{\Gamma(l+2)}\right]^2\left(\frac{\omega}{E_{n\ell_i}}\right)^{-1}\\
    &-c_{lm}\left[(C_1)^{l\ell_i (n+l)/2}_{m m_i}(C_1)^{l\ell_i (n+l)/2}_{m m_i}+(C_2)^{l\ell_i (n+l)/2}_{m m_i}(C_2)^{l\ell_i (n+l)/2}_{m m_i}\right]\mu^3\left(K_{n\ell_i}\right)^{-2l-2}\mathcal{F}^{n\ell_i l}\\
    &+\sum_{k=(n+l)/2+1}^{l}\hspace{-0.5cm}c_{lm}\left[(C_1)^{l\ell_i k}_{m m_i}(C_1)^{l\ell_i k}_{m m_i}+(C_2)^{l\ell_i k}_{m m_i}(C_2)^{l\ell_i k}_{m m_i}\right]\mu^3\left(K_{n\ell_i}\right)^{-2l-2}\mathcal{C}_k^{n\ell_i l}+\mathcal{O}\left(\frac{\omega}{E_{n\ell_i}}\right),
\end{split}
\end{align}

\begin{align}
\begin{split}
    \hat{u}_{(2),sC}^{lm}(\omega)&=\sum_{k=0}^{l-1}\Bigg\{c_{lm}\left[(C_1)^{l\ell_i k}_{m m_i}(C_1)^{l\ell_i k}_{m m_i}-(C_2)^{l\ell_i k}_{m m_i}(C_2)^{l\ell_i k}_{m m_i}\right]\mu^3\left(K_{n\ell_i}\right)^{-2l-2}\mathcal{A}_k^{l\ell_i l}\left(\frac{\omega}{E_{n\ell_i}}\right)^{-1}\\
    &+c_{lm}[(C_1)^{l\ell_i k}_{m m_i}(C_1)^{l\ell_i k}_{m m_i}+(C_2)^{l\ell_i k}_{m m_i}(C_2)^{l\ell_i k}_{m m_i}]\mu^3\left(K_{n\ell_i}\right)^{-2l-2}\mathcal{B}_k^{l\ell_i l}\Bigg\}\\
    &+c_{lm}\left[(C_1)^{l\ell_i l}_{m m_i}(C_1)^{l\ell_i l}_{m m_i}-(C_2)^{l\ell_i l}_{m m_i}(C_2)^{l\ell_i l}_{m m_i}\right]\frac{8\pi\mu^3}{2l+1}\frac{\left(K_{n\ell_i}\right)^{-2l-2}}{n!(n+\ell_i+1)(n+2\ell_i+1)!}\\
    &\times\frac{1}{(n+\ell_i+1)\Gamma(2+2n+2\ell_i)}\left[\frac{\Gamma(n+2\ell_i+l+3)\Gamma(n+l+2)}{\Gamma(l+2)}\right]^2\left(\frac{\omega}{E_{n\ell_i}}\right)^{-1}\\
    &-c_{lm}\left[(C_1)^{l\ell_i l}_{m m_i}(C_1)^{l\ell_i l}_{m m_i}+(C_2)^{l\ell_i l}_{m m_i}(C_2)^{l\ell_i l}_{m m_i}\right]\mu^3\left(K_{n\ell_i}\right)^{-2l-2}\mathcal{F}^{l\ell_i l}+\mathcal{O}\left(\frac{\omega}{E_{n\ell_i}}\right),
\end{split}
\end{align}

\begin{align}
\begin{split}
    \hat{u}_{(2),sD}^{lm}(\omega)&=\sum_{k=0}^{l}\Bigg\{c_{lm}\left[(C_1)^{l\ell_i k}_{m m_i}(C_1)^{l\ell_i k}_{m m_i}-(C_2)^{l\ell_i k}_{m m_i}(C_2)^{l\ell_i k}_{m m_i}\right]\mu^3\left(K_{n\ell_i}\right)^{-2l-2}\mathcal{A}_k^{n\ell_i l}\left(\frac{\omega}{E_{n\ell_i}}\right)^{-1}\\
    &+c_{lm}[(C_1)^{l\ell_i k}_{m m_i}(C_1)^{l\ell_i k}_{m m_i}+(C_2)^{l\ell_i k}_{m m_i}(C_2)^{l\ell_i k}_{m m_i}]\mu^3\left(K_{n\ell_i}\right)^{-2l-2}\mathcal{B}_k^{n\ell_i l}\Bigg\}+\mathcal{O}\left(\frac{\omega}{E_{n\ell_i}}\right),
\end{split}
\end{align}

\begin{align}\label{usE}
    \hat{u}_{(2),sE}^{lm}(\omega)&=\sum_{k=0}^{\ell_i}c_{lm}\left[(C_1)^{l\ell_i k}_{m m_i}(C_1)^{l\ell_i k}_{m m_i}+(C_2)^{l\ell_i k}_{m m_i}(C_2)^{l\ell_i k}_{m m_i}\right]\mu^3\left(K_{n\ell_i}\right)^{-2l-2}\mathcal{G}_k^{n\ell_i l}+\mathcal{O}\left(\frac{\omega}{E_{n\ell_i}}\right),
\end{align}

\begin{align}\label{usF}
\begin{split}
    \hat{u}_{(2),sF}^{lm}(\omega)&=c_{lm}\left[(C_1)^{l\ell_i 0}_{m m_i}(C_1)^{l\ell_i 0}_{m m_i}-(C_2)^{l\ell_i 0}_{m m_i}(C_2)^{l\ell_i 0}_{m m_i}\right]\frac{8\pi\mu^3}{2l+1}\frac{\left(K_{n\ell_i}\right)^{-2l-2}}{n!(n+\ell_i+1)(n+2\ell_i+1)!}\\
    &\times\frac{1}{(n+\ell_i+1)\Gamma(2+2n+2\ell_i)}\left[\frac{\Gamma(n+2\ell_i+l+3)\Gamma(n+l+2)}{\Gamma(l+2)}\right]^2\left(\frac{\omega}{E_{n\ell_i}}\right)^{-1}-c_{lm}\Big[(C_1)^{l\ell_i 0}_{m m_i}(C_1)^{l\ell_i 0}_{m m_i}\\
    &+(C_2)^{l\ell_i 0}_{m m_i}(C_2)^{l\ell_i 0}_{m m_i}\Big]\mu^3\left(K_{n\ell_i}\right)^{-2l-2}\mathcal{F}^{n\ell_i l}\\
    &+\sum_{k=1}^{\ell_i}c_{lm}\left[(C_1)^{l\ell_i k}_{m m_i}(C_1)^{l\ell_i k}_{m m_i}+(C_2)^{l\ell_i k}_{m m_i}(C_2)^{l\ell_i k}_{m m_i}\right]\mu^3\left(K_{n\ell_i}\right)^{-2l-2}\mathcal{G}_k^{n\ell_i l}+\mathcal{O}\left(\frac{\omega}{E_{n\ell_i}}\right),
\end{split}
\end{align}

\begin{align}
\begin{split}
    \hat{u}_{(2),sG}^{lm}(\omega)&=\sum_{k=0}^{(2\ell_i+n-l-1)/2}\Bigg\{c_{lm}\left[(C_1)^{l\ell_i k}_{m m_i}(C_1)^{l\ell_i k}_{m m_i}-(C_2)^{l\ell_i k}_{m m_i}(C_2)^{l\ell_i k}_{m m_i}\right]\mu^3\left(K_{n\ell_i}\right)^{-2l-2}\mathcal{H}_k^{n\ell_i l}\left(\frac{\omega}{E_{n\ell_i}}\right)^{-1}\\
    &+c_{lm}\left[(C_1)^{l\ell_i k}_{m m_i}(C_1)^{l\ell_i k}_{m m_i}+(C_2)^{l\ell_i k}_{m m_i}(C_2)^{l\ell_i k}_{m m_i}\right]\mu^3\left(K_{n\ell_i}\right)^{-2l-2}\mathcal{J}_k^{n\ell_i l}\Bigg\}\\
    &+\sum_{k=(2\ell_i+n-l+1)/2}^{\ell_i}\hspace{-0.8cm}c_{lm}\left[(C_1)^{l\ell_i k}_{m m_i}(C_1)^{l\ell_i k}_{m m_i}+(C_2)^{l\ell_i k}_{m m_i}(C_2)^{l\ell_i k}_{m m_i}\right]\mu^3\left(K_{n\ell_i}\right)^{-2l-2}\mathcal{G}_k^{n\ell_i l}+\mathcal{O}\left(\frac{\omega}{E_{n\ell_i}}\right),
\end{split}
\end{align}

\begin{align}
\begin{split}
    \hat{u}_{(2),sH}^{lm}(\omega)&=\sum_{k=0}^{\ell_i+(n-l)/2-1}\Bigg\{c_{lm}\left[(C_1)^{l\ell_i k}_{m m_i}(C_1)^{l\ell_i k}_{m m_i}-(C_2)^{l\ell_i k}_{m m_i}(C_2)^{l\ell_i k}_{m m_i}\right]\mu^3\left(K_{n\ell_i}\right)^{-2l-2}\mathcal{H}_k^{n\ell_i l}\left(\frac{\omega}{E_{n\ell_i}}\right)^{-1}\\
    &+c_{lm}\left[(C_1)^{l\ell_i k}_{m m_i}(C_1)^{l\ell_i k}_{m m_i}+(C_2)^{l\ell_i k}_{m m_i}(C_2)^{l\ell_i k}_{m m_i}\right]\mu^3\left(K_{n\ell_i}\right)^{-2l-2}\mathcal{J}_k^{n\ell_i l}\Bigg\}\\
    &+c_{lm}\left[(C_1)^{l\ell_i \ell_i+(n-l)/2}_{m m_i}(C_1)^{l\ell_i \ell_i+(n-l)/2}_{m m_i}-(C_2)^{l\ell_i \ell_i+(n-l)/2}_{m m_i}(C_2)^{l\ell_i \ell_i+(n-l)/2}_{m m_i}\right]\frac{8\pi\mu^3}{2l+1}\\
    &\times\frac{\left(K_{n\ell_i}\right)^{-2l-2}}{n!(n+\ell_i+1)(n+2\ell_i+1)!}\frac{1}{(n+\ell_i+1)\Gamma(2+2n+2\ell_i)}\left[\frac{\Gamma(n+2\ell_i+l+3)\Gamma(n+l+2)}{\Gamma(l+2)}\right]^2\left(\frac{\omega}{E_{n\ell_i}}\right)^{-1}\\
    &-c_{lm}\left[(C_1)^{l\ell_i \ell_i+(n-l)/2}_{m m_i}(C_1)^{l\ell_i \ell_i+(n-l)/2}_{m m_i}+(C_2)^{l\ell_i \ell_i+(n-l)/2}_{m m_i}(C_2)^{l\ell_i \ell_i+(n-l)/2}_{m m_i}\right]\mu^3\left(K_{n\ell_i}\right)^{-2l-2}\mathcal{F}^{n\ell_i l}\\
    &+\sum_{k=\ell_i+(n-l)/2+1}^{\ell_i}\hspace{-0.8cm}c_{lm}\left[(C_1)^{l\ell_i k}_{m m_i}(C_1)^{l\ell_i k}_{m m_i}+(C_2)^{l\ell_i k}_{m m_i}(C_2)^{l\ell_i k}_{m m_i}\right]\mu^3\left(K_{n\ell_i}\right)^{-2l-2}\mathcal{G}_k^{n\ell_i l}+\mathcal{O}\left(\frac{\omega}{E_{n\ell_i}}\right),
\end{split}
\end{align}

\begin{align}
\begin{split}
    \hat{u}_{(2),sI}^{lm}(\omega)&=\sum_{k=0}^{\ell_i-1}\Bigg\{c_{lm}\left[(C_1)^{l\ell_i k}_{m m_i}(C_1)^{l\ell_i k}_{m m_i}-(C_2)^{l\ell_i k}_{m m_i}(C_2)^{l\ell_i k}_{m m_i}\right]\mu^3\left(K_{n\ell_i}\right)^{-2l-2}\mathcal{H}_k^{n\ell_i l}\left(\frac{\omega}{E_{n\ell_i}}\right)^{-1}\\
    &+c_{lm}\left[(C_1)^{l\ell_i k}_{m m_i}(C_1)^{l\ell_i k}_{m m_i}+(C_2)^{l\ell_i k}_{m m_i}(C_2)^{l\ell_i k}_{m m_i}\right]\mu^3\left(K_{n\ell_i}\right)^{-2l-2}\mathcal{J}_k^{n\ell_i l}\Bigg\}\\
    &+c_{lm}\left[(C_1)^{l\ell_i \ell_i}_{m m_i}(C_1)^{l\ell_i \ell_i}_{m m_i}-(C_2)^{l\ell_i \ell_i}_{m m_i}(C_2)^{l\ell_i \ell_i}_{m m_i}\right]\frac{8\pi\mu^3}{2l+1}\frac{\left(K_{n\ell_i}\right)^{-2l-2}}{n!(n+\ell_i+1)(n+2\ell_i+1)!}\\
    &\times\frac{1}{(n+\ell_i+1)\Gamma(2+2n+2\ell_i)}\left[\frac{\Gamma(n+2\ell_i+l+3)\Gamma(n+l+2)}{\Gamma(l+2)}\right]^2\left(\frac{\omega}{E_{n\ell_i}}\right)^{-1}\\
    &-c_{lm}\left[(C_1)^{l\ell_i \ell_i}_{m m_i}(C_1)^{l\ell_i \ell_i}_{m m_i}+(C_2)^{l\ell_i \ell_i}_{m m_i}(C_2)^{l\ell_i \ell_i}_{m m_i}\right]\mu^3\left(K_{n\ell_i}\right)^{-2l-2}\mathcal{F}^{n\ell_i l}+\mathcal{O}\left(\frac{\omega}{E_{n\ell_i}}\right),
\end{split}
\end{align}

\begin{align}
\begin{split}
    \hat{u}_{(2),sJ}^{lm}(\omega)&=\sum_{k=0}^{\ell_i}\Bigg\{c_{lm}\left[(C_1)^{l\ell_i k}_{m m_i}(C_1)^{l\ell_i k}_{m m_i}-(C_2)^{l\ell_i k}_{m m_i}(C_2)^{l\ell_i k}_{m m_i}\right]\mu^3\left(K_{n\ell_i}\right)^{-2l-2}\mathcal{H}_k^{n\ell_i l}\left(\frac{\omega}{E_{n\ell_i}}\right)^{-1}\\
    &+c_{lm}\left[(C_1)^{l\ell_i k}_{m m_i}(C_1)^{l\ell_i k}_{m m_i}+(C_2)^{l\ell_i k}_{m m_i}(C_2)^{l\ell_i k}_{m m_i}\right]\mu^3\left(K_{n\ell_i}\right)^{-2l-2}\mathcal{J}_k^{n\ell_i l}\Bigg\}+\mathcal{O}\left(\frac{\omega}{E_{n\ell_i}}\right).
\end{split}
\label{ultimaus}
\end{align}
%\end{widetext}
%
In the expressions above we defined the following real numbers 
%
%\begin{widetext}
{\small
\begin{align}
\begin{split}
    \mathcal{A}_{k}^{n\ell_i l}&\equiv \frac{8\pi}{2l+1}\frac{1}{n!(n+\ell_i+1)(n+2\ell_i+1)!}\frac{1}{\Gamma(2+2\ell_i-2l+4k)(n+\ell_i+1)(n+l-2k)!}\Bigg\{\sum_{s=0}^{2n+l-2k}\sum_{p=0}^s(-1)^s\binom{n+l-2k}{p}\\
    &\times \binom{n}{s-p}\frac{(2\ell_i+2+s-p)_{n-s+p}}{(2\ell_i-2l+4k+2)_p}(2\ell_i+2k+s+2)!\sum_{j=0}^{2n+l-2k}\sum_{v=0}^j(-1)^{j}\binom{n+l-2k}{v}\binom{n}{j-v} \\
    &\times (2\ell_i-2l+4k+2+v)_{n+l-2k-v}(2\ell_i+2+j-v)_{n-j+v}\Bigg[(2\ell_i+2k+2+j)!-\sum_{q=0}^{2\ell_i+2k+s+2}\frac{(2\ell_i+2k+2+q+j)!}{2^{2\ell_i+2k+3+q+j}q!}\Bigg]\\
    &+\sum_{s=0}^{2n+l-2k}\sum_{p=0}^s(-1)^s\binom{n+l-2k}{p}\binom{n}{s-p} (2\ell_i-2l+4k+2+p)_{n+l-2k-p}  (2\ell_i+2+s-p)_{n-s+p}(2\ell_i+2k+s+2)!\\
    &\times \sum_{j=0}^{2n+l-2k}\sum_{v=0}^j(-1)^j\binom{n+l-2k}{v}\binom{n}{j-v}\frac{(2\ell_i+2+j-v)_{n-j+v}}{(2\ell_i-2l+4k+2)_{v}}\sum_{q=0}^{2\ell_i+2k+s+2}\frac{(2\ell_i+2k+2+q+j)!}{2^{2\ell_i+2k+3+q+j}q!}\Bigg\},
\end{split}
\label{Acal}
\end{align}

\begin{align}
\begin{split}
    \mathcal{B}_{k}^{n\ell_i l}&\equiv \frac{4\pi}{2l+1}\frac{(-1)^{n+l}}{n!(n+\ell_i+1)(n+2\ell_i+1)!}\frac{1}{\Gamma(2+2\ell_i-2l+4k)(n+\ell_i+1)(n+l-2k)!}\\
    &\times\int_0^{\infty}e^{-y}y^{2\ell_i+2k+2}U(-n,2\ell_i+2,y)\Big\{\frac{1}{n+\ell_i+1}U(-n-l+2k,2\ell_i-2l+4k+2,y)\Bigg(2\ell_i-2l+4k+\frac{5}{2}-\frac{y}{2}\\
    &+\psi(n+l-2k+1)\Bigg)\sum_{s=0}^{2n+l-2k}\sum_{p=0}^s(-1)^{n+s}\binom{n+l-2k}{p}\binom{n}{s-p}\frac{(2\ell_i+2+s-p)_{n-s+p}}{(2\ell_i-2l+4k+2)_p}\Bigg[(2\ell_i+2k+s+2)!\\
    &-\sum_{q=0}^{2\ell_i+2k+s+2}\frac{(2\ell_i+2k+s+2)!}{q!}y^q e^{-y}\Bigg]-\frac{1}{2(n+\ell_i+1)}U(-n-l+2k,2\ell_i-2l+4k+2,y)\sum_{s=0}^{2n+l-2k}\sum_{p=0}^s(-1)^{n+s}\\
    &\times\binom{n+l-2k}{p}\binom{n}{s-p}\frac{(2\ell_i+2+s-p)_{n-s+p}}{(2\ell_i-2l+4k+2)_p}\Bigg[(2\ell_i+2k+s+3)!-\sum_{q=0}^{2\ell_i+2k+s+3}\frac{(2\ell_i+2k+s+3)!}{q!}y^q e^{-y}\Bigg]\\
    &+\frac{-n-l+2k}{(2\ell_i-2l+4k+2)(n+\ell_i+1)} U(-n-l+2k,2\ell_i-2l+4k+2,y)\sum_{s=0}^{2n+l-2k-1}\sum_{p=0}^s(-1)^{n+s}\binom{n+l-2k-1}{p}\\
    &\times\binom{n}{s-p}\frac{(2\ell_i+2+s-p)_{n-s+p}}{(2\ell_i-2l+4k+3)_p}\Bigg[(2\ell_i+2k+s+3)!-\sum_{q=0}^{2\ell_i+2k+s+3}\frac{(2\ell_i+2k+s+3)!}{q!}y^q e^{-y}\Bigg]\\
    &+U(-n-l+2k,2\ell_i-2l+4k+2,y)\int_0^{y} e^{-x}x^{2\ell_i+2k+2}\frac{\partial}{\partial a}M(a,2\ell_i-2l+4k+2,x)\Big|_{a=-n-l+2k}U(-n,2\ell_i+2,x)dx\\
    &+\Bigg[\frac{n+l-2k}{n+\ell_i+1}y\, U(-n-l+2k+1,2\ell_i-2l+4k+3,y)+\frac{\partial}{\partial a}U(a,2\ell_i-2l+4k+2,y)\Big|_{a=-n-l+2k}\Bigg]\\
    &\times\sum_{s=0}^{2n+l-2k}\sum_{p=0}^s(-1)^{n+s}\binom{n+l-2k}{p}\binom{n}{s-p}\frac{(2\ell_i+2+s-p)_{n-s+p}}{(2\ell_i-2l+4k+2)_p}\Bigg[(2\ell_i+2k+s+2)!\\
    &-\sum_{q=0}^{2\ell_i+2k+s+2}\frac{(2\ell_i+2k+s+2)!}{q!}y^q e^{-y}\Bigg]+\frac{1}{n+\ell_i+1}M(-n-l+2k,2\ell_i-2l+4k+2,y)\Bigg(2\ell_i-2l+4k+\frac{5}{2}-\frac{y}{2}\\
    &+\psi(n+l-2k+1)\Bigg)\sum_{s=0}^{2n+l-2k}\sum_{p=0}^s(-1)^{l+s}\binom{n+l-2k}{p}\binom{n}{s-p}(2\ell_i-2l+4k+2+p)_{n+l-2k-p}(2\ell_i+2+s-p)_{n-s+p}\\
    &\times\sum_{q=0}^{2\ell_i+2k+s+2}\frac{(2\ell_i+2k+s+2)!}{q!}y^q e^{-y}-\frac{1}{2(n+\ell_i+1)}M(-n-l+2k,2\ell_i-2l+4k+2,y)\\
    &\times\sum_{s=0}^{2n+l-2k}\sum_{p=0}^s(-1)^{l+s}\binom{n+l-2k}{p}\binom{n}{s-p}(2\ell_i-2l+4k+2+p)_{n+l-2k-p}(2\ell_i+2+s-p)_{n-s+p}\\
    &\times\sum_{q=0}^{2\ell_i+2k+s+3}\frac{(2\ell_i+2k+s+3)!}{q!}y^q e^{-y}+\Bigg[\frac{-n-l+2k}{n+\ell_i+1}y\, M(-n-l+2k+1,2\ell_i-2l+4k+3,y)\\
    &+\frac{\partial}{\partial a}M(a,2\ell_i-2l+4k+2,y)\Big|_{a=-n-l+2k}\Bigg]\sum_{s=0}^{2n+l-2k}\sum_{p=0}^s(-1)^{l+s}\binom{n+l-2k}{p}\binom{n}{s-p}(2\ell_i-2l+4k+2+p)_{n+l-2k-p}\\
    &\times (2\ell_i+2+s-p)_{n-s+p}\sum_{q=0}^{2\ell_i+2k+s+2}\frac{(2\ell_i+2k+s+2)!}{q!}y^q e^{-y}+\frac{n+l-2k}{n+\ell_i+1}M(-n-l+2k,2\ell_i-2l+4k+2,y)\\
    &\times\sum_{s=0}^{2n+l-2k-1}\sum_{p=0}^s(-1)^{l+s+1}\binom{n+l-2k-1}{p}\binom{n}{s-p}(2\ell_i-2l+4k+3+p)_{n+l-2k-1-p}(2\ell_i+2+s-p)_{n-s+p}\\
    &\times \sum_{q=0}^{2\ell_i+2k+s+3}\frac{(2\ell_i+2k+s+3)!}{q!}y^q e^{-y}+M(-n-l+2k,2\ell_i-2l+4k+2,y) \int_{y}^{\infty} e^{-x}x^{2\ell_i+2k+2}\\
    &\times\frac{\partial}{\partial a}U(a,2\ell_i-2l+4k+2,x)\Big|_{a=-n-l+2k}U(-n,2\ell_i+2,x)dx\Big\}dy,
\end{split}
\label{Bcal}
\end{align}

\begin{align}
\begin{split}
    \mathcal{C}_{k}^{n\ell_i l}&\equiv \frac{4\pi}{2l+1}\frac{1}{n!(n+\ell_i+1)(n+2\ell_i+1)!}\frac{\Gamma(2k-n-l)}{\Gamma(2+2\ell_i-2l+4k)}\int_0^{\infty}e^{-y} y^{2\ell_i+2k+2}U(-n,2\ell_i+2,y)\\
    &\times\Bigg[U(2k-n-l,2\ell_i-2l+4k+2,y)\int_0^{y} e^{-x}x^{2\ell_i+2k+2}M(2k-n-l,2\ell_i-2l+4k+2,x) U(-n,2\ell_i+2,x)dx\\
    &+ M(2k-n-l,2\ell_i-2l+4k+2,y)\int_{y}^{\infty} e^{-x} x^{2\ell_i+2k+2} U(2k-n-l,2\ell_i-2l+4k+2,x) U(-n,2\ell_i+2,x)dx\Bigg]dy,
\end{split}
\label{Ccal}
\end{align}

\begin{align}
\begin{split}
    \mathcal{F}^{n\ell_i l}&\equiv \frac{4\pi}{2l+1}\frac{1}{n!(n+\ell_i+1)(n+2\ell_i+1)!}\frac{1}{\Gamma(2+2n+2\ell_i)}\int_0^{\infty}e^{-y}y^{n+2\ell_i+l+2}U(-n,2\ell_i+2,y)\\
    &\times\Bigg\{\Big[-5-4n-4\ell_i+(2+2n+2\ell_i)\gamma+y\Big]\frac{\Gamma(n+2\ell_i+l+3)\Gamma(n+l+2)}{(2+2n+2\ell_i)\Gamma(l+2)}+\frac{\Gamma(n+2\ell_i+l+4)\Gamma(n+l+3)}{(2+2n+2\ell_i)\Gamma(l+3)}\\
    &-\frac{1}{2+2n+2\ell_i}\int_0^{y} e^{-x}x^{n+2\ell_i+l+3} {}_2F_2(1,1;2,2n+2\ell_i+3;x)U(-n,2\ell_i+2,x)\,dx\\
    &-\int_{y}^{\infty}e^{-x}x^{n+2\ell_i+l+2}\left[\sum_{s=1}^{2n+2\ell_i+1}\binom{2n+2\ell_i+1}{-s+2n+2\ell_i+1}\Gamma(s)x^{-s}-\log(x)\right] U(-n,2\ell_i+2,x)dx\\
    &+(-1)^{n+1}\Bigg[\sum_{s=1}^{2n+2\ell_i+1}\binom{2n+2\ell_i+1}{-s+2n+2\ell_i+1}\Gamma(s)y^{-s}-\log(y)\Bigg]\sum_{p=0}^n\binom{n}{p}(2\ell_i+2+p)_{n-p}(-1)^p\\
    &\times\Bigg[(n+2\ell_i+l+p+2)!-\sum_{q=0}^{n+2\ell_i+l+p+2}\frac{(n+2\ell_i+l+p+2)!}{q!}y^q e^{-y}\Bigg]+(-1)^{n+1}\frac{y}{2+2n+2\ell_i}\\
    &\times{}_2F_2(1,1;2,2n+2\ell_i+3;y)\sum_{p=0}^n\binom{n}{p}(2\ell_i+2+p)_{n-p} (-1)^p\sum_{q=0}^{n+2\ell_i+l+p+2}\frac{(n+2\ell_i+l+p+2)!}{q!}y^q e^{-y}\Bigg\}dy,
\end{split}
\label{Fcal}
\end{align}

\begin{align}
\begin{split}
    &\mathcal{G}_{k}^{n\ell_i l}\equiv \frac{4\pi}{2l+1}\frac{1}{n!(n+\ell_i+1)(n+2\ell_i+1)!}\frac{\Gamma(l+2k-n-2\ell_i)}{\Gamma(2+2l-2\ell_i+4k)}\int_0^{\infty}e^{-y}y^{2l+2k+2}U(-n,2\ell_i+2,y)\\
    &\times \Bigg[U(l+2k-n-2\ell_i,2l-2\ell_i+4k+2,y)\int_0^{y} e^{-x}x^{2l+2k+2} M(l+2k-n-2\ell_i,2l-2\ell_i+4k+2,x) U(-n,2\ell_i+2,x)\, dx\\
    &+ M(l+2k-n-2\ell_i,2l-2\ell_i+4k+2,y)\int_{y}^{\infty}e^{-x} x^{2l+2k+2} U(l+2k-n-2\ell_i,2l-2\ell_i+4k+2,x)U(-n,2\ell_i+2,x)\, dx\Bigg] dy,
\end{split}
\label{Gcal}
\end{align}

\begin{align}
\begin{split}
    \mathcal{H}_{k}^{n\ell_i l}&\equiv \frac{8\pi}{2l+1}\frac{1}{n!(n+\ell_i+1)(n+2\ell_i+1)!}\frac{1}{\Gamma(2+2l-2\ell_i+4k)(n+\ell_i+1)(n+2\ell_i-l-2k)!}\\
    &\times\Bigg\{\sum_{s=0}^{2n+2\ell_i-l-2k}\sum_{p=0}^s(-1)^{s}\binom{n+2\ell_i-l-2k}{p}\binom{n}{s-p}\frac{(2\ell_i+2+s-p)_{n-s+p}}{(2l-2\ell_i+4k+2)_p}(2l+2k+s+2)!\\
    &\times \Bigg[ \sum_{j=0}^{n+2\ell_i-l-2k}(-1)^j\binom{n+2\ell_i-l-2k}{j}(2l-2\ell_i+4k+2+j)_{n+2\ell_i-l-2k-j}\sum_{v=0}^{n}(-1)^v\binom{n}{v}(2\ell_i+2+v)_{n-v}\\
    &\times (2l+2k+2+j+v)!-\sum_{j=0}^{n+2\ell_i+l+s+2}\sum_{q=0}^j\binom{n+2\ell_i-l-2k}{q}(2l-2\ell_i+4k+2+q)_{n+2\ell_i-l-2k-q}\frac{(-1)^q}{(j-q)!}\\
    &\times \sum_{v=0}^{n}(-1)^v\binom{n}{v}(2\ell_i+2+v)_{n-v}\frac{(2l+2k+2+j+v)!}{2^{2l+2k+3+j+v}}\Bigg]+\hspace{-0.3cm}\sum_{s=0}^{2n+2\ell_i-l-2k}\sum_{p=0}^s(-1)^s\binom{n+2\ell_i-l-2k}{p}\binom{n}{s-p}\\
    &\times (2l-2\ell_i+4k+2+p)_{n+2\ell_i-2k-p} (2\ell_i+2+s-p)_{n-s+p}(2l+2k+s+2)! \sum_{j=0}^{n+2\ell_i+l+s+2}\sum_{q=0}^j\binom{n+2\ell_i-l-2k}{q}\\
    &\times\frac{(-1)^q}{(2l-2\ell_i+4k+2)_q}\frac{1}{(j-q)!}\sum_{v=0}^{n}(-1)^v\binom{n}{v}(2\ell_i+2+v)_{n-v}\frac{(2l+2k+2+j+v)!}{2^{2l+2k+3+j+v}}\Bigg\},
\end{split}
\label{Hcal}
\end{align}

\begin{align}
    \mathcal{J}_{k}^{n\ell_i l}&\equiv \frac{4\pi}{2l+1}\frac{(-1)^l}{n!(n+\ell_i+1)(n+2\ell_i+1)!}\frac{1}{\Gamma(2+2l-2\ell_i+4k)(n+2\ell_i-l-2k)!}\int_0^{\infty} e^{-y}y^{2l+2k+2}U(-n,2\ell_i+2,y)\notag\\
    &\times\Bigg\{\frac{1}{n+\ell_i+1}U(-n-2\ell_i+l+2k,2l-2\ell_i+4k+2,y)\Bigg(2l-2\ell_i+4k+\frac{5}{2}-\frac{y}{2}+\psi(n+2\ell_i-l-2k+1)\Bigg)\notag\\
    &\times\sum_{s=0}^{2n+2\ell_i-l-2k}\sum_{p=0}^s(-1)^{s}\binom{n+2\ell_i-l-2k}{p}\binom{n}{s-p}\frac{(2\ell_i+2+s-p)_{n-s+p}}{(2l-2\ell_i+4k+2)_p}\notag\\
    &\times\Bigg[(2l+2k+s+2)!-\sum_{q=0}^{2l+2k+s+2}\frac{(2l+2k+s+2)!}{q!}y^q e^{-y}\Bigg]-\frac{1}{2(n+\ell_i+1)}U(-n-2\ell_i+l+2k,2l-2\ell_i+4k+2,y)\notag\\
    &\times\sum_{s=0}^{2n+2\ell_i-l-2k}\sum_{p=0}^s(-1)^{s}\binom{n+2\ell_i-l-2k}{p}\binom{n}{s-p}\frac{(2\ell_i+2+s-p)_{n-s+p}}{(2l-2\ell_i+4k+2)_p}\Bigg[(2l+2k+s+3)!\notag\\
    &-\sum_{q=0}^{2l+2k+s+3}\frac{(2l+2k+s+3)!}{q!}y^q e^{-y}\Bigg]+\frac{-n-2\ell_i+l+2k}{(2l-2\ell_i+4k+2)(n+\ell_i+1)} U(-n-2\ell_i+l+2k,2l-2\ell_i+4k+2,y)\notag\\
    &\times\sum_{s=0}^{2n+2\ell_i-l-2k-1}\sum_{p=0}^s(-1)^{s}\binom{n+2\ell_i-l-2k-1}{p}\binom{n}{s-p}\frac{(2\ell_i+2+s-p)_{n-s+p}}{(2l-2\ell_i+4k+3)_p}\Bigg[(2l+2k+s+3)!\notag\\
    &-\sum_{q=0}^{2l+2k+s+3}\frac{(2l+2k+s+3)!}{q!}y^q e^{-y}\Bigg]+(-1)^n U(-n-2\ell_i+l+2k,2l-2\ell_i+4k+2,y)\int_0^{y} e^{-x}x^{2l+2k+2}\notag\\
    &\times\frac{\partial}{\partial a}M(a,2l-2\ell_i+4k+2,x)\Big|_{a=-n-2\ell_i+l+2k}U(-n,2\ell_i+2,x)dx+\Bigg[\frac{n+2\ell_i-l-2k}{n+\ell_i+1}y\notag\\
    &\times U(-n-2\ell_i+l+2k+1,2l-2\ell_i+4k+3,y)+\frac{\partial}{\partial a}U(a,2l-2\ell_i+4k+2,y)\Big|_{a=-n-2\ell_i+l+2k}\Bigg]\notag\\
    &\times\sum_{s=0}^{2n+2\ell_i-l-2k}\sum_{p=0}^s(-1)^{s}\binom{n+2\ell_i-l-2k}{p}\binom{n}{s-p}\frac{(2\ell_i+2+s-p)_{n-s+p}}{(2l-2\ell_i+4k+2)_p}\Bigg[(2l+2k+s+2)!\notag\\
    &-\sum_{q=0}^{2l+2k+s+2}\frac{(2l+2k+s+2)!}{q!}y^q e^{-y}\Bigg]+\frac{1}{n+\ell_i+1} M(-n-2\ell_i+l+2k,2l-2\ell_i+4k+2,y)\Bigg(2l-2\ell_i+4k+\frac{5}{2}-\frac{y}{2}\notag\\
    &+\psi(n+2\ell_i-l-2k+1)\Bigg)\sum_{s=0}^{2n+2\ell_i-l-2k}\sum_{p=0}^s(-1)^{n+l+s}\binom{n+2\ell_i-l-2k}{p}\binom{n}{s-p}(2l-2\ell_i+4k+2+p)_{n+2\ell_i-l-2k-p}\notag\\
    &\times (2\ell_i+2+s-p)_{n-s+p}\sum_{q=0}^{2l+2k+s+2}\frac{(2l+2k+s+2)!}{q!}y^q e^{-y}-\frac{1}{2(n+\ell_i+1)} M(-n-2\ell_i+l+2k,2l-2\ell_i+4k+2,y)\notag\\
    &\times \sum_{s=0}^{2n+2\ell_i-l-2k}\sum_{p=0}^s(-1)^{n+l+s}\binom{n+2\ell_i-l-2k}{p}\binom{n}{s-p}(2l-2\ell_i+4k+2+p)_{n+2\ell_i-l-2k-p}(2\ell_i+2+s-p)_{n-s+p}\notag\\
    &\times\sum_{q=0}^{2l+2k+s+3}\frac{(2l+2k+s+3)!}{q!}y^q e^{-y}+\Bigg[\frac{-n-2\ell_i+l+2k}{n+\ell_i+1}y\, M(-n-2\ell_i+l+2k+1,2l-2\ell_i+4k+3,y)\notag\\
    &+\frac{\partial}{\partial a}M(a,2l-2\ell_i+4k+2,y)\Big|_{a=-n-2\ell_i+l+2k}\Bigg]\sum_{s=0}^{2n+2\ell_i-l-2k}\sum_{p=0}^s(-1)^{n+l+s}\binom{n+2\ell_i-l-2k}{p}\binom{n}{s-p}\notag\\
    &\times (2l-2\ell_i+4k+2+p)_{n+2\ell_i-l-2k-p}(2\ell_i+2+s-p)_{n-s+p}\sum_{q=0}^{2l+2k+s+2}\frac{(2l+2k+s+2)!}{q!}y^q e^{-y}\notag\\
    &+\frac{n+2\ell_i-l-2k}{n+\ell_i+1}M(-n-2\ell_i+l+2k,2l-2\ell_i+4k+2,y)\sum_{s=0}^{2n+2\ell_i-l-2k-1}\sum_{p=0}^s(-1)^{n+l+1+s}\binom{n+2\ell_i-l-2k-1}{p}\notag\\
    &\times\binom{n}{s-p}(2l-2\ell_i+4k+3+p)_{n+2\ell_i-l-2k-1-p}(2\ell_i+2+s-p)_{n-s+p}\sum_{q=0}^{2l+2k+s+3}\frac{(2l+2k+s+3)!}{q!}y^q e^{-y}\notag\\
    &+(-1)^n M(-n-2\ell_i+l+2k,2l-2\ell_i+4k+2,y)\int_{y}^{\infty} e^{-x}x^{2l+2k+2}\frac{\partial}{\partial a}U(a,2l-2\ell_i+4k+2,x)\Big|_{a=-n-2\ell_i+l+2k}\notag\\
    &\times U(-n,2\ell_i+2,x)dx\Bigg\}dy. \label{Jcal}
\end{align}
}
\end{widetext}
When calculating $\mathcal{A}_{k}^{n\ell_i l}$, the following integrals (which were determined with the help of Eqs.~\eqref{Mmenos}, \eqref{Umenos}, and \eqref{0inf}, as well as the Cauchy product) were used
\begin{widetext}
\begin{align}
\begin{split}
    &\int_0^{\infty}e^{-K_{n\ell_i}r'/2}r'^{\ell_i+2k+2}R_{n\ell_i}(r')U\left(-n-l+2k,2\ell_i-2l+4k+2,K_{n\ell_i}r'\right)\\
    &\times\left[1-\sum_{q=0}^{2\ell_i+2k+s+2}\frac{\left(K_{n\ell_i}r'\right)^q}{q!} e^{-K_{n\ell_i}r'}\right]dr'=\frac{(-1)^{n+l}(K_{n\ell_i})^{1/2}}{\sqrt{2n!(n+\ell_i+1)(n+2\ell_i+1)!}}\sum_{j=0}^{2n+l-2k}\sum_{v=0}^j(-1)^{j}\binom{n+l-2k}{v}\binom{n}{j-v}\\
    &\times (2\ell_i-2l+4k+2+v)_{n+l-2k-v}(2\ell_i+2+j-v)_{n-j+v}\left(K_{n\ell_i}\right)^{-\ell_i-2k-2}\\
    &\times\left[(2\ell_i+2k+2+j)!-\sum_{q=0}^{2\ell_i+2k+s+2}\frac{(2\ell_i+2k+2+q+j)!}{2^{2\ell_i+2k+3+q+j}q!}\right],\quad \text{for }2\leq l\leq \ell_i,\ 0\leq k<\frac{n+l}{2},\ 0\leq s\leq 2n+l-2k,
\end{split}
\end{align}
\begin{align}
\begin{split}
    &\int_0^{\infty}e^{-K_{n\ell_i}r'/2}r'^{\ell_i+2k+2}R_{n\ell_i}(r')M\left(-n-l+2k,2\ell_i-2l+4k+2,K_{n\ell_i}r'\right)\sum_{q=0}^{2\ell_i+2k+s+2}\frac{\left(K_{n\ell_i}r'\right)^q}{q!}e^{-K_{n\ell_i}r'}dr'\\
    &=\frac{(K_{n\ell_i})^{1/2}}{\sqrt{2n!(n+\ell_i+1)(n+2\ell_i+1)!}}\sum_{j=0}^{2n+l-2k}\sum_{v=0}^j(-1)^j\binom{n+l-2k}{v}\binom{n}{j-v}\frac{(2\ell_i+2+j-v)_{n-j+v}}{(2\ell_i-2l+4k+2)_{v}}\left(K_{n\ell_i}\right)^{-\ell_i-2k-2}\\
    &\times\sum_{q=0}^{2\ell_i+2k+s+2}\frac{(2\ell_i+2k+2+q+j)!}{2^{2\ell_i+2k+3+q+j}q!},\qquad \text{for }2\leq l\leq \ell_i,\ 0\leq k<\frac{n+l}{2},\ 0\leq s\leq 2n+l-2k.
\end{split}
\end{align}
\end{widetext}
When calculating $\hat{u}_{(2),sB}^{lm};\hat{u}_{(2),sC}^{lm};\hat{u}_{(2),sF}^{lm};\hat{u}_{(2),sH}^{lm};\hat{u}_{(2),sI}^{lm}$, the integral in Eq.~\eqref{nexttolastInt} was used to compute the terms of order $(\omega/E_{n\ell_i})^{-1}$ which are not inside any summation sign (i.e. corresponding to a particular value of $k$).\par
On the other hand, when calculating $\mathcal{H}_{k}^{n\ell_i l}$, the following integrals (again with the help of Eqs.~\eqref{Mmenos}, \eqref{Umenos}, \eqref{0inf}, and the Cauchy product) were used
\begin{widetext}
\begin{align}
\begin{split}
    &\int_0^{\infty}e^{-K_{n\ell_i}r'/2}r'^{2l-\ell_i+2k+2}R_{n\ell_i}(r') U\left(-n-2\ell_i+l+2k,2l-2\ell_i+4k+2,K_{n\ell_i}r'\right)\\
    &\times\left[1-\sum_{q=0}^{2l+2k+s+2}\frac{\left(K_{n\ell_i}r'\right)^q}{q!} e^{-K_{n\ell_i}r'}\right]dr'=\frac{(-1)^{n+l}(8\mu|E_{n\ell_i}|)^{1/4}}{\sqrt{2n!(n+\ell_i+1)(n+2\ell_i+1)!}}\left(K_{n\ell_i}\right)^{-2l+\ell_i-2k-2}\\
    &\times \sum_{j=0}^{n+2\ell_i-l-2k}\hspace{-0.4cm}(-1)^j\binom{n+2\ell_i-l-2k}{j}(2l-2\ell_i+4k+2+j)_{n+2\ell_i-l-2k-j}\sum_{v=0}^{n}(-1)^v\binom{n}{v}\\
    &\times (2\ell_i+2+v)_{n-v}(2l+2k+2+j+v)!-\frac{(-1)^{n+l}(8\mu|E_{n\ell_i}|)^{1/4}}{\sqrt{2n!(n+\ell_i+1)(n+2\ell_i+1)!}}\left(K_{n\ell_i}\right)^{-2l+\ell_i-2k-2}\\
    &\times \sum_{j=0}^{n+2\ell_i+l+s+2}\sum_{q=0}^j\binom{n+2\ell_i-l-2k}{q}(2l-2\ell_i+4k+2+q)_{n+2\ell_i-l-2k-q}\\
    &\times\frac{(-1)^q}{(j-q)!}\sum_{v=0}^{n}(-1)^v\binom{n}{v}(2\ell_i+2+v)_{n-v}\frac{(2l+2k+2+j+v)!}{2^{2l+2k+3+j+v}},\\ 
    &\text{for }l>\ell_i,\ 0\leq n<l,\ \ell_i>\frac{l-n}{2},\ 0\leq k<\ell_i+\frac{n-l}{2},\ 0\leq s\leq 2n+2\ell_i-l-2k\\
    &\text{or }l>\ell_i,\ n=l,\ 0\leq k<\ell_i,\ 0\leq s\leq 2n+2\ell_i-l-2k\ \text{or }l>\ell_i,\ n>l,\ 0\leq k\leq\ell_i,\ 0\leq s\leq 2n+2\ell_i-l-2k,
\end{split}
\end{align}
\begin{align}
\begin{split}
    &\int_0^{\infty}e^{-K_{n\ell_i}r'/2}r'^{2l-\ell_i+2k+2}R_{n\ell_i}(r') M\left(-n-2\ell_i+l+2k,2l-2\ell_i+4k+2,K_{n\ell_i}r'\right)\sum_{q=0}^{2l+2k+s+2}\frac{\left(K_{n\ell_i}r'\right)^q}{q!}e^{-K_{n\ell_i}r'}dr'\\
    &=\frac{(8\mu|E_{n\ell_i}|)^{1/4}}{\sqrt{2n!(n+\ell_i+1)(n+2\ell_i+1)!}}\left(K_{n\ell_i}\right)^{-2l+\ell_i-2k-2}\sum_{j=0}^{n+2\ell_i+l+s+2}\sum_{q=0}^j\binom{n+2\ell_i-l-2k}{q}\frac{(-1)^q}{(2l-2\ell_i+4k+2)_q}\\
    &\times\frac{1}{(j-q)!}\sum_{v=0}^{n}(-1)^v\binom{n}{v}(2\ell_i+2+v)_{n-v}\frac{(2l+2k+2+j+v)!}{2^{2l+2k+3+j+v}},\\ 
    &\text{for }l>\ell_i,\ 0\leq n<l,\ \ell_i>\frac{l-n}{2},\ 0\leq k<\ell_i+\frac{n-l}{2},\ 0\leq s\leq 2n+2\ell_i-l-2k\\
    &\text{or }l>\ell_i,\ n=l,\ 0\leq k<\ell_i,\ 0\leq s\leq 2n+2\ell_i-l-2k\ \text{or }l>\ell_i,\ n>l,\ 0\leq k\leq\ell_i,\ 0\leq s\leq 2n+2\ell_i-l-2k.
\end{split}
\end{align}
\end{widetext}
Finally, note that some integrals in the equations above were left uncomputed because we were not able to compute them analytically. In the \textit{Mathematica} package that we provide in Ref.~\cite{github}, those integrals are computed numerically using the \textit{Mathematica} built-in function \texttt{NIntegrate}, given a set of values for $n,\ell_i,m_i,l$ that the user provides.

With the expressions above we can now provide a proof of Eq.~\eqref{eq:u_sym}. The coefficients in front of the delta function in Eq.~\eqref{solucoesu2}, may be collectively written as\footnote{This is true except for $\hat{u}_{(2),sE}^{lm}$, for which case the terms of order $\omega^{-1}$ are never present, see Eq.~\eqref{usE}. In that case the same analysis still applies but without those terms in the expressions below.}
\begin{widetext}
\begin{align}
\begin{split}
    \hat{u}_{(2),sX}^{lm}(\omega)&=\sum_k c_{lm}\left[(C_1)^{l\ell_i k}_{m m_i}(C_1)^{l\ell_i k}_{m m_i}-(C_2)^{l\ell_i k}_{m m_i}(C_2)^{l\ell_i k}_{m m_i}\right](\text{const. independent of $m$})\left(\frac{\omega}{E_{n\ell_i}}\right)^{-1}\\
    &+\sum_k c_{lm}\left[(C_1)^{l\ell_i k}_{m m_i}(C_1)^{l\ell_i k}_{m m_i}+(C_2)^{l\ell_i k}_{m m_i}(C_2)^{l\ell_i k}_{m m_i}\right](\text{const. independent of $m$})+\mathcal{O}\left(\frac{\omega}{E_{n\ell_i}}\right)\,.
\end{split}
\end{align}
%\end{widetext}
%
Using Eqs.~\eqref{symmetry1} and~\eqref{symmetry4} one gets $(C_2)^{l\ell_i k}_{m m_i}=(-1)^{m_i}(C_1)^{l\ell_i k}_{-m, m_i}$. Hence, Eqs.~\eqref{solucoesu2} and~\eqref{cl-m} imply
%\begin{widetext}
\begin{align}
\begin{split}
    &\int_{-\infty}^\infty\frac{d\omega}{2\pi}(-1)^m\hat{u}_{(2)}^{l,-m}(\omega,r)e^{i\omega t}=\int_{-\infty}^\infty\frac{d\omega}{2\pi}(-1)^m\Bigg\{\frac{1}{r^{l+1}}\sum_k c_{l,-m}\left[(C_1)^{l\ell_i k}_{-m, m_i}(C_1)^{l\ell_i k}_{-m, m_i}-(C_2)^{l\ell_i k}_{-m, m_i}(C_2)^{l\ell_i k}_{-m, m_i}\right]\\
    &\times (\text{const. independent of $m$})\left(\frac{\omega}{E_{n\ell_i}}\right)^{-1}+\frac{1}{r^{l+1}}\sum_k c_{l,-m}\left[(C_1)^{l\ell_i k}_{-m, m_i}(C_1)^{l\ell_i k}_{-m, m_i}+(C_2)^{l\ell_i k}_{-m, m_i}(C_2)^{l\ell_i k}_{-m, m_i}\right]\\
    &\times (\text{const. independent of $m$})+\mathcal{O}\left(\frac{\omega}{E_{n\ell_i}}\right)\Bigg\}e^{i\omega t}\delta(\omega+m\Omegaorb)\\
    &=\frac{1}{2\pi}\Bigg\{\frac{1}{r^{l+1}}\sum_k c_{lm}\left[(C_2)^{l\ell_i k}_{m m_i}(C_2)^{l\ell_i k}_{m m_i}-(C_1)^{l\ell_i k}_{m m_i}(C_1)^{l\ell_i k}_{m m_i}\right] (\text{const. independent of $m$})\frac{E_{n\ell_i}}{(-m\Omegaorb)}\\
    &+\frac{1}{r^{l+1}}\sum_k c_{lm}\left[(C_1)^{l\ell_i k}_{m m_i}(C_1)^{l\ell_i k}_{m m_i}+(C_2)^{l\ell_i k}_{m m_i}(C_2)^{l\ell_i k}_{m m_i}\right] (\text{const. independent of $m$})+\mathcal{O}\left(\frac{\Omegaorb}{E_{n\ell_i}}\right)\Bigg\}e^{-im\Omegaorb t}\\
    &=\int_{-\infty}^\infty\frac{d\omega}{2\pi}\Bigg\{\frac{1}{r^{l+1}}\sum_k c_{lm}\left[(C_1)^{l\ell_i k}_{m m_i}(C_1)^{l\ell_i k}_{m m_i}-(C_2)^{l\ell_i k}_{m m_i}(C_2)^{l\ell_i k}_{m m_i}\right] (\text{const. independent of $m$})\left(\frac{\omega}{E_{n\ell_i}}\right)^{-1}\\
    &+\frac{1}{r^{l+1}}\sum_k c_{lm}\left[(C_1)^{l\ell_i k}_{m m_i}(C_1)^{l\ell_i k}_{m m_i}+(C_2)^{l\ell_i k}_{m m_i}(C_2)^{l\ell_i k}_{m m_i}\right] (\text{const. independent of $m$})+\mathcal{O}\left(\frac{\omega}{E_{n\ell_i}}\right)\Bigg\}e^{-i\omega t}\delta(\omega-m\Omegaorb)\\
    &=\int_{-\infty}^\infty\frac{d\omega}{2\pi}\hat{u}_{(2)}^{lm}(\omega,r)e^{-i\omega t}.
\end{split}
\end{align}
%\end{widetext}
On the other hand, we get from Eqs.~\eqref{u0sol}, \eqref{Alm}, and \eqref{cl-m}
%\begin{widetext}
\begin{align}
\begin{split}
    &\int_{-\infty}^\infty\frac{d\omega}{2\pi}(-1)^m\hat{u}^{l,-m}_{(0)}(\omega,r)e^{i\omega t}=\int_{-\infty}^\infty\frac{d\omega}{2\pi}(-1)^m c_{l,-m}e^{i\omega t}\delta(\omega+m\Omegaorb)r^l=\frac{1}{2\pi}c_{lm}e^{-im\Omegaorb t}r^l\\
    &=\int_{-\infty}^\infty\frac{d\omega}{2\pi}c_{lm}e^{-i\omega t}\delta(\omega-m\Omegaorb)r^l=\int_{-\infty}^\infty\frac{d\omega}{2\pi}\hat{u}^{lm}_{(0)}(\omega,r)e^{-i\omega t}.
\end{split}
\end{align}
%\end{widetext}
Using these two equations results in
%\begin{widetext}
\begin{equation}
%\begin{split}
    \int_{-\infty}^\infty\frac{d\omega}{2\pi}(-1)^m(\hat{u}^{l,-m}_{(0)}(\omega,r)+\epsilon^2\hat{u}^{l,-m}_{(2)}(\omega,r))e^{i\omega t}
    =\int_{-\infty}^\infty\frac{d\omega}{2\pi}[(\hat{u}^{lm}_{(0)}(\omega,r)+\epsilon^2\hat{u}^{lm}_{(2)}(\omega,r))e^{-i\omega t}].
%\end{split}
\end{equation}
\end{widetext}

%%%%%%%%%%%%%%%%%%%%%%%%%%%%%%%%%%%%%%%%%
\bibliography{article_v3.bbl}

%merlin.mbs apsrev4-1.bst 2010-07-25 4.21a (PWD, AO, DPC) hacked
%Control: key (0)
%Control: author (8) initials jnrlst
%Control: editor formatted (1) identically to author
%Control: production of article title (0) allowed
%Control: page (1) range
%Control: year (1) truncated
%Control: production of eprint (0) enabled
\begin{thebibliography}{80}%
\makeatletter
\providecommand \@ifxundefined [1]{%
 \@ifx{#1\undefined}
}%
\providecommand \@ifnum [1]{%
 \ifnum #1\expandafter \@firstoftwo
 \else \expandafter \@secondoftwo
 \fi
}%
\providecommand \@ifx [1]{%
 \ifx #1\expandafter \@firstoftwo
 \else \expandafter \@secondoftwo
 \fi
}%
\providecommand \natexlab [1]{#1}%
\providecommand \enquote  [1]{``#1''}%
\providecommand \bibnamefont  [1]{#1}%
\providecommand \bibfnamefont [1]{#1}%
\providecommand \citenamefont [1]{#1}%
\providecommand \href@noop [0]{\@secondoftwo}%
\providecommand \href [0]{\begingroup \@sanitize@url \@href}%
\providecommand \@href[1]{\@@startlink{#1}\@@href}%
\providecommand \@@href[1]{\endgroup#1\@@endlink}%
\providecommand \@sanitize@url [0]{\catcode `\\12\catcode `\$12\catcode
  `\&12\catcode `\#12\catcode `\^12\catcode `\_12\catcode `\%12\relax}%
\providecommand \@@startlink[1]{}%
\providecommand \@@endlink[0]{}%
\providecommand \url  [0]{\begingroup\@sanitize@url \@url }%
\providecommand \@url [1]{\endgroup\@href {#1}{\urlprefix }}%
\providecommand \urlprefix  [0]{URL }%
\providecommand \Eprint [0]{\href }%
\providecommand \doibase [0]{http://dx.doi.org/}%
\providecommand \selectlanguage [0]{\@gobble}%
\providecommand \bibinfo  [0]{\@secondoftwo}%
\providecommand \bibfield  [0]{\@secondoftwo}%
\providecommand \translation [1]{[#1]}%
\providecommand \BibitemOpen [0]{}%
\providecommand \bibitemStop [0]{}%
\providecommand \bibitemNoStop [0]{.\EOS\space}%
\providecommand \EOS [0]{\spacefactor3000\relax}%
\providecommand \BibitemShut  [1]{\csname bibitem#1\endcsname}%
\let\auto@bib@innerbib\@empty
%</preamble>
\bibitem [{\citenamefont {Love}(1911)}]{Love:1911}%
  \BibitemOpen
  \bibfield  {author} {\bibinfo {author} {\bibfnamefont {A.~E.~H.}\
  \bibnamefont {Love}},\ }\href
  {https://www.cambridge.org/us/universitypress/subjects/history/history-science-general-interest/some-problems-geodynamics?format=PB&isbn=9781107536470}
  {\emph {\bibinfo {title} {{Some Problems of Geodynamics}}}}\ (\bibinfo
  {publisher} {Cambridge University Press},\ \bibinfo {year}
  {1911})\BibitemShut {NoStop}%
\bibitem [{\citenamefont {Poisson}\ and\ \citenamefont
  {Will}(2014)}]{Poisson:2014}%
  \BibitemOpen
  \bibfield  {author} {\bibinfo {author} {\bibfnamefont {E.}~\bibnamefont
  {Poisson}}\ and\ \bibinfo {author} {\bibfnamefont {C.~M.}\ \bibnamefont
  {Will}},\ }\href {\doibase 10.1017/CBO9781139507486} {\emph {\bibinfo {title}
  {Gravity: Newtonian, Post-Newtonian, Relativistic}}}\ (\bibinfo  {publisher}
  {Cambridge University Press},\ \bibinfo {year} {2014})\BibitemShut {NoStop}%
\bibitem [{\citenamefont {Hinderer}(2008)}]{Hinderer:2007}%
  \BibitemOpen
  \bibfield  {author} {\bibinfo {author} {\bibfnamefont {T.}~\bibnamefont
  {Hinderer}},\ }\bibfield  {title} {\enquote {\bibinfo {title} {{Tidal Love
  numbers of neutron stars}},}\ }\href {\doibase 10.1086/533487} {\bibfield
  {journal} {\bibinfo  {journal} {Astrophys. J.}\ }\textbf {\bibinfo {volume}
  {677}},\ \bibinfo {pages} {1216--1220} (\bibinfo {year} {2008})},\ \bibinfo
  {note} {[Erratum: Astrophys. J. 697, 964, (2009)]},\ \Eprint
  {http://arxiv.org/abs/0711.2420} {arXiv:0711.2420 [astro-ph]} \BibitemShut
  {NoStop}%
\bibitem [{\citenamefont {Flanagan}\ and\ \citenamefont
  {Hinderer}(2008)}]{Flanagan:2007}%
  \BibitemOpen
  \bibfield  {author} {\bibinfo {author} {\bibfnamefont {E.~E.}\ \bibnamefont
  {Flanagan}}\ and\ \bibinfo {author} {\bibfnamefont {T.}~\bibnamefont
  {Hinderer}},\ }\bibfield  {title} {\enquote {\bibinfo {title} {{Constraining
  neutron star tidal Love numbers with gravitational wave detectors}},}\ }\href
  {\doibase 10.1103/PhysRevD.77.021502} {\bibfield  {journal} {\bibinfo
  {journal} {Phys. Rev. D}\ }\textbf {\bibinfo {volume} {77}},\ \bibinfo
  {pages} {021502} (\bibinfo {year} {2008})},\ \Eprint
  {http://arxiv.org/abs/0709.1915} {arXiv:0709.1915 [astro-ph]} \BibitemShut
  {NoStop}%
\bibitem [{\citenamefont {Chatziioannou}(2020)}]{Chatziioannou:2020pqz}%
  \BibitemOpen
  \bibfield  {author} {\bibinfo {author} {\bibfnamefont {K.}~\bibnamefont
  {Chatziioannou}},\ }\bibfield  {title} {\enquote {\bibinfo {title} {{Neutron
  star tidal deformability and equation of state constraints}},}\ }\href
  {\doibase 10.1007/s10714-020-02754-3} {\bibfield  {journal} {\bibinfo
  {journal} {Gen. Rel. Grav.}\ }\textbf {\bibinfo {volume} {52}},\ \bibinfo
  {pages} {109} (\bibinfo {year} {2020})},\ \Eprint
  {http://arxiv.org/abs/2006.03168} {arXiv:2006.03168 [gr-qc]} \BibitemShut
  {NoStop}%
\bibitem [{\citenamefont {Fang}\ and\ \citenamefont
  {Lovelace}(2005)}]{Fang:2005}%
  \BibitemOpen
  \bibfield  {author} {\bibinfo {author} {\bibfnamefont {H.}~\bibnamefont
  {Fang}}\ and\ \bibinfo {author} {\bibfnamefont {G.}~\bibnamefont
  {Lovelace}},\ }\bibfield  {title} {\enquote {\bibinfo {title} {{Tidal
  coupling of a Schwarzschild black hole and circularly orbiting moon}},}\
  }\href {\doibase 10.1103/PhysRevD.72.124016} {\bibfield  {journal} {\bibinfo
  {journal} {Phys. Rev. D}\ }\textbf {\bibinfo {volume} {72}},\ \bibinfo
  {pages} {124016} (\bibinfo {year} {2005})},\ \Eprint
  {http://arxiv.org/abs/gr-qc/0505156} {arXiv:gr-qc/0505156} \BibitemShut
  {NoStop}%
\bibitem [{\citenamefont {Binnington}\ and\ \citenamefont
  {Poisson}(2009)}]{Binnington:2009}%
  \BibitemOpen
  \bibfield  {author} {\bibinfo {author} {\bibfnamefont {T.}~\bibnamefont
  {Binnington}}\ and\ \bibinfo {author} {\bibfnamefont {E.}~\bibnamefont
  {Poisson}},\ }\bibfield  {title} {\enquote {\bibinfo {title} {{Relativistic
  theory of tidal Love numbers}},}\ }\href {\doibase
  10.1103/PhysRevD.80.084018} {\bibfield  {journal} {\bibinfo  {journal} {Phys.
  Rev. D}\ }\textbf {\bibinfo {volume} {80}},\ \bibinfo {pages} {084018}
  (\bibinfo {year} {2009})},\ \Eprint {http://arxiv.org/abs/0906.1366}
  {arXiv:0906.1366 [gr-qc]} \BibitemShut {NoStop}%
\bibitem [{\citenamefont {Pani}\ \emph
  {et~al.}(2015{\natexlab{a}})\citenamefont {Pani}, \citenamefont {Gualtieri},
  \citenamefont {Maselli},\ and\ \citenamefont {Ferrari}}]{Pani:2015hfa}%
  \BibitemOpen
  \bibfield  {author} {\bibinfo {author} {\bibfnamefont {P.}~\bibnamefont
  {Pani}}, \bibinfo {author} {\bibfnamefont {L.}~\bibnamefont {Gualtieri}},
  \bibinfo {author} {\bibfnamefont {A.}~\bibnamefont {Maselli}}, \ and\
  \bibinfo {author} {\bibfnamefont {V.}~\bibnamefont {Ferrari}},\ }\bibfield
  {title} {\enquote {\bibinfo {title} {{Tidal deformations of a spinning
  compact object}},}\ }\href {\doibase 10.1103/PhysRevD.92.024010} {\bibfield
  {journal} {\bibinfo  {journal} {Phys. Rev. D}\ }\textbf {\bibinfo {volume}
  {92}},\ \bibinfo {pages} {024010} (\bibinfo {year} {2015}{\natexlab{a}})},\
  \Eprint {http://arxiv.org/abs/1503.07365} {arXiv:1503.07365 [gr-qc]}
  \BibitemShut {NoStop}%
\bibitem [{\citenamefont {Landry}\ and\ \citenamefont
  {Poisson}(2015)}]{Landry:2015zfa}%
  \BibitemOpen
  \bibfield  {author} {\bibinfo {author} {\bibfnamefont {P.}~\bibnamefont
  {Landry}}\ and\ \bibinfo {author} {\bibfnamefont {E.}~\bibnamefont
  {Poisson}},\ }\bibfield  {title} {\enquote {\bibinfo {title} {{Tidal
  deformation of a slowly rotating material body. External metric}},}\ }\href
  {\doibase 10.1103/PhysRevD.91.104018} {\bibfield  {journal} {\bibinfo
  {journal} {Phys. Rev. D}\ }\textbf {\bibinfo {volume} {91}},\ \bibinfo
  {pages} {104018} (\bibinfo {year} {2015})},\ \Eprint
  {http://arxiv.org/abs/1503.07366} {arXiv:1503.07366 [gr-qc]} \BibitemShut
  {NoStop}%
\bibitem [{\citenamefont {Chia}(2021)}]{Chia:2020}%
  \BibitemOpen
  \bibfield  {author} {\bibinfo {author} {\bibfnamefont {H.~S.}\ \bibnamefont
  {Chia}},\ }\bibfield  {title} {\enquote {\bibinfo {title} {{Tidal deformation
  and dissipation of rotating black holes}},}\ }\href {\doibase
  10.1103/PhysRevD.104.024013} {\bibfield  {journal} {\bibinfo  {journal}
  {Phys. Rev. D}\ }\textbf {\bibinfo {volume} {104}},\ \bibinfo {pages}
  {024013} (\bibinfo {year} {2021})},\ \Eprint
  {http://arxiv.org/abs/2010.07300} {arXiv:2010.07300 [gr-qc]} \BibitemShut
  {NoStop}%
\bibitem [{\citenamefont {Charalambous}\ \emph {et~al.}(2021)\citenamefont
  {Charalambous}, \citenamefont {Dubovsky},\ and\ \citenamefont
  {Ivanov}}]{Charalambous:2021}%
  \BibitemOpen
  \bibfield  {author} {\bibinfo {author} {\bibfnamefont {P.}~\bibnamefont
  {Charalambous}}, \bibinfo {author} {\bibfnamefont {S.}~\bibnamefont
  {Dubovsky}}, \ and\ \bibinfo {author} {\bibfnamefont {M.~M.}\ \bibnamefont
  {Ivanov}},\ }\bibfield  {title} {\enquote {\bibinfo {title} {{On the
  Vanishing of Love Numbers for Kerr Black Holes}},}\ }\href {\doibase
  10.1007/JHEP05(2021)038} {\bibfield  {journal} {\bibinfo  {journal} {JHEP}\
  }\textbf {\bibinfo {volume} {05}},\ \bibinfo {pages} {038} (\bibinfo {year}
  {2021})},\ \Eprint {http://arxiv.org/abs/2102.08917} {arXiv:2102.08917
  [hep-th]} \BibitemShut {NoStop}%
\bibitem [{\citenamefont {Nair}\ \emph {et~al.}(2024)\citenamefont {Nair},
  \citenamefont {Chakraborty},\ and\ \citenamefont {Sarkar}}]{Nair:2024}%
  \BibitemOpen
  \bibfield  {author} {\bibinfo {author} {\bibfnamefont {S.}~\bibnamefont
  {Nair}}, \bibinfo {author} {\bibfnamefont {S.}~\bibnamefont {Chakraborty}}, \
  and\ \bibinfo {author} {\bibfnamefont {S.}~\bibnamefont {Sarkar}},\
  }\bibfield  {title} {\enquote {\bibinfo {title} {{Asymptotically de Sitter
  black holes have nonzero tidal Love numbers}},}\ }\href {\doibase
  10.1103/PhysRevD.109.064025} {\bibfield  {journal} {\bibinfo  {journal}
  {Phys. Rev. D}\ }\textbf {\bibinfo {volume} {109}},\ \bibinfo {pages}
  {064025} (\bibinfo {year} {2024})},\ \Eprint
  {http://arxiv.org/abs/2401.06467} {arXiv:2401.06467 [gr-qc]} \BibitemShut
  {NoStop}%
\bibitem [{\citenamefont {Mendes}\ and\ \citenamefont
  {Yang}(2017)}]{Mendes:2016vdr}%
  \BibitemOpen
  \bibfield  {author} {\bibinfo {author} {\bibfnamefont {R.~F.~P.}\
  \bibnamefont {Mendes}}\ and\ \bibinfo {author} {\bibfnamefont
  {H.}~\bibnamefont {Yang}},\ }\bibfield  {title} {\enquote {\bibinfo {title}
  {{Tidal deformability of boson stars and dark matter clumps}},}\ }\href
  {\doibase 10.1088/1361-6382/aa842d} {\bibfield  {journal} {\bibinfo
  {journal} {Class. Quant. Grav.}\ }\textbf {\bibinfo {volume} {34}},\ \bibinfo
  {pages} {185001} (\bibinfo {year} {2017})},\ \Eprint
  {http://arxiv.org/abs/1606.03035} {arXiv:1606.03035 [astro-ph.CO]}
  \BibitemShut {NoStop}%
\bibitem [{\citenamefont {Cardoso}\ \emph {et~al.}(2017)\citenamefont {Cardoso}
  \emph {et~al.}}]{Cardoso:2017}%
  \BibitemOpen
  \bibfield  {author} {\bibinfo {author} {\bibfnamefont {V.}~\bibnamefont
  {Cardoso}} \emph {et~al.},\ }\bibfield  {title} {\enquote {\bibinfo {title}
  {{Testing strong-field gravity with tidal Love numbers}},}\ }\href {\doibase
  10.1103/PhysRevD.95.084014} {\bibfield  {journal} {\bibinfo  {journal} {Phys.
  Rev. D}\ }\textbf {\bibinfo {volume} {95}},\ \bibinfo {pages} {084014}
  (\bibinfo {year} {2017})},\ \bibinfo {note} {[Addendum: Phys.Rev.D 95, 089901
  (2017)]},\ \Eprint {http://arxiv.org/abs/1701.01116} {arXiv:1701.01116
  [gr-qc]} \BibitemShut {NoStop}%
\bibitem [{\citenamefont {Maselli}\ \emph {et~al.}(2018)\citenamefont
  {Maselli}, \citenamefont {Pani}, \citenamefont {Cardoso}, \citenamefont
  {Abdelsalhin}, \citenamefont {Gualtieri},\ and\ \citenamefont
  {Ferrari}}]{Maselli:2017cmm}%
  \BibitemOpen
  \bibfield  {author} {\bibinfo {author} {\bibfnamefont {A.}~\bibnamefont
  {Maselli}}, \bibinfo {author} {\bibfnamefont {P.}~\bibnamefont {Pani}},
  \bibinfo {author} {\bibfnamefont {V.}~\bibnamefont {Cardoso}}, \bibinfo
  {author} {\bibfnamefont {T.}~\bibnamefont {Abdelsalhin}}, \bibinfo {author}
  {\bibfnamefont {L.}~\bibnamefont {Gualtieri}}, \ and\ \bibinfo {author}
  {\bibfnamefont {V.}~\bibnamefont {Ferrari}},\ }\bibfield  {title} {\enquote
  {\bibinfo {title} {{Probing Planckian corrections at the horizon scale with
  LISA binaries}},}\ }\href {\doibase 10.1103/PhysRevLett.120.081101}
  {\bibfield  {journal} {\bibinfo  {journal} {Phys. Rev. Lett.}\ }\textbf
  {\bibinfo {volume} {120}},\ \bibinfo {pages} {081101} (\bibinfo {year}
  {2018})},\ \Eprint {http://arxiv.org/abs/1703.10612} {arXiv:1703.10612
  [gr-qc]} \BibitemShut {NoStop}%
\bibitem [{\citenamefont {Sennett}\ \emph {et~al.}(2017)\citenamefont
  {Sennett}, \citenamefont {Hinderer}, \citenamefont {Steinhoff}, \citenamefont
  {Buonanno},\ and\ \citenamefont {Ossokine}}]{Sennett:2017etc}%
  \BibitemOpen
  \bibfield  {author} {\bibinfo {author} {\bibfnamefont {N.}~\bibnamefont
  {Sennett}}, \bibinfo {author} {\bibfnamefont {T.}~\bibnamefont {Hinderer}},
  \bibinfo {author} {\bibfnamefont {J.}~\bibnamefont {Steinhoff}}, \bibinfo
  {author} {\bibfnamefont {A.}~\bibnamefont {Buonanno}}, \ and\ \bibinfo
  {author} {\bibfnamefont {S.}~\bibnamefont {Ossokine}},\ }\bibfield  {title}
  {\enquote {\bibinfo {title} {{Distinguishing Boson Stars from Black Holes and
  Neutron Stars from Tidal Interactions in Inspiraling Binary Systems}},}\
  }\href {\doibase 10.1103/PhysRevD.96.024002} {\bibfield  {journal} {\bibinfo
  {journal} {Phys. Rev. D}\ }\textbf {\bibinfo {volume} {96}},\ \bibinfo
  {pages} {024002} (\bibinfo {year} {2017})},\ \Eprint
  {http://arxiv.org/abs/1704.08651} {arXiv:1704.08651 [gr-qc]} \BibitemShut
  {NoStop}%
\bibitem [{\citenamefont {Cardoso}\ and\ \citenamefont
  {Pani}(2019)}]{Cardoso:2019rvt}%
  \BibitemOpen
  \bibfield  {author} {\bibinfo {author} {\bibfnamefont {V.}~\bibnamefont
  {Cardoso}}\ and\ \bibinfo {author} {\bibfnamefont {P.}~\bibnamefont {Pani}},\
  }\bibfield  {title} {\enquote {\bibinfo {title} {{Testing the nature of dark
  compact objects: a status report}},}\ }\href {\doibase
  10.1007/s41114-019-0020-4} {\bibfield  {journal} {\bibinfo  {journal} {Living
  Rev. Rel.}\ }\textbf {\bibinfo {volume} {22}},\ \bibinfo {pages} {4}
  (\bibinfo {year} {2019})},\ \Eprint {http://arxiv.org/abs/1904.05363}
  {arXiv:1904.05363 [gr-qc]} \BibitemShut {NoStop}%
\bibitem [{\citenamefont {Herdeiro}\ \emph {et~al.}(2020)\citenamefont
  {Herdeiro}, \citenamefont {Panotopoulos},\ and\ \citenamefont
  {Radu}}]{Herdeiro:2020kba}%
  \BibitemOpen
  \bibfield  {author} {\bibinfo {author} {\bibfnamefont {C.~A.~R.}\
  \bibnamefont {Herdeiro}}, \bibinfo {author} {\bibfnamefont {G.}~\bibnamefont
  {Panotopoulos}}, \ and\ \bibinfo {author} {\bibfnamefont {E.}~\bibnamefont
  {Radu}},\ }\bibfield  {title} {\enquote {\bibinfo {title} {{Tidal Love
  numbers of Proca stars}},}\ }\href {\doibase 10.1088/1475-7516/2020/08/029}
  {\bibfield  {journal} {\bibinfo  {journal} {JCAP}\ }\textbf {\bibinfo
  {volume} {08}},\ \bibinfo {pages} {029} (\bibinfo {year} {2020})},\ \Eprint
  {http://arxiv.org/abs/2006.11083} {arXiv:2006.11083 [gr-qc]} \BibitemShut
  {NoStop}%
\bibitem [{\citenamefont {Cardoso}\ \emph {et~al.}(2018)\citenamefont {Cardoso}
  \emph {et~al.}}]{Cardoso:2018ptl}%
  \BibitemOpen
  \bibfield  {author} {\bibinfo {author} {\bibfnamefont {V.}~\bibnamefont
  {Cardoso}} \emph {et~al.},\ }\bibfield  {title} {\enquote {\bibinfo {title}
  {{Black Holes in an Effective Field Theory Extension of General
  Relativity}},}\ }\href {\doibase 10.1103/PhysRevLett.121.251105} {\bibfield
  {journal} {\bibinfo  {journal} {Phys. Rev. Lett.}\ }\textbf {\bibinfo
  {volume} {121}},\ \bibinfo {pages} {251105} (\bibinfo {year} {2018})},\
  \bibinfo {note} {[Erratum: Phys.Rev.Lett. 131, 109903 (2023)]},\ \Eprint
  {http://arxiv.org/abs/1808.08962} {arXiv:1808.08962 [gr-qc]} \BibitemShut
  {NoStop}%
\bibitem [{\citenamefont {De~Luca}\ \emph
  {et~al.}(2023{\natexlab{a}})\citenamefont {De~Luca}, \citenamefont {Khoury},\
  and\ \citenamefont {Wong}}]{DeLuca:2022tkm}%
  \BibitemOpen
  \bibfield  {author} {\bibinfo {author} {\bibfnamefont {V.}~\bibnamefont
  {De~Luca}}, \bibinfo {author} {\bibfnamefont {J.}~\bibnamefont {Khoury}}, \
  and\ \bibinfo {author} {\bibfnamefont {S.~S.~C.}\ \bibnamefont {Wong}},\
  }\bibfield  {title} {\enquote {\bibinfo {title} {{Implications of the weak
  gravity conjecture for tidal Love numbers of black holes}},}\ }\href
  {\doibase 10.1103/PhysRevD.108.044066} {\bibfield  {journal} {\bibinfo
  {journal} {Phys. Rev. D}\ }\textbf {\bibinfo {volume} {108}},\ \bibinfo
  {pages} {044066} (\bibinfo {year} {2023}{\natexlab{a}})},\ \Eprint
  {http://arxiv.org/abs/2211.14325} {arXiv:2211.14325 [hep-th]} \BibitemShut
  {NoStop}%
\bibitem [{\citenamefont {Baumann}\ \emph
  {et~al.}(2019{\natexlab{a}})\citenamefont {Baumann}, \citenamefont {Chia},\
  and\ \citenamefont {Porto}}]{Baumann:2018}%
  \BibitemOpen
  \bibfield  {author} {\bibinfo {author} {\bibfnamefont {D.}~\bibnamefont
  {Baumann}}, \bibinfo {author} {\bibfnamefont {H.~S.}\ \bibnamefont {Chia}}, \
  and\ \bibinfo {author} {\bibfnamefont {R.~A.}\ \bibnamefont {Porto}},\
  }\bibfield  {title} {\enquote {\bibinfo {title} {{Probing Ultralight Bosons
  with Binary Black Holes}},}\ }\href {\doibase 10.1103/PhysRevD.99.044001}
  {\bibfield  {journal} {\bibinfo  {journal} {Phys. Rev. D}\ }\textbf {\bibinfo
  {volume} {99}},\ \bibinfo {pages} {044001} (\bibinfo {year}
  {2019}{\natexlab{a}})},\ \Eprint {http://arxiv.org/abs/1804.03208}
  {arXiv:1804.03208 [gr-qc]} \BibitemShut {NoStop}%
\bibitem [{\citenamefont {Cardoso}\ and\ \citenamefont
  {Duque}(2020)}]{Duque:2019}%
  \BibitemOpen
  \bibfield  {author} {\bibinfo {author} {\bibfnamefont {V.}~\bibnamefont
  {Cardoso}}\ and\ \bibinfo {author} {\bibfnamefont {F.}~\bibnamefont
  {Duque}},\ }\bibfield  {title} {\enquote {\bibinfo {title} {{Environmental
  effects in gravitational-wave physics: Tidal deformability of black holes
  immersed in matter}},}\ }\href {\doibase 10.1103/PhysRevD.101.064028}
  {\bibfield  {journal} {\bibinfo  {journal} {Phys. Rev. D}\ }\textbf {\bibinfo
  {volume} {101}},\ \bibinfo {pages} {064028} (\bibinfo {year} {2020})},\
  \Eprint {http://arxiv.org/abs/1912.07616} {arXiv:1912.07616 [gr-qc]}
  \BibitemShut {NoStop}%
\bibitem [{\citenamefont {De~Luca}\ and\ \citenamefont
  {Pani}(2021)}]{DeLuca:2021}%
  \BibitemOpen
  \bibfield  {author} {\bibinfo {author} {\bibfnamefont {V.}~\bibnamefont
  {De~Luca}}\ and\ \bibinfo {author} {\bibfnamefont {P.}~\bibnamefont {Pani}},\
  }\bibfield  {title} {\enquote {\bibinfo {title} {{Tidal deformability of
  dressed black holes and tests of ultralight bosons in extended mass
  ranges}},}\ }\href {\doibase 10.1088/1475-7516/2021/08/032} {\bibfield
  {journal} {\bibinfo  {journal} {JCAP}\ }\textbf {\bibinfo {volume} {08}},\
  \bibinfo {pages} {032} (\bibinfo {year} {2021})},\ \Eprint
  {http://arxiv.org/abs/2106.14428} {arXiv:2106.14428 [gr-qc]} \BibitemShut
  {NoStop}%
\bibitem [{\citenamefont {De~Luca}\ \emph
  {et~al.}(2023{\natexlab{b}})\citenamefont {De~Luca}, \citenamefont
  {Maselli},\ and\ \citenamefont {Pani}}]{DeLuca:2022xlz}%
  \BibitemOpen
  \bibfield  {author} {\bibinfo {author} {\bibfnamefont {V.}~\bibnamefont
  {De~Luca}}, \bibinfo {author} {\bibfnamefont {A.}~\bibnamefont {Maselli}}, \
  and\ \bibinfo {author} {\bibfnamefont {P.}~\bibnamefont {Pani}},\ }\bibfield
  {title} {\enquote {\bibinfo {title} {{Modeling frequency-dependent tidal
  deformability for environmental black hole mergers}},}\ }\href {\doibase
  10.1103/PhysRevD.107.044058} {\bibfield  {journal} {\bibinfo  {journal}
  {Phys. Rev. D}\ }\textbf {\bibinfo {volume} {107}},\ \bibinfo {pages}
  {044058} (\bibinfo {year} {2023}{\natexlab{b}})},\ \Eprint
  {http://arxiv.org/abs/2212.03343} {arXiv:2212.03343 [gr-qc]} \BibitemShut
  {NoStop}%
\bibitem [{\citenamefont {Katagiri}\ \emph {et~al.}(2023)\citenamefont
  {Katagiri}, \citenamefont {Nakano},\ and\ \citenamefont
  {Omukai}}]{Katagiri:2023yzm}%
  \BibitemOpen
  \bibfield  {author} {\bibinfo {author} {\bibfnamefont {T.}~\bibnamefont
  {Katagiri}}, \bibinfo {author} {\bibfnamefont {H.}~\bibnamefont {Nakano}}, \
  and\ \bibinfo {author} {\bibfnamefont {K.}~\bibnamefont {Omukai}},\
  }\bibfield  {title} {\enquote {\bibinfo {title} {{Stability of relativistic
  tidal response against small potential modification}},}\ }\href {\doibase
  10.1103/PhysRevD.108.084049} {\bibfield  {journal} {\bibinfo  {journal}
  {Phys. Rev. D}\ }\textbf {\bibinfo {volume} {108}},\ \bibinfo {pages}
  {084049} (\bibinfo {year} {2023})},\ \Eprint
  {http://arxiv.org/abs/2304.04551} {arXiv:2304.04551 [gr-qc]} \BibitemShut
  {NoStop}%
\bibitem [{\citenamefont {Cannizzaro}\ \emph {et~al.}(2024)\citenamefont
  {Cannizzaro}, \citenamefont {De~Luca},\ and\ \citenamefont
  {Pani}}]{Cannizzaro:2024fpz}%
  \BibitemOpen
  \bibfield  {author} {\bibinfo {author} {\bibfnamefont {E.}~\bibnamefont
  {Cannizzaro}}, \bibinfo {author} {\bibfnamefont {V.}~\bibnamefont {De~Luca}},
  \ and\ \bibinfo {author} {\bibfnamefont {P.}~\bibnamefont {Pani}},\
  }\bibfield  {title} {\enquote {\bibinfo {title} {{Tidal deformability of
  black holes surrounded by thin accretion disks}},}\ }\href {\doibase
  10.1103/PhysRevD.110.123004} {\bibfield  {journal} {\bibinfo  {journal}
  {Phys. Rev. D}\ }\textbf {\bibinfo {volume} {110}},\ \bibinfo {pages}
  {123004} (\bibinfo {year} {2024})},\ \Eprint
  {http://arxiv.org/abs/2408.14208} {arXiv:2408.14208 [astro-ph.HE]}
  \BibitemShut {NoStop}%
\bibitem [{\citenamefont {Barranco}\ \emph {et~al.}(2012)\citenamefont
  {Barranco}, \citenamefont {Bernal}, \citenamefont {Degollado}, \citenamefont
  {Diez-Tejedor}, \citenamefont {Megevand}, \citenamefont {Alcubierre},
  \citenamefont {Nunez},\ and\ \citenamefont {Sarbach}}]{Barranco:2012qs}%
  \BibitemOpen
  \bibfield  {author} {\bibinfo {author} {\bibfnamefont {J.}~\bibnamefont
  {Barranco}}, \bibinfo {author} {\bibfnamefont {A.}~\bibnamefont {Bernal}},
  \bibinfo {author} {\bibfnamefont {J.~C.}\ \bibnamefont {Degollado}}, \bibinfo
  {author} {\bibfnamefont {A.}~\bibnamefont {Diez-Tejedor}}, \bibinfo {author}
  {\bibfnamefont {M.}~\bibnamefont {Megevand}}, \bibinfo {author}
  {\bibfnamefont {M.}~\bibnamefont {Alcubierre}}, \bibinfo {author}
  {\bibfnamefont {D.}~\bibnamefont {Nunez}}, \ and\ \bibinfo {author}
  {\bibfnamefont {O.}~\bibnamefont {Sarbach}},\ }\bibfield  {title} {\enquote
  {\bibinfo {title} {{Schwarzschild black holes can wear scalar wigs}},}\
  }\href {\doibase 10.1103/PhysRevLett.109.081102} {\bibfield  {journal}
  {\bibinfo  {journal} {Phys. Rev. Lett.}\ }\textbf {\bibinfo {volume} {109}},\
  \bibinfo {pages} {081102} (\bibinfo {year} {2012})},\ \Eprint
  {http://arxiv.org/abs/1207.2153} {arXiv:1207.2153 [gr-qc]} \BibitemShut
  {NoStop}%
\bibitem [{\citenamefont {Cardoso}\ \emph
  {et~al.}(2022{\natexlab{a}})\citenamefont {Cardoso}, \citenamefont {Ikeda},
  \citenamefont {Zhong},\ and\ \citenamefont {Zilh\~ao}}]{Cardoso:2022vpj}%
  \BibitemOpen
  \bibfield  {author} {\bibinfo {author} {\bibfnamefont {V.}~\bibnamefont
  {Cardoso}}, \bibinfo {author} {\bibfnamefont {T.}~\bibnamefont {Ikeda}},
  \bibinfo {author} {\bibfnamefont {Z.}~\bibnamefont {Zhong}}, \ and\ \bibinfo
  {author} {\bibfnamefont {M.}~\bibnamefont {Zilh\~ao}},\ }\bibfield  {title}
  {\enquote {\bibinfo {title} {{Piercing of a boson star by a black hole}},}\
  }\href {\doibase 10.1103/PhysRevD.106.044030} {\bibfield  {journal} {\bibinfo
   {journal} {Phys. Rev. D}\ }\textbf {\bibinfo {volume} {106}},\ \bibinfo
  {pages} {044030} (\bibinfo {year} {2022}{\natexlab{a}})},\ \Eprint
  {http://arxiv.org/abs/2206.00021} {arXiv:2206.00021 [gr-qc]} \BibitemShut
  {NoStop}%
\bibitem [{\citenamefont {Cardoso}\ \emph
  {et~al.}(2022{\natexlab{b}})\citenamefont {Cardoso} \emph
  {et~al.}}]{Cardoso:2022nzc}%
  \BibitemOpen
  \bibfield  {author} {\bibinfo {author} {\bibfnamefont {V.}~\bibnamefont
  {Cardoso}} \emph {et~al.},\ }\bibfield  {title} {\enquote {\bibinfo {title}
  {{Parasitic black holes: The swallowing of a fuzzy dark matter soliton}},}\
  }\href {\doibase 10.1103/PhysRevD.106.L121302} {\bibfield  {journal}
  {\bibinfo  {journal} {Phys. Rev. D}\ }\textbf {\bibinfo {volume} {106}},\
  \bibinfo {pages} {L121302} (\bibinfo {year} {2022}{\natexlab{b}})},\ \Eprint
  {http://arxiv.org/abs/2207.09469} {arXiv:2207.09469 [gr-qc]} \BibitemShut
  {NoStop}%
\bibitem [{\citenamefont {Arvanitaki}\ \emph {et~al.}(2010)\citenamefont
  {Arvanitaki} \emph {et~al.}}]{Arvanitaki:2009fg}%
  \BibitemOpen
  \bibfield  {author} {\bibinfo {author} {\bibfnamefont {A.}~\bibnamefont
  {Arvanitaki}} \emph {et~al.},\ }\bibfield  {title} {\enquote {\bibinfo
  {title} {{String Axiverse}},}\ }\href {\doibase 10.1103/PhysRevD.81.123530}
  {\bibfield  {journal} {\bibinfo  {journal} {Phys. Rev. D}\ }\textbf {\bibinfo
  {volume} {81}},\ \bibinfo {pages} {123530} (\bibinfo {year} {2010})},\
  \Eprint {http://arxiv.org/abs/0905.4720} {arXiv:0905.4720 [hep-th]}
  \BibitemShut {NoStop}%
\bibitem [{\citenamefont {Brito}\ \emph
  {et~al.}(2015{\natexlab{a}})\citenamefont {Brito}, \citenamefont {Cardoso},\
  and\ \citenamefont {Pani}}]{Brito:2014wla}%
  \BibitemOpen
  \bibfield  {author} {\bibinfo {author} {\bibfnamefont {R.}~\bibnamefont
  {Brito}}, \bibinfo {author} {\bibfnamefont {V.}~\bibnamefont {Cardoso}}, \
  and\ \bibinfo {author} {\bibfnamefont {P.}~\bibnamefont {Pani}},\ }\bibfield
  {title} {\enquote {\bibinfo {title} {{Black holes as particle detectors:
  evolution of superradiant instabilities}},}\ }\href {\doibase
  10.1088/0264-9381/32/13/134001} {\bibfield  {journal} {\bibinfo  {journal}
  {Class. Quant. Grav.}\ }\textbf {\bibinfo {volume} {32}},\ \bibinfo {pages}
  {134001} (\bibinfo {year} {2015}{\natexlab{a}})},\ \Eprint
  {http://arxiv.org/abs/1411.0686} {arXiv:1411.0686 [gr-qc]} \BibitemShut
  {NoStop}%
\bibitem [{\citenamefont {East}\ and\ \citenamefont
  {Pretorius}(2017)}]{East:2017}%
  \BibitemOpen
  \bibfield  {author} {\bibinfo {author} {\bibfnamefont {W.~E.}\ \bibnamefont
  {East}}\ and\ \bibinfo {author} {\bibfnamefont {F.}~\bibnamefont
  {Pretorius}},\ }\bibfield  {title} {\enquote {\bibinfo {title} {{Superradiant
  Instability and Backreaction of Massive Vector Fields around Kerr Black
  Holes}},}\ }\href {\doibase 10.1103/PhysRevLett.119.041101} {\bibfield
  {journal} {\bibinfo  {journal} {Phys. Rev. Lett.}\ }\textbf {\bibinfo
  {volume} {119}},\ \bibinfo {pages} {041101} (\bibinfo {year} {2017})},\
  \Eprint {http://arxiv.org/abs/1704.04791} {arXiv:1704.04791 [gr-qc]}
  \BibitemShut {NoStop}%
\bibitem [{\citenamefont {East}(2018)}]{East:2018glu}%
  \BibitemOpen
  \bibfield  {author} {\bibinfo {author} {\bibfnamefont {W.~E.}\ \bibnamefont
  {East}},\ }\bibfield  {title} {\enquote {\bibinfo {title} {{Massive Boson
  Superradiant Instability of Black Holes: Nonlinear Growth, Saturation, and
  Gravitational Radiation}},}\ }\href {\doibase 10.1103/PhysRevLett.121.131104}
  {\bibfield  {journal} {\bibinfo  {journal} {Phys. Rev. Lett.}\ }\textbf
  {\bibinfo {volume} {121}},\ \bibinfo {pages} {131104} (\bibinfo {year}
  {2018})},\ \Eprint {http://arxiv.org/abs/1807.00043} {arXiv:1807.00043
  [gr-qc]} \BibitemShut {NoStop}%
\bibitem [{\citenamefont {Brito}\ \emph
  {et~al.}(2015{\natexlab{b}})\citenamefont {Brito}, \citenamefont {Cardoso},\
  and\ \citenamefont {Pani}}]{Brito:2015oca}%
  \BibitemOpen
  \bibfield  {author} {\bibinfo {author} {\bibfnamefont {R.}~\bibnamefont
  {Brito}}, \bibinfo {author} {\bibfnamefont {V.}~\bibnamefont {Cardoso}}, \
  and\ \bibinfo {author} {\bibfnamefont {P.}~\bibnamefont {Pani}},\ }\bibfield
  {title} {\enquote {\bibinfo {title} {{Superradiance}: {New Frontiers in Black
  Hole Physics}},}\ }\href {\doibase 10.1007/978-3-319-19000-6} {\bibfield
  {journal} {\bibinfo  {journal} {Lect. Notes Phys.}\ }\textbf {\bibinfo
  {volume} {906}},\ \bibinfo {pages} {pp.1--237} (\bibinfo {year}
  {2015}{\natexlab{b}})},\ \Eprint {http://arxiv.org/abs/1501.06570}
  {arXiv:1501.06570 [gr-qc]} \BibitemShut {NoStop}%
\bibitem [{\citenamefont {Arvanitaki}\ and\ \citenamefont
  {Dubovsky}(2011)}]{Arvanitaki:2010sy}%
  \BibitemOpen
  \bibfield  {author} {\bibinfo {author} {\bibfnamefont {A.}~\bibnamefont
  {Arvanitaki}}\ and\ \bibinfo {author} {\bibfnamefont {S.}~\bibnamefont
  {Dubovsky}},\ }\bibfield  {title} {\enquote {\bibinfo {title} {{Exploring the
  String Axiverse with Precision Black Hole Physics}},}\ }\href {\doibase
  10.1103/PhysRevD.83.044026} {\bibfield  {journal} {\bibinfo  {journal} {Phys.
  Rev. D}\ }\textbf {\bibinfo {volume} {83}},\ \bibinfo {pages} {044026}
  (\bibinfo {year} {2011})},\ \Eprint {http://arxiv.org/abs/1004.3558}
  {arXiv:1004.3558 [hep-th]} \BibitemShut {NoStop}%
\bibitem [{\citenamefont {Baumann}\ \emph
  {et~al.}(2019{\natexlab{b}})\citenamefont {Baumann} \emph
  {et~al.}}]{Baumann:2019eav}%
  \BibitemOpen
  \bibfield  {author} {\bibinfo {author} {\bibfnamefont {D.}~\bibnamefont
  {Baumann}} \emph {et~al.},\ }\bibfield  {title} {\enquote {\bibinfo {title}
  {{The Spectra of Gravitational Atoms}},}\ }\href {\doibase
  10.1088/1475-7516/2019/12/006} {\bibfield  {journal} {\bibinfo  {journal}
  {JCAP}\ }\textbf {\bibinfo {volume} {12}},\ \bibinfo {pages} {006} (\bibinfo
  {year} {2019}{\natexlab{b}})},\ \Eprint {http://arxiv.org/abs/1908.10370}
  {arXiv:1908.10370 [gr-qc]} \BibitemShut {NoStop}%
\bibitem [{\citenamefont {Baumann}\ \emph {et~al.}(2020)\citenamefont {Baumann}
  \emph {et~al.}}]{Baumann:2019ztm}%
  \BibitemOpen
  \bibfield  {author} {\bibinfo {author} {\bibfnamefont {D.}~\bibnamefont
  {Baumann}} \emph {et~al.},\ }\bibfield  {title} {\enquote {\bibinfo {title}
  {{Gravitational Collider Physics}},}\ }\href {\doibase
  10.1103/PhysRevD.101.083019} {\bibfield  {journal} {\bibinfo  {journal}
  {Phys. Rev. D}\ }\textbf {\bibinfo {volume} {101}},\ \bibinfo {pages}
  {083019} (\bibinfo {year} {2020})},\ \Eprint
  {http://arxiv.org/abs/1912.04932} {arXiv:1912.04932 [gr-qc]} \BibitemShut
  {NoStop}%
\bibitem [{\citenamefont {Cardoso}\ \emph {et~al.}(2020)\citenamefont
  {Cardoso}, \citenamefont {Duque},\ and\ \citenamefont
  {Ikeda}}]{Cardoso:2020hca}%
  \BibitemOpen
  \bibfield  {author} {\bibinfo {author} {\bibfnamefont {V.}~\bibnamefont
  {Cardoso}}, \bibinfo {author} {\bibfnamefont {F.}~\bibnamefont {Duque}}, \
  and\ \bibinfo {author} {\bibfnamefont {T.}~\bibnamefont {Ikeda}},\ }\bibfield
   {title} {\enquote {\bibinfo {title} {{Tidal effects and disruption in
  superradiant clouds: a numerical investigation}},}\ }\href {\doibase
  10.1103/PhysRevD.101.064054} {\bibfield  {journal} {\bibinfo  {journal}
  {Phys. Rev. D}\ }\textbf {\bibinfo {volume} {101}},\ \bibinfo {pages}
  {064054} (\bibinfo {year} {2020})},\ \Eprint
  {http://arxiv.org/abs/2001.01729} {arXiv:2001.01729 [gr-qc]} \BibitemShut
  {NoStop}%
\bibitem [{\citenamefont {Baumann}\ \emph {et~al.}(2022)\citenamefont {Baumann}
  \emph {et~al.}}]{Baumann:2021fkf}%
  \BibitemOpen
  \bibfield  {author} {\bibinfo {author} {\bibfnamefont {D.}~\bibnamefont
  {Baumann}} \emph {et~al.},\ }\bibfield  {title} {\enquote {\bibinfo {title}
  {{Ionization of gravitational atoms}},}\ }\href {\doibase
  10.1103/PhysRevD.105.115036} {\bibfield  {journal} {\bibinfo  {journal}
  {Phys. Rev. D}\ }\textbf {\bibinfo {volume} {105}},\ \bibinfo {pages}
  {115036} (\bibinfo {year} {2022})},\ \Eprint
  {http://arxiv.org/abs/2112.14777} {arXiv:2112.14777 [gr-qc]} \BibitemShut
  {NoStop}%
\bibitem [{\citenamefont {Tomaselli}\ \emph {et~al.}(2023)\citenamefont
  {Tomaselli}, \citenamefont {Spieksma},\ and\ \citenamefont
  {Bertone}}]{Tomaselli:2023ysb}%
  \BibitemOpen
  \bibfield  {author} {\bibinfo {author} {\bibfnamefont {G.~M.}\ \bibnamefont
  {Tomaselli}}, \bibinfo {author} {\bibfnamefont {T.~F.~M.}\ \bibnamefont
  {Spieksma}}, \ and\ \bibinfo {author} {\bibfnamefont {G.}~\bibnamefont
  {Bertone}},\ }\bibfield  {title} {\enquote {\bibinfo {title} {{Dynamical
  friction in gravitational atoms}},}\ }\href {\doibase
  10.1088/1475-7516/2023/07/070} {\bibfield  {journal} {\bibinfo  {journal}
  {JCAP}\ }\textbf {\bibinfo {volume} {07}},\ \bibinfo {pages} {070} (\bibinfo
  {year} {2023})},\ \Eprint {http://arxiv.org/abs/2305.15460} {arXiv:2305.15460
  [gr-qc]} \BibitemShut {NoStop}%
\bibitem [{\citenamefont {Brito}\ and\ \citenamefont
  {Shah}(2023)}]{Brito:2023pyl}%
  \BibitemOpen
  \bibfield  {author} {\bibinfo {author} {\bibfnamefont {R.}~\bibnamefont
  {Brito}}\ and\ \bibinfo {author} {\bibfnamefont {S.}~\bibnamefont {Shah}},\
  }\bibfield  {title} {\enquote {\bibinfo {title} {{Extreme mass-ratio
  inspirals into black holes surrounded by scalar clouds}},}\ }\href {\doibase
  10.1103/PhysRevD.108.084019} {\bibfield  {journal} {\bibinfo  {journal}
  {Phys. Rev. D}\ }\textbf {\bibinfo {volume} {108}},\ \bibinfo {pages}
  {084019} (\bibinfo {year} {2023})},\ \Eprint
  {http://arxiv.org/abs/2307.16093} {arXiv:2307.16093 [gr-qc]} \BibitemShut
  {NoStop}%
\bibitem [{\citenamefont {Duque}\ \emph {et~al.}(2024)\citenamefont {Duque},
  \citenamefont {Macedo}, \citenamefont {Vicente},\ and\ \citenamefont
  {Cardoso}}]{Duque:2023seg}%
  \BibitemOpen
  \bibfield  {author} {\bibinfo {author} {\bibfnamefont {F.}~\bibnamefont
  {Duque}}, \bibinfo {author} {\bibfnamefont {C.~F.~B.}\ \bibnamefont
  {Macedo}}, \bibinfo {author} {\bibfnamefont {R.}~\bibnamefont {Vicente}}, \
  and\ \bibinfo {author} {\bibfnamefont {V.}~\bibnamefont {Cardoso}},\
  }\bibfield  {title} {\enquote {\bibinfo {title} {{Extreme-Mass-Ratio
  Inspirals in Ultralight Dark Matter}},}\ }\href {\doibase
  10.1103/PhysRevLett.133.121404} {\bibfield  {journal} {\bibinfo  {journal}
  {Phys. Rev. Lett.}\ }\textbf {\bibinfo {volume} {133}},\ \bibinfo {pages}
  {121404} (\bibinfo {year} {2024})},\ \Eprint
  {http://arxiv.org/abs/2312.06767} {arXiv:2312.06767 [gr-qc]} \BibitemShut
  {NoStop}%
\bibitem [{\citenamefont {Tomaselli}\ \emph {et~al.}(2024)\citenamefont
  {Tomaselli}, \citenamefont {Spieksma},\ and\ \citenamefont
  {Bertone}}]{Tomaselli:2024bdd}%
  \BibitemOpen
  \bibfield  {author} {\bibinfo {author} {\bibfnamefont {G.~M.}\ \bibnamefont
  {Tomaselli}}, \bibinfo {author} {\bibfnamefont {T.~F.~M.}\ \bibnamefont
  {Spieksma}}, \ and\ \bibinfo {author} {\bibfnamefont {G.}~\bibnamefont
  {Bertone}},\ }\bibfield  {title} {\enquote {\bibinfo {title} {{Resonant
  history of gravitational atoms in black hole binaries}},}\ }\href {\doibase
  10.1103/PhysRevD.110.064048} {\bibfield  {journal} {\bibinfo  {journal}
  {Phys. Rev. D}\ }\textbf {\bibinfo {volume} {110}},\ \bibinfo {pages}
  {064048} (\bibinfo {year} {2024})},\ \Eprint
  {http://arxiv.org/abs/2403.03147} {arXiv:2403.03147 [gr-qc]} \BibitemShut
  {NoStop}%
\bibitem [{\citenamefont {Bo\v{s}kovi\'c}\ \emph {et~al.}(2024)\citenamefont
  {Bo\v{s}kovi\'c}, \citenamefont {Koschnitzke},\ and\ \citenamefont
  {Porto}}]{Boskovic:2024fga}%
  \BibitemOpen
  \bibfield  {author} {\bibinfo {author} {\bibfnamefont {M.}~\bibnamefont
  {Bo\v{s}kovi\'c}}, \bibinfo {author} {\bibfnamefont {M.}~\bibnamefont
  {Koschnitzke}}, \ and\ \bibinfo {author} {\bibfnamefont {R.~A.}\ \bibnamefont
  {Porto}},\ }\bibfield  {title} {\enquote {\bibinfo {title} {{Signatures of
  Ultralight Bosons in the Orbital Eccentricity of Binary Black Holes}},}\
  }\href {\doibase 10.1103/PhysRevLett.133.121401} {\bibfield  {journal}
  {\bibinfo  {journal} {Phys. Rev. Lett.}\ }\textbf {\bibinfo {volume} {133}},\
  \bibinfo {pages} {121401} (\bibinfo {year} {2024})},\ \Eprint
  {http://arxiv.org/abs/2403.02415} {arXiv:2403.02415 [gr-qc]} \BibitemShut
  {NoStop}%
\bibitem [{git()}]{github}%
  \BibitemOpen
  \href@noop {} {}\bibinfo {howpublished}
  {\url{https://github.com/richbrito/Tidal_Grav_Atoms/}}\BibitemShut {NoStop}%
\bibitem [{\citenamefont {Damour}\ and\ \citenamefont
  {Nagar}(2009)}]{Damour:2009vw}%
  \BibitemOpen
  \bibfield  {author} {\bibinfo {author} {\bibfnamefont {T.}~\bibnamefont
  {Damour}}\ and\ \bibinfo {author} {\bibfnamefont {A.}~\bibnamefont {Nagar}},\
  }\bibfield  {title} {\enquote {\bibinfo {title} {{Relativistic tidal
  properties of neutron stars}},}\ }\href {\doibase 10.1103/PhysRevD.80.084035}
  {\bibfield  {journal} {\bibinfo  {journal} {Phys. Rev. D}\ }\textbf {\bibinfo
  {volume} {80}},\ \bibinfo {pages} {084035} (\bibinfo {year} {2009})},\
  \Eprint {http://arxiv.org/abs/0906.0096} {arXiv:0906.0096 [gr-qc]}
  \BibitemShut {NoStop}%
\bibitem [{\citenamefont {Thorne}(1980)}]{Thorne:1980ru}%
  \BibitemOpen
  \bibfield  {author} {\bibinfo {author} {\bibfnamefont {K.~S.}\ \bibnamefont
  {Thorne}},\ }\bibfield  {title} {\enquote {\bibinfo {title} {{Multipole
  Expansions of Gravitational Radiation}},}\ }\href {\doibase
  10.1103/RevModPhys.52.299} {\bibfield  {journal} {\bibinfo  {journal} {Rev.
  Mod. Phys.}\ }\textbf {\bibinfo {volume} {52}},\ \bibinfo {pages} {299--339}
  (\bibinfo {year} {1980})}\BibitemShut {NoStop}%
\bibitem [{\citenamefont {Zhang}(1986)}]{Zhang:1986}%
  \BibitemOpen
  \bibfield  {author} {\bibinfo {author} {\bibfnamefont {X.~H.}\ \bibnamefont
  {Zhang}},\ }\bibfield  {title} {\enquote {\bibinfo {title} {{Multipole
  expansions of the general-relativistic gravitational field of the external
  universe}},}\ }\href {\doibase 10.1103/PhysRevD.34.991} {\bibfield  {journal}
  {\bibinfo  {journal} {Phys. Rev. D}\ }\textbf {\bibinfo {volume} {34}},\
  \bibinfo {pages} {991--1004} (\bibinfo {year} {1986})}\BibitemShut {NoStop}%
\bibitem [{\citenamefont {Mayerson}(2023)}]{Mayerson:2022}%
  \BibitemOpen
  \bibfield  {author} {\bibinfo {author} {\bibfnamefont {D.~R.}\ \bibnamefont
  {Mayerson}},\ }\bibfield  {title} {\enquote {\bibinfo {title} {{Gravitational
  multipoles in general stationary spacetimes}},}\ }\href {\doibase
  10.21468/SciPostPhys.15.4.154} {\bibfield  {journal} {\bibinfo  {journal}
  {SciPost Phys.}\ }\textbf {\bibinfo {volume} {15}},\ \bibinfo {pages} {154}
  (\bibinfo {year} {2023})},\ \Eprint {http://arxiv.org/abs/2210.05687}
  {arXiv:2210.05687 [gr-qc]} \BibitemShut {NoStop}%
\bibitem [{\citenamefont {Herdeiro}\ and\ \citenamefont
  {Radu}(2014)}]{Herdeiro:2014goa}%
  \BibitemOpen
  \bibfield  {author} {\bibinfo {author} {\bibfnamefont {C.~A.~R.}\
  \bibnamefont {Herdeiro}}\ and\ \bibinfo {author} {\bibfnamefont
  {E.}~\bibnamefont {Radu}},\ }\bibfield  {title} {\enquote {\bibinfo {title}
  {{Kerr black holes with scalar hair}},}\ }\href {\doibase
  10.1103/PhysRevLett.112.221101} {\bibfield  {journal} {\bibinfo  {journal}
  {Phys. Rev. Lett.}\ }\textbf {\bibinfo {volume} {112}},\ \bibinfo {pages}
  {221101} (\bibinfo {year} {2014})},\ \Eprint {http://arxiv.org/abs/1403.2757}
  {arXiv:1403.2757 [gr-qc]} \BibitemShut {NoStop}%
\bibitem [{\citenamefont {Yoshino}\ and\ \citenamefont
  {Kodama}(2014)}]{Yoshino:2013}%
  \BibitemOpen
  \bibfield  {author} {\bibinfo {author} {\bibfnamefont {H.}~\bibnamefont
  {Yoshino}}\ and\ \bibinfo {author} {\bibfnamefont {H.}~\bibnamefont
  {Kodama}},\ }\bibfield  {title} {\enquote {\bibinfo {title} {{Gravitational
  radiation from an axion cloud around a black hole: Superradiant phase}},}\
  }\href {\doibase 10.1093/ptep/ptu029} {\bibfield  {journal} {\bibinfo
  {journal} {PTEP}\ }\textbf {\bibinfo {volume} {2014}},\ \bibinfo {pages}
  {043E02} (\bibinfo {year} {2014})},\ \Eprint {http://arxiv.org/abs/1312.2326}
  {arXiv:1312.2326 [gr-qc]} \BibitemShut {NoStop}%
\bibitem [{\citenamefont {Arvanitaki}\ \emph {et~al.}(2015)\citenamefont
  {Arvanitaki}, \citenamefont {Baryakhtar},\ and\ \citenamefont
  {Huang}}]{Arvanitaki:2014wva}%
  \BibitemOpen
  \bibfield  {author} {\bibinfo {author} {\bibfnamefont {A.}~\bibnamefont
  {Arvanitaki}}, \bibinfo {author} {\bibfnamefont {M.}~\bibnamefont
  {Baryakhtar}}, \ and\ \bibinfo {author} {\bibfnamefont {X.}~\bibnamefont
  {Huang}},\ }\bibfield  {title} {\enquote {\bibinfo {title} {{Discovering the
  QCD Axion with Black Holes and Gravitational Waves}},}\ }\href {\doibase
  10.1103/PhysRevD.91.084011} {\bibfield  {journal} {\bibinfo  {journal} {Phys.
  Rev. D}\ }\textbf {\bibinfo {volume} {91}},\ \bibinfo {pages} {084011}
  (\bibinfo {year} {2015})},\ \Eprint {http://arxiv.org/abs/1411.2263}
  {arXiv:1411.2263 [hep-ph]} \BibitemShut {NoStop}%
\bibitem [{\citenamefont {Brito}\ \emph {et~al.}(2017)\citenamefont {Brito}
  \emph {et~al.}}]{Brito:2017zvb}%
  \BibitemOpen
  \bibfield  {author} {\bibinfo {author} {\bibfnamefont {R.}~\bibnamefont
  {Brito}} \emph {et~al.},\ }\bibfield  {title} {\enquote {\bibinfo {title}
  {{Gravitational wave searches for ultralight bosons with LIGO and LISA}},}\
  }\href {\doibase 10.1103/PhysRevD.96.064050} {\bibfield  {journal} {\bibinfo
  {journal} {Phys. Rev. D}\ }\textbf {\bibinfo {volume} {96}},\ \bibinfo
  {pages} {064050} (\bibinfo {year} {2017})},\ \Eprint
  {http://arxiv.org/abs/1706.06311} {arXiv:1706.06311 [gr-qc]} \BibitemShut
  {NoStop}%
\bibitem [{\citenamefont {Detweiler}(1980)}]{Detweiler:1980}%
  \BibitemOpen
  \bibfield  {author} {\bibinfo {author} {\bibfnamefont {S.~L.}\ \bibnamefont
  {Detweiler}},\ }\bibfield  {title} {\enquote {\bibinfo {title} {{Klein-Gordon
  equation and rotating black holes}},}\ }\href {\doibase
  10.1103/PhysRevD.22.2323} {\bibfield  {journal} {\bibinfo  {journal} {Phys.
  Rev. D}\ }\textbf {\bibinfo {volume} {22}},\ \bibinfo {pages} {2323--2326}
  (\bibinfo {year} {1980})}\BibitemShut {NoStop}%
\bibitem [{\citenamefont {Dolan}(2007)}]{Dolan:2007mj}%
  \BibitemOpen
  \bibfield  {author} {\bibinfo {author} {\bibfnamefont {S.~R.}\ \bibnamefont
  {Dolan}},\ }\bibfield  {title} {\enquote {\bibinfo {title} {{Instability of
  the massive Klein-Gordon field on the Kerr spacetime}},}\ }\href {\doibase
  10.1103/PhysRevD.76.084001} {\bibfield  {journal} {\bibinfo  {journal} {Phys.
  Rev. D}\ }\textbf {\bibinfo {volume} {76}},\ \bibinfo {pages} {084001}
  (\bibinfo {year} {2007})},\ \Eprint {http://arxiv.org/abs/0705.2880}
  {arXiv:0705.2880 [gr-qc]} \BibitemShut {NoStop}%
\bibitem [{\citenamefont {Berti}\ \emph {et~al.}(2006)\citenamefont {Berti},
  \citenamefont {Cardoso},\ and\ \citenamefont {Casals}}]{Berti:2005gp}%
  \BibitemOpen
  \bibfield  {author} {\bibinfo {author} {\bibfnamefont {E.}~\bibnamefont
  {Berti}}, \bibinfo {author} {\bibfnamefont {V.}~\bibnamefont {Cardoso}}, \
  and\ \bibinfo {author} {\bibfnamefont {M.}~\bibnamefont {Casals}},\
  }\bibfield  {title} {\enquote {\bibinfo {title} {{Eigenvalues and
  eigenfunctions of spin-weighted spheroidal harmonics in four and higher
  dimensions}},}\ }\href {\doibase 10.1103/PhysRevD.73.109902} {\bibfield
  {journal} {\bibinfo  {journal} {Phys. Rev. D}\ }\textbf {\bibinfo {volume}
  {73}},\ \bibinfo {pages} {024013} (\bibinfo {year} {2006})},\ \bibinfo {note}
  {[Erratum: Phys.Rev.D 73, 109902 (2006)]},\ \Eprint
  {http://arxiv.org/abs/gr-qc/0511111} {arXiv:gr-qc/0511111} \BibitemShut
  {NoStop}%
\bibitem [{\citenamefont {Herdeiro}\ and\ \citenamefont
  {Radu}(2017)}]{Herdeiro:2017phl}%
  \BibitemOpen
  \bibfield  {author} {\bibinfo {author} {\bibfnamefont {C.~A.~R.}\
  \bibnamefont {Herdeiro}}\ and\ \bibinfo {author} {\bibfnamefont
  {E.}~\bibnamefont {Radu}},\ }\bibfield  {title} {\enquote {\bibinfo {title}
  {{Dynamical Formation of Kerr Black Holes with Synchronized Hair: An Analytic
  Model}},}\ }\href {\doibase 10.1103/PhysRevLett.119.261101} {\bibfield
  {journal} {\bibinfo  {journal} {Phys. Rev. Lett.}\ }\textbf {\bibinfo
  {volume} {119}},\ \bibinfo {pages} {261101} (\bibinfo {year} {2017})},\
  \Eprint {http://arxiv.org/abs/1706.06597} {arXiv:1706.06597 [gr-qc]}
  \BibitemShut {NoStop}%
\bibitem [{\citenamefont {Herdeiro}\ \emph {et~al.}(2022)\citenamefont
  {Herdeiro}, \citenamefont {Radu},\ and\ \citenamefont
  {Santos}}]{Herdeiro:2021znw}%
  \BibitemOpen
  \bibfield  {author} {\bibinfo {author} {\bibfnamefont {C.~A.~R.}\
  \bibnamefont {Herdeiro}}, \bibinfo {author} {\bibfnamefont {E.}~\bibnamefont
  {Radu}}, \ and\ \bibinfo {author} {\bibfnamefont {N.~M.}\ \bibnamefont
  {Santos}},\ }\bibfield  {title} {\enquote {\bibinfo {title} {{A bound on
  energy extraction (and hairiness) from superradiance}},}\ }\href {\doibase
  10.1016/j.physletb.2021.136835} {\bibfield  {journal} {\bibinfo  {journal}
  {Phys. Lett. B}\ }\textbf {\bibinfo {volume} {824}},\ \bibinfo {pages}
  {136835} (\bibinfo {year} {2022})},\ \Eprint
  {http://arxiv.org/abs/2111.03667} {arXiv:2111.03667 [gr-qc]} \BibitemShut
  {NoStop}%
\bibitem [{\citenamefont {Ficarra}\ \emph {et~al.}(2019)\citenamefont
  {Ficarra}, \citenamefont {Pani},\ and\ \citenamefont
  {Witek}}]{Ficarra:2018rfu}%
  \BibitemOpen
  \bibfield  {author} {\bibinfo {author} {\bibfnamefont {G.}~\bibnamefont
  {Ficarra}}, \bibinfo {author} {\bibfnamefont {P.}~\bibnamefont {Pani}}, \
  and\ \bibinfo {author} {\bibfnamefont {H.}~\bibnamefont {Witek}},\ }\bibfield
   {title} {\enquote {\bibinfo {title} {{Impact of multiple modes on the
  black-hole superradiant instability}},}\ }\href {\doibase
  10.1103/PhysRevD.99.104019} {\bibfield  {journal} {\bibinfo  {journal} {Phys.
  Rev. D}\ }\textbf {\bibinfo {volume} {99}},\ \bibinfo {pages} {104019}
  (\bibinfo {year} {2019})},\ \Eprint {http://arxiv.org/abs/1812.02758}
  {arXiv:1812.02758 [gr-qc]} \BibitemShut {NoStop}%
\bibitem [{\citenamefont {Barranco}\ \emph {et~al.}(2017)\citenamefont
  {Barranco}, \citenamefont {Bernal}, \citenamefont {Degollado}, \citenamefont
  {Diez-Tejedor}, \citenamefont {Megevand}, \citenamefont {Nunez},\ and\
  \citenamefont {Sarbach}}]{Barranco:2017aes}%
  \BibitemOpen
  \bibfield  {author} {\bibinfo {author} {\bibfnamefont {J.}~\bibnamefont
  {Barranco}}, \bibinfo {author} {\bibfnamefont {A.}~\bibnamefont {Bernal}},
  \bibinfo {author} {\bibfnamefont {J.~C.}\ \bibnamefont {Degollado}}, \bibinfo
  {author} {\bibfnamefont {A.}~\bibnamefont {Diez-Tejedor}}, \bibinfo {author}
  {\bibfnamefont {M.}~\bibnamefont {Megevand}}, \bibinfo {author}
  {\bibfnamefont {D.}~\bibnamefont {Nunez}}, \ and\ \bibinfo {author}
  {\bibfnamefont {O.}~\bibnamefont {Sarbach}},\ }\bibfield  {title} {\enquote
  {\bibinfo {title} {{Self-gravitating black hole scalar wigs}},}\ }\href
  {\doibase 10.1103/PhysRevD.96.024049} {\bibfield  {journal} {\bibinfo
  {journal} {Phys. Rev. D}\ }\textbf {\bibinfo {volume} {96}},\ \bibinfo
  {pages} {024049} (\bibinfo {year} {2017})},\ \Eprint
  {http://arxiv.org/abs/1704.03450} {arXiv:1704.03450 [gr-qc]} \BibitemShut
  {NoStop}%
\bibitem [{\citenamefont {Annulli}\ \emph {et~al.}(2020)\citenamefont
  {Annulli}, \citenamefont {Cardoso},\ and\ \citenamefont
  {Vicente}}]{Annulli:2020}%
  \BibitemOpen
  \bibfield  {author} {\bibinfo {author} {\bibfnamefont {L.}~\bibnamefont
  {Annulli}}, \bibinfo {author} {\bibfnamefont {V.}~\bibnamefont {Cardoso}}, \
  and\ \bibinfo {author} {\bibfnamefont {R.}~\bibnamefont {Vicente}},\
  }\bibfield  {title} {\enquote {\bibinfo {title} {{Response of ultralight dark
  matter to supermassive black holes and binaries}},}\ }\href {\doibase
  10.1103/PhysRevD.102.063022} {\bibfield  {journal} {\bibinfo  {journal}
  {Phys. Rev. D}\ }\textbf {\bibinfo {volume} {102}},\ \bibinfo {pages}
  {063022} (\bibinfo {year} {2020})},\ \Eprint
  {http://arxiv.org/abs/2009.00012} {arXiv:2009.00012 [gr-qc]} \BibitemShut
  {NoStop}%
\bibitem [{\citenamefont {Pani}\ \emph
  {et~al.}(2015{\natexlab{b}})\citenamefont {Pani}, \citenamefont {Gualtieri},\
  and\ \citenamefont {Ferrari}}]{Pani:2015nua}%
  \BibitemOpen
  \bibfield  {author} {\bibinfo {author} {\bibfnamefont {P.}~\bibnamefont
  {Pani}}, \bibinfo {author} {\bibfnamefont {L.}~\bibnamefont {Gualtieri}}, \
  and\ \bibinfo {author} {\bibfnamefont {V.}~\bibnamefont {Ferrari}},\
  }\bibfield  {title} {\enquote {\bibinfo {title} {{Tidal Love numbers of a
  slowly spinning neutron star}},}\ }\href {\doibase
  10.1103/PhysRevD.92.124003} {\bibfield  {journal} {\bibinfo  {journal} {Phys.
  Rev. D}\ }\textbf {\bibinfo {volume} {92}},\ \bibinfo {pages} {124003}
  (\bibinfo {year} {2015}{\natexlab{b}})},\ \Eprint
  {http://arxiv.org/abs/1509.02171} {arXiv:1509.02171 [gr-qc]} \BibitemShut
  {NoStop}%
\bibitem [{\citenamefont {Pnigouras}\ \emph {et~al.}(2024)\citenamefont
  {Pnigouras}, \citenamefont {Gittins}, \citenamefont {Nanda}, \citenamefont
  {Andersson},\ and\ \citenamefont {Jones}}]{Pnigouras:2022zpx}%
  \BibitemOpen
  \bibfield  {author} {\bibinfo {author} {\bibfnamefont {P.}~\bibnamefont
  {Pnigouras}}, \bibinfo {author} {\bibfnamefont {F.}~\bibnamefont {Gittins}},
  \bibinfo {author} {\bibfnamefont {A.}~\bibnamefont {Nanda}}, \bibinfo
  {author} {\bibfnamefont {N.}~\bibnamefont {Andersson}}, \ and\ \bibinfo
  {author} {\bibfnamefont {D.~I.}\ \bibnamefont {Jones}},\ }\bibfield  {title}
  {\enquote {\bibinfo {title} {{Rotating Love: The dynamical tides of spinning
  Newtonian stars}},}\ }\href {\doibase 10.1093/mnras/stad3593} {\bibfield
  {journal} {\bibinfo  {journal} {Mon. Not. Roy. Astron. Soc.}\ }\textbf
  {\bibinfo {volume} {527}},\ \bibinfo {pages} {8409--8428} (\bibinfo {year}
  {2024})},\ \Eprint {http://arxiv.org/abs/2205.07577} {arXiv:2205.07577
  [gr-qc]} \BibitemShut {NoStop}%
\bibitem [{\citenamefont {Arana}(2023)}]{Arana_thesis}%
  \BibitemOpen
  \bibfield  {author} {\bibinfo {author} {\bibfnamefont {R.}~\bibnamefont
  {Arana}},\ }\emph {\bibinfo {title} {Tidal deformability of gravitational
  atoms}},\ \href
  {https://scholar.tecnico.ulisboa.pt/records/M7zc04RE5lIYmTteN5CNo4IVURpesgJIv7ji}
  {\bibinfo {type} {Master's thesis}},\ \bibinfo  {school} {Instituto Superior
  T\'{e}cnico} (\bibinfo {year} {2023})\BibitemShut {NoStop}%
\bibitem [{\citenamefont {Katagiri}\ \emph
  {et~al.}(2024{\natexlab{a}})\citenamefont {Katagiri}, \citenamefont {Yagi},\
  and\ \citenamefont {Cardoso}}]{Katagiri:2024sep}%
  \BibitemOpen
  \bibfield  {author} {\bibinfo {author} {\bibfnamefont {T.}~\bibnamefont
  {Katagiri}}, \bibinfo {author} {\bibfnamefont {K.}~\bibnamefont {Yagi}}, \
  and\ \bibinfo {author} {\bibfnamefont {V.}~\bibnamefont {Cardoso}},\
  }\bibfield  {title} {\enquote {\bibinfo {title} {{On relativistic dynamical
  tides: subtleties and calibration}},}\ }\href@noop {} {\  (\bibinfo {year}
  {2024}{\natexlab{a}})},\ \Eprint {http://arxiv.org/abs/2409.18034}
  {arXiv:2409.18034 [gr-qc]} \BibitemShut {NoStop}%
\bibitem [{\citenamefont {Katagiri}\ \emph
  {et~al.}(2024{\natexlab{b}})\citenamefont {Katagiri}, \citenamefont
  {Cardoso}, \citenamefont {Ikeda},\ and\ \citenamefont
  {Yagi}}]{Katagiri:2024oct}%
  \BibitemOpen
  \bibfield  {author} {\bibinfo {author} {\bibfnamefont {T.}~\bibnamefont
  {Katagiri}}, \bibinfo {author} {\bibfnamefont {V.}~\bibnamefont {Cardoso}},
  \bibinfo {author} {\bibfnamefont {T.}~\bibnamefont {Ikeda}}, \ and\ \bibinfo
  {author} {\bibfnamefont {K.}~\bibnamefont {Yagi}},\ }\bibfield  {title}
  {\enquote {\bibinfo {title} {{Tidal response beyond vacuum General Relativity
  with a canonical definition}},}\ }\href@noop {} {\  (\bibinfo {year}
  {2024}{\natexlab{b}})},\ \Eprint {http://arxiv.org/abs/2410.02531}
  {arXiv:2410.02531 [gr-qc]} \BibitemShut {NoStop}%
\bibitem [{\citenamefont {Landry}(2017)}]{Landry:2017piv}%
  \BibitemOpen
  \bibfield  {author} {\bibinfo {author} {\bibfnamefont {P.}~\bibnamefont
  {Landry}},\ }\bibfield  {title} {\enquote {\bibinfo {title} {{Tidal
  deformation of a slowly rotating material body: Interior metric and Love
  numbers}},}\ }\href {\doibase 10.1103/PhysRevD.95.124058} {\bibfield
  {journal} {\bibinfo  {journal} {Phys. Rev. D}\ }\textbf {\bibinfo {volume}
  {95}},\ \bibinfo {pages} {124058} (\bibinfo {year} {2017})},\ \Eprint
  {http://arxiv.org/abs/1703.08168} {arXiv:1703.08168 [gr-qc]} \BibitemShut
  {NoStop}%
\bibitem [{\citenamefont {Geroch}(1970)}]{Geroch:1970}%
  \BibitemOpen
  \bibfield  {author} {\bibinfo {author} {\bibfnamefont {R.~P.}\ \bibnamefont
  {Geroch}},\ }\bibfield  {title} {\enquote {\bibinfo {title} {{Multipole
  moments. II. Curved space}},}\ }\href {\doibase 10.1063/1.1665427} {\bibfield
   {journal} {\bibinfo  {journal} {J. Math. Phys.}\ }\textbf {\bibinfo {volume}
  {11}},\ \bibinfo {pages} {2580--2588} (\bibinfo {year} {1970})}\BibitemShut
  {NoStop}%
\bibitem [{\citenamefont {Hansen}(1974)}]{Hansen:1974}%
  \BibitemOpen
  \bibfield  {author} {\bibinfo {author} {\bibfnamefont {R.~O.}\ \bibnamefont
  {Hansen}},\ }\bibfield  {title} {\enquote {\bibinfo {title} {{Multipole
  moments of stationary space-times}},}\ }\href {\doibase 10.1063/1.1666501}
  {\bibfield  {journal} {\bibinfo  {journal} {J. Math. Phys.}\ }\textbf
  {\bibinfo {volume} {15}},\ \bibinfo {pages} {46--52} (\bibinfo {year}
  {1974})}\BibitemShut {NoStop}%
\bibitem [{\citenamefont {G{\"u}rsel}(1983)}]{Gursel:1983}%
  \BibitemOpen
  \bibfield  {author} {\bibinfo {author} {\bibfnamefont {Y.}~\bibnamefont
  {G{\"u}rsel}},\ }\bibfield  {title} {\enquote {\bibinfo {title} {Multipole
  moments for stationary systems: The equivalence of the geroch-hansen
  formulation and the thorne formulation},}\ }\href {\doibase
  10.1007/BF01031881} {\bibfield  {journal} {\bibinfo  {journal} {General
  Relativity and Gravitation}\ }\textbf {\bibinfo {volume} {15}},\ \bibinfo
  {pages} {737--754} (\bibinfo {year} {1983})}\BibitemShut {NoStop}%
\bibitem [{{\relax DLMF}()}]{NIST:DLMF}%
  \BibitemOpen
  {\relax DLMF},\ \href {https://dlmf.nist.gov/} {\enquote {\bibinfo {title}
  {{\it NIST Digital Library of Mathematical Functions}},}\ }\bibinfo
  {howpublished} {\url{https://dlmf.nist.gov/}, Release 1.1.12 of 2023-12-15},\
  \bibinfo {note} {~F.~W.~J. Olver {\it et al}}\BibitemShut {NoStop}%
\bibitem [{\citenamefont {Ancarani}\ and\ \citenamefont
  {Gasaneo}(2008)}]{Ancarani:2008}%
  \BibitemOpen
  \bibfield  {author} {\bibinfo {author} {\bibfnamefont {L.~U.}\ \bibnamefont
  {Ancarani}}\ and\ \bibinfo {author} {\bibfnamefont {G.}~\bibnamefont
  {Gasaneo}},\ }\bibfield  {title} {\enquote {\bibinfo {title} {Derivatives of
  any order of the confluent hypergeometric function {${}_1F_1(a,b,z)$} with
  respect to the parameter {$a$} or {$b$}},}\ }\href {\doibase
  10.1063/1.2939395} {\bibfield  {journal} {\bibinfo  {journal} {Journal of
  Mathematical Physics}\ }\textbf {\bibinfo {volume} {49}},\ \bibinfo {pages}
  {063508} (\bibinfo {year} {2008})}\BibitemShut {NoStop}%
\bibitem [{Wolfram Research({\natexlab{a}})}]{weisstein3}%
  \BibitemOpen
  Wolfram Research,\ \href@noop {} {\enquote {\bibinfo {title} {{\it The
  Mathematical Functions Site}},}\ }\bibinfo {howpublished}
  {\url{http://functions.wolfram.com/07.33.03.0029.01}} (\bibinfo {year}
  {2007}{\natexlab{a}}),\ \bibinfo {note} {last acessed 2024-01-03}\BibitemShut
  {NoStop}%
\bibitem [{Wolfram Research({\natexlab{b}})}]{weisstein1}%
  \BibitemOpen
  Wolfram Research,\ \href@noop {} {\enquote {\bibinfo {title} {{\it The
  Mathematical Functions Site}},}\ }\bibinfo {howpublished}
  {\url{http://functions.wolfram.com/06.06.03.0009.01}} (\bibinfo {year}
  {2001}{\natexlab{b}}),\ \bibinfo {note} {last acessed 2024-01-03}\BibitemShut
  {NoStop}%
\bibitem [{Wolfram Research({\natexlab{c}})}]{weisstein2}%
  \BibitemOpen
  Wolfram Research,\ \href@noop {} {\enquote {\bibinfo {title} {{\it The
  Mathematical Functions Site}},}\ }\bibinfo {howpublished}
  {\url{http://functions.wolfram.com/06.06.26.0015.01}} (\bibinfo {year}
  {2001}{\natexlab{c}}),\ \bibinfo {note} {last acessed 2024-01-03}\BibitemShut
  {NoStop}%
\bibitem [{\citenamefont {Messiah}(1961)}]{Messiah:1961}%
  \BibitemOpen
  \bibfield  {author} {\bibinfo {author} {\bibfnamefont {A.}~\bibnamefont
  {Messiah}},\ }\href@noop {} {\emph {\bibinfo {title} {Quantum Mechanics, Vol.
  2}}},\ Quantum Mechanics\ (\bibinfo  {publisher} {North-Holland},\ \bibinfo
  {year} {1961})\BibitemShut {NoStop}%
\bibitem [{\citenamefont {Spiers}\ \emph {et~al.}(2024)\citenamefont {Spiers},
  \citenamefont {Pound},\ and\ \citenamefont {Wardell}}]{Spiers:2023}%
  \BibitemOpen
  \bibfield  {author} {\bibinfo {author} {\bibfnamefont {A.}~\bibnamefont
  {Spiers}}, \bibinfo {author} {\bibfnamefont {A.}~\bibnamefont {Pound}}, \
  and\ \bibinfo {author} {\bibfnamefont {B.}~\bibnamefont {Wardell}},\
  }\bibfield  {title} {\enquote {\bibinfo {title} {{Second-order perturbations
  of the Schwarzschild spacetime: Practical, covariant, and gauge-invariant
  formalisms}},}\ }\href {\doibase 10.1103/PhysRevD.110.064030} {\bibfield
  {journal} {\bibinfo  {journal} {Phys. Rev. D}\ }\textbf {\bibinfo {volume}
  {110}},\ \bibinfo {pages} {064030} (\bibinfo {year} {2024})},\ \Eprint
  {http://arxiv.org/abs/2306.17847} {arXiv:2306.17847 [gr-qc]} \BibitemShut
  {NoStop}%
\bibitem [{Wolfram Research({\natexlab{d}})}]{weisstein4}%
  \BibitemOpen
  Wolfram Research,\ \href@noop {} {\enquote {\bibinfo {title} {{\it The
  Mathematical Functions Site}},}\ }\bibinfo {howpublished}
  {\url{http://functions.wolfram.com/03.02.26.0003.01}} (\bibinfo {year}
  {2001}{\natexlab{d}}),\ \bibinfo {note} {last acessed 2024-01-03}\BibitemShut
  {NoStop}%
\bibitem [{Wolfram Research({\natexlab{e}})}]{weisstein5}%
  \BibitemOpen
  Wolfram Research,\ \href@noop {} {\enquote {\bibinfo {title} {{\it The
  Mathematical Functions Site}},}\ }\bibinfo {howpublished}
  {\url{http://functions.wolfram.com/03.04.26.0003.01}} (\bibinfo {year}
  {2001}{\natexlab{e}}),\ \bibinfo {note} {last acessed 2024-01-03}\BibitemShut
  {NoStop}%
\bibitem [{\citenamefont {Jeffrey}\ \emph {et~al.}(2007)\citenamefont {Jeffrey}
  \emph {et~al.}}]{Jeffrey:2007}%
  \BibitemOpen
  \bibinfo {editor} {\bibfnamefont {A.}~\bibnamefont {Jeffrey}} \emph
  {et~al.},\ eds.,\ \enquote {\bibinfo {title} {{6--7 - Definite Integrals of
  Special Functions}},}\ in\ \href
  {https://www.sciencedirect.com/science/article/pii/B9780080471112500157}
  {\emph {\bibinfo {booktitle} {Table of Integrals, Series, and Products
  (Seventh Edition)}}}\ (\bibinfo  {publisher} {Academic Press},\ \bibinfo
  {address} {Boston},\ \bibinfo {year} {2007})\ pp.\ \bibinfo {pages}
  {631--857}\BibitemShut {NoStop}%
\end{thebibliography}%

\end{document}